\documentclass[longauth]{aa} 
\usepackage{graphicx}
\usepackage{txfonts}
\usepackage{textcomp}
\usepackage{amssymb, amsmath}
\usepackage[dvipsnames]{xcolor}
\usepackage[utf8]{inputenc}
\usepackage{array}
\usepackage{multirow}
\usepackage{datatool}
\usepackage{natbib}
\usepackage{booktabs}
\usepackage{url}
\bibpunct{(}{)}{;}{a}{}{,}{}
\usepackage[colorlinks=true,allcolors=blue]{hyperref}

\DTLsetseparator{ }
\DTLloaddb{NGC2146}{numbers/NGC2146variables.txt}
\DTLloaddb{NGC2976}{numbers/NGC2976variables.txt}
\DTLloaddb{global2146}{numbers/NGC2146-globalsed.txt}
\DTLloaddb{global2976}{numbers/NGC2976-globalsed.txt}

\usepackage{xcolor}
\newcommand{\reviseII}[1]{#1}
\newcommand{\revise}[1]{#1}
 
\begin{document}
\begin{filecontents*}{numbers/NGC2146variables.txt}
variable	value
COcontr_mean	17
COcontr_std	6
COcontr_global	20


mdens_max_c	2.6
mdens_max_c_er	0.2
mdens_max_c_el	0.2
mdens_mean_c	8.2
mdens_std_c	5.9


mdens_max_w	1.1
mdens_max_w_er	0.7
mdens_max_w_el	0.5
mdens_mean_w	2.2
mdens_std_w	2.5


temp_min_c	25
temp_min_c_er	2
temp_min_c_el	2
temp_max_c	35
temp_max_c_er	1
temp_max_c_el	1
temp_mean_c	30.5
temp_std_c	2.6


temp_min_w	61
temp_min_w_er	11
temp_min_w_el	8
temp_max_w	74
temp_max_w_er	4
temp_max_w_el	5
temp_mean_w	69.3
temp_std_w	2.9


thfr_mean	0.09
thfr_std	0.04
nonthfr_mean	0.91
nonthfr_std	0.04

SFRcalibpow_1	1.5
eSFRcalibpow_1	0.1
SFRcalibpow_2	1.4
eSFRcalibpow_2	0.1
rP_SFR1	0.97
erP_SFR1	0.02
rP_SFR2	0.99
erP_SFR2	0.01
H2calibpow_1	1.2
eH2calibpow_1	0.1
H2calibpow_2	1.0
eH2calibpow_2	0.1
rP_H1	0.97
erP_H1	0.02
rP_H2	0.96
erP_H2	0.02


M_c	3.33
M_c_le	0.72
M_c_re	0.73
M_w	1.71
M_w_le	0.91
M_w_re	1.00
T_c	32.3
T_c_le	1.1
T_c_re	1.1
T_w	87.6
T_w_le	6.9
T_w_re	7.2
beta	1.72
beta_le	0.07
beta_re	0.07
A1	0.016
A1_le	0.004
A1_re	0.004
A2	1.42
A2_le	0.06
A2_re	0.06
alpha	0.94
alpha_le	0.06
alpha_re	0.06
relres_1	0
relres_2	4
thfr6	0.21
thfr6_le	0.06
thfr6_re	0.06
thfr21	0.08
thfr21_le	0.02
thfr21_re	0.02
dustcontr_1	88.4
ffcontr_1	10.0
synccontr_1	1.6
dustcontr_2	47.6
ffcontr_2	42.2
synccontr_2	10.2
\end{filecontents*}

\begin{filecontents*}{numbers/NGC2976variables.txt}
variable	value
COcontr_mean	5
COcontr_std	2
COcontr_global	5


mdens_max_c	1.9
mdens_max_c_er	0.2
mdens_max_c_el	0.2
mdens_mean_c	9.7
mdens_std_c	2.5


mdens_max_w	6.1
mdens_max_w_er	4.5
mdens_max_w_el	5.4
mdens_mean_w	1.0
mdens_std_w	1.0


temp_min_c	22
temp_min_c_er	2
temp_min_c_el	3
temp_max_c	29
temp_max_c_er	1
temp_max_c_el	1
temp_mean_c	26.0
temp_std_c	1.3


temp_min_w	56
temp_min_w_er	11
temp_min_w_el	5
temp_max_w	70
temp_max_w_er	6
temp_max_w_el	7
temp_mean_w	64.0
temp_std_w	2.6


thfr_mean	0.65
thfr_std	0.33
nonthfr_mean	0.35
nonthfr_std	0.33

H2mass	4.1
HImass	3.9
gasmass	0.8
DGRmean	0.006
DGRstd	0.001
globalDGR	0.006

dustHI_a	0.88
dustHI_ae	0.13
dustHI_b	2.40
dustHI_be	0.55
dustHI_r	0.80
dustHI_re	0.07
dustH2_a	2.04
dustH2_ae	0.21
dustH2_b	-2.58
dustH2_be	0.90
dustH2_r	0.88
dustH2_re	0.04
HIH2_a	1.40
HIH2_ae	0.30
HIH2_b	-2.52
HIH2_be	1.87
HIH2_r	0.67
HIH2_re	0.11

variable	value
M_c	7.56
M_c_le	2.28
M_c_re	2.33
M_w	7.35
M_w_le	5.03
M_w_re	5.30
T_c	27.4
T_c_le	0.8
T_c_re	0.8
T_w	85.1
T_w_le	7.9
T_w_re	9.0
beta	1.25
beta_le	0.06
beta_re	0.06
A1	0.003
A1_le	0.001
A1_re	0.001
A2	0.079
A2_le	0.006
A2_re	0.007
alpha	1.14
alpha_le	0.31
alpha_re	0.37
relres_1	4
relres_2	6
thfr6	0.58
thfr6_le	0.27
thfr6_re	0.20
thfr21	0.27
thfr21_le	0.12
thfr21_re	0.07
dustcontr_1	97.6
ffcontr_1	2.3
synccontr_1	0.1
dustcontr_2	87.3
ffcontr_2	12.4
synccontr_2	0.3
SFRcalibpow_1	2.9
eSFRcalibpow_1	1.0
SFRcalibpow_2	2.3
eSFRcalibpow_2	0.4
rP_SFR1	0.88
erP_SFR1	0.04
rP_SFR2	0.89
erP_SFR2	0.04
H2calibpow_1	2.2
eH2calibpow_1	0.9
H2calibpow_2	2.0
eH2calibpow_2	0.6
rP_H1	0.78
erP_H1	0.08
rP_H2	0.76
erP_H2	0.08

\end{filecontents*}

\begin{filecontents*}{numbers/NGC2146-globalsed.txt}
w	 intgflux 	 rms 	 er
70 	193 	3721 	15
100 	231 	3652 	18
160 	160 	921 	13
250 	55 	142 	4
350 	22 	74 	1
500 	7 	53 	1
1 	0.48 	2.2 	0.03
2 	0.13 	0.5 	0.01
138 	0.3 	177 	0.1
24 	15 	5 	1
18 	0.94 	0.05 	0.05
21 	1.11 	0.07 	0.06
6 	0.43 	0.50 	0.02
28 	0.224 	0.0 	0.006
1-mJy	 529
2-mJy	 130
1-mJy-er	 38
2-mJy-er	 9
\end{filecontents*}

\begin{filecontents*}{numbers/NGC2976-globalsed.txt}
w	 intgflux 	 rms 	 er
70 	22 	678 	2
100 	39 	934 	4
160 	43 	360 	4
250 	22 	79 	2
350 	11 	18 	1
500 	5 	14 	0.3
1 	0.45 	2 	0.03
2 	0.08 	0.4 	0.01
138 	0.2 	68 	0.1
24 	1.3 	0.9 	0.1
18 	0.062 	0.04 	0.003
21 	0.070 	0.03 	0.004
6 	0.030 	2.0 	0.005
28 	0.021 	0.0 	0.003
1-mJy	 465
2-mJy	 84
1-mJy-er	 34
2-mJy-er	 8

\end{filecontents*}

 

 \title{Interpreting Millimeter Emission from IMEGIN galaxies NGC~2146 and NGC~2976}
\author{ {G.~Ejlali}\inst{\ref{IPM}}
 \and F.~S.~Tabatabaei \inst{\ref{IPM}}
 \and H.~Roussel \inst{\ref{IAP}}
 \and R.~Adam \inst{\ref{OCA}}
 \and P.~Ade \inst{\ref{Cardiff}}
 \and H.~Ajeddig \inst{\ref{CEA}}
 \and P.~Andr\'e \inst{\ref{CEA}}
 \and E.~Artis \inst{\ref{LPSC},\ref{Garching}}
 \and H.~Aussel \inst{\ref{CEA}}
 \and M.~Baes \inst{\ref{Belguim3}}
 \and A.~Beelen \inst{\ref{LAM}}
 \and A.~Beno\^it \inst{\ref{Neel}}
 \and S.~Berta \inst{\ref{IRAMF}}
 \and L.~Bing \inst{\ref{LAM}}
 \and O.~Bourrion \inst{\ref{LPSC}}
 \and M.~Calvo \inst{\ref{Neel}}
 \and A.~Catalano \inst{\ref{LPSC}}
 \and I.~De Looze \inst{\ref{Belguim3}} 
 \and M.~De~Petris \inst{\ref{Roma}}
 \and F.-X.~D\'esert \inst{\ref{IPAG}}
 \and S.~Doyle \inst{\ref{Cardiff}}
 \and E.~F.~C.~Driessen \inst{\ref{IRAMF}}
 \and F.~Galliano \inst{\ref{CEA}}
 \and A.~Gomez \inst{\ref{CAB}} 
 \and J.~Goupy \inst{\ref{Neel}}
 \and A.~P.~Jones \inst{\ref{IAS}}
 \and C.~Hanser \inst{\ref{LPSC}}
 \and A.~Hughes \inst{\ref{IRAP}}
 \and S.~Katsioli \inst{\ref{Athens_obs},\ref{Athens_univ}}
 \and F.~K\'eruzor\'e \inst{\ref{Argonne}}
 \and C.~Kramer \inst{\ref{IRAMF}}
 \and B.~Ladjelate \inst{\ref{IRAME}} 
 \and G.~Lagache \inst{\ref{LAM}}
 \and S.~Leclercq \inst{\ref{IRAMF}}
 \and J.-F.~Lestrade \inst{\ref{LERMA}}
 \and J.~F.~Mac\'ias-P\'erez \inst{\ref{LPSC}}
 \and S.~C.~Madden \inst{\ref{CEA}}
 \and A.~Maury \inst{\ref{CEA}}
 \and P.~Mauskopf \inst{\ref{Cardiff},\ref{Arizona}}
 \and F.~Mayet \inst{\ref{LPSC}}
 \and A.~Monfardini \inst{\ref{Neel}}
 \and A.~Moyer-Anin \inst{\ref{LPSC}}
 \and M.~Mu\~noz-Echeverr\'ia \inst{\ref{IRAP}}
 \and A.~Nersesian \inst{\ref{Belguim1},\ref{Belguim2}}
 \and L.~Pantoni \inst{\ref{CEA}}
 \and D.~Paradis \inst{\ref{IRAP}}
 \and L.~Perotto \inst{\ref{LPSC}}
 \and G.~Pisano \inst{\ref{Roma}}
 \and N.~Ponthieu \inst{\ref{IPAG}}
 \and V.~Rev\'eret \inst{\ref{CEA}}
 \and A.~J.~Rigby \inst{\ref{Leeds}}
 \and A.~Ritacco \inst{\ref{INAF},\ref{ENS}}
 \and C.~Romero \inst{\ref{Pennsylvanie}}
 \and F.~Ruppin \inst{\ref{IP2I}}
 \and K.~Schuster \inst{\ref{IRAMF}}
 \and A.~Sievers \inst{\ref{IRAME}}
 \and M.~W.~L.~Smith \inst{\ref{Cardiff}}
 \and J.~Tedros \inst{\ref{IRAME}}
 \and C.~Tucker \inst{\ref{Cardiff}}
 \and E.~M.~Xilouris \inst{\ref{Athens_obs}}
 \and N.~Ysard \inst{\ref{IAS},\ref{IRAP}}
 \and R.~Zylka \inst{\ref{IRAMF}}
}\offprints{G. Ejlali, \email{gejlali@ipm.ir}}

\institute{
 Institute for Research in Fundamental Sciences (IPM), School of Astronomy, Tehran, Iran
 \label{IPM}
 \and
 Institut d'Astrophysique de Paris, Sorbonne Universit\'e, CNRS (UMR7095), 98 bis boulevard Arago, 75014 Paris, France
 \label{IAP}
 \and
 Universit\'e C\^ote d'Azur, Observatoire de la C\^ote d'Azur, CNRS, Laboratoire Lagrange, France 
 \label{OCA}
 \and
 School of Physics and Astronomy, Cardiff University, Queen's Buildings, The Parade, Cardiff, CF24 3AA, UK
 \label{Cardiff}
 \and
 Universit\'e Paris-Saclay, Universit\'e Paris Cit\'e, CEA, CNRS, AIM, 91191, Gif-sur-Yvette, France
 \label{CEA}
 \and
 Universit\'e Grenoble Alpes, CNRS, Grenoble INP, LPSC-IN2P3, 38000 Grenoble, France
 \label{LPSC}
 \and	
 Max Planck Institute for Extraterrestrial Physics, Giessenbach-strasse 1, 85748 Garching, Germany
 \label{Garching}
 \and
 University of Ghent, Department of Physics and Astronomy, Ghent, Belgium
 \label{Belguim3}
 \and
 Aix Marseille Univ, CNRS, CNES, LAM (Laboratoire d'Astrophysique de Marseille), Marseille, France
 \label{LAM}
 \and
 Universit\'e Grenoble Alpes, CNRS, Institut N\'eel, France
 \label{Neel}
 \and
 Institut de RadioAstronomie Millim\'etrique (IRAM), Grenoble, France
 \label{IRAMF}
 \and 
 Dipartimento di Fisica, Sapienza Universit\`a di Roma, Piazzale Aldo Moro 5, I-00185 Roma, Italy
 \label{Roma}
 \and
 Univ. Grenoble Alpes, CNRS, IPAG, 38000 Grenoble, France
 \label{IPAG}
 \and
 Centro de Astrobiolog\'ia (CSIC-INTA), Torrej\'on de Ardoz, 28850 Madrid, Spain
 \label{CAB}
 \and
 Institut d'Astrophysique Spatiale (IAS), CNRS, Universit\'e Paris Sud, Orsay, France
 \label{IAS}
 \and
 IRAP, Universit\'e de Toulouse, CNRS, UPS, IRAP, Toulouse Cedex 4, France
 \label{IRAP}
 \and
 National Observatory of Athens, Institute for Astronomy, Astrophysics, Space Applications and Remote Sensing, Ioannou Metaxa and Vasileos Pavlou GR-15236, Athens, Greece
 \label{Athens_obs}
 \and
 Department of Astrophysics, Astronomy \& Mechanics, Faculty of Physics, University of Athens, Panepistimiopolis, GR-15784 Zografos, Athens, Greece
 \label{Athens_univ}
 \and
 High Energy Physics Division, Argonne National Laboratory, 9700 South Cass Avenue, Lemont, IL 60439, USA
 \label{Argonne}
 \and 
 Instituto de Radioastronom\'ia Milim\'etrica (IRAM), Granada, Spain
 \label{IRAME}
 \and
 LERMA, Observatoire de Paris, PSL Research Univ., CNRS, Sorbonne Univ., UPMC, 75014 Paris, France 
 \label{LERMA}
 \and
 School of Earth and Space Exploration and Department of Physics, Arizona State University, Tempe, AZ 85287, USA
 \label{Arizona}
 \and
 STAR Institute, Universit\'e de Li{\`e}ge, Quartier Agora, All\'ee du six Aout 19c, B-4000 Liege, Belgium
 \label{Belguim1}
 \and
 Sterrenkundig Observatorium Universiteit Gent, Krijgslaan 281 S9, B-9000 Gent, Belgium
 \label{Belguim2}
 \and
 School of Physics and Astronomy, University of Leeds, Leeds LS2 9JT, UK
 \label{Leeds}
 \and
 INAF-Osservatorio Astronomico di Cagliari, Via della Scienza 5, 09047 Selargius, IT
 \label{INAF}
 \and 
 Laboratoire de Physique de l'\'Ecole Normale Sup\'erieure, ENS, PSL Research University, CNRS, Sorbonne Universit\'e, Universit\'e de Paris, 75005 Paris, France
 \label{ENS}
 \and 
 Department of Physics and Astronomy, University of Pennsylvania, 209 South 33rd Street, Philadelphia, PA, 19104, USA
 \label{Pennsylvanie}
 \and
 University of Lyon, UCB Lyon 1, CNRS/IN2P3, IP2I, 69622 Villeurbanne, France
 \label{IP2I}
} 

\abstract{The millimeter continuum emission from galaxies provides important information about cold dust, its distribution, heating, and role in their InterStellar Medium (ISM). This emission also carries an unknown portion of the free-free and synchrotron radiation. \revise{The IRAM 30\,m Guaranteed Time Large Project, Interpreting Millimeter Emission of Galaxies with IRAM and NIKA2 (IMEGIN) provides a unique opportunity to study the origin of the millimeter emission on angular resolutions of $<18\arcsec$ in a sample of nearby galaxies. As a pilot study, we present millimeter observations of two IMEGIN galaxies, NGC~2146 (starburst) and NGC~2976 (peculiar dwarf) at 1.15\,mm and 2\,mm}. \revise{Combined with the data taken with \textit{Spitzer}, \textit{Herschel}, Plank, WSRT, and the 100\,m Effelsberg telescopes, we model the infrared-to-radio Spectral Energy Distribution (SED) of these galaxies, both globally and at resolved scales, using a Bayesian approach to 1) dissect different components of the millimeter emission, 2) investigate the physical properties of dust, and 3) explore correlations between millimeter emission, gas, and Star Formation Rate (SFR).} We find that cold dust is responsible for most of the 1.15\,mm emission in both galaxies and at 2\,mm in NGC~2976. The free-free emission emits more importantly in NGC~2146 at 2\,mm. \revise{The cold dust emissivity index is flatter in the dwarf galaxy ($\beta=1.3\pm0.1$) compared to the starburst galaxy ($\beta=1.7\pm0.1$). Mapping the dust-to-gas ratio, we find that it changes between 0.004 and 0.01 with a mean of $0.006\pm0.001$ in the dwarf galaxy. In addition, no global balance holds between the formation and dissociation of H$_2$ in this galaxy. We find tight correlations between the millimeter emission and both the SFR and molecular gas mass in both galaxies.} }


\titlerunning{Interpreting Millimeter Emission from IMEGIN galaxies NGC~2146 and NGC~2976}
\maketitle

\section{Introduction}

Dust is one of the most important constituents of the ISM of galaxies, catalyzing the formation of molecules needed to form protostellar cores~\citep{Gould&Salpeter1963, Perets2006, Bron+2014}. 
By absorbing starlight and re-emitting it at infrared to millimeter wavelengths ($\lambda\,{\sim}$5\,$\mu$m-3\,mm), dust significantly impacts the heating and cooling processes of the ISM, and thus has a profound influence on the overall physics of a galaxy~\citep{Tielens1997, Smith2007, Pope2008, Galametz2013, Nersesian2019}, as well as its appearance in the ultraviolet to infrared wavelength range. Modeling the Spectral Energy Distribution (SED) of dust allows us to infer the physical properties of the grains, such as their temperature, mass, composition, and size distribution. Understanding these properties and their relationship with other components of the ISM, such as gas and star formation tracers, is essential to understand the evolution of the galaxy~\citep{ Draine+2007, Galametz2009, Kramer+2010, Aniano+2012, Tabatabaei+2014, Orellana2017, Aniano+2020, Galliano+2018}.

Dust in galaxies emits across a range of temperatures. \revise{The warm, more emissive component of dust, emits mostly in the mid-infrared (MIR) wavelength range. In contrast, the colder and more massive dust component emits at longer wavelengths in the Far-Infrared (FIR), up to 3\,mm.} This cold (15-30\,K) dust component is heated by the diffuse InterStellar Radiation Field (ISRF) mainly fed by photons from the global stellar population. However, the emission from this cold component is often intermixed with emission from warmer dust peaking at $\lambda\,{\sim}70\,\mu$m, which is heated by star-forming regions, and hence it is often poorly constrained. The submillimeter/millimeter waveband (500-3000\,$\mu$m) is crucial for detecting this cold dust component and thus for making accurate estimates of the total dust mass~\citep{Boquien2011, Tabatabaei_2013b}.

Space telescopes such as IRAS, ISO, \textit{Spitzer}, and \textit{Herschel} have allowed us to study dust emission in galaxies up to $500\,\mu$m. These telescopes have provided sensitive MIR to FIR observations to constrain the dust properties, particularly around the peak of the dust SED (100-200\,$\mu$m in star-forming galaxies). However, studying longer millimeter wavelengths, well beyond the peak, is crucial for modeling the mass and temperature of cold dust~\citep[e.g.][]{Galliano2003, Galametz2011, RemyRuyer2013, Hunt2015}. To fully sample the dust SED, reliable data in the millimeter range is needed. The Planck space telescope has observed the Milky Way and other galaxies at millimeter wavelengths, but its resolution (300$^{\prime\prime}$ at best) does not allow for the study of resolved ISM in most galaxies. 
As a result, even for many nearby galaxies, we lack high-resolution observations at the low-frequency end of their dust SED. This coarse sampling of the Rayleigh-Jeans tail of dust emission in most galaxies has resulted in poor constraints on the content and physical properties of cold dust in those objects.

The New IRAM KID Array (NIKA2) on the IRAM 30\,m telescope, with an instantaneous Field-of-View (FoV) of 6.5$^{\prime}$ and angular resolutions of 11.1$^{\prime\prime}$ and 17.6$^{\prime\prime}$ at 260 and 150\,GHz, respectively, is an ideal instrument for studying the dusty ISM of nearby galaxies in detail~\citep{Adam2017, Perotto_2020}. The Guaranteed-Time Large Project of the NIKA2 collaboration, Interpreting Millimeter Emission of Galaxies with IRAM and NIKA2 (IMEGIN), led by S. Madden, has observed a sample of 22 nearby galaxies with varying ranges of mass, morphological types, SFR, and ISM properties with distances less than 25\,Mpc. This sample is expected to provide essential data for studying the thermal emission of cold dust and its relation to other components of the ISM in nearby galaxies. The first IMEGIN paper was focused on the edge-on galaxy NGC~891~\citep{Katsioli+2023}. This paper is the second IMEGIN paper that presents and analyzes the NIKA2 observations of the two galaxies, NGC~2146 and NGC~2976.
\reviseII{These two galaxies, which differ extensively in terms of mass, SFR, and ISM properties, were among the first IMEGIN galaxies to have their observations and data processing completed, which is the reason for selecting these galaxies for this paper.}

NGC~2146 is a barred spiral galaxy with a morphology type of SB(s)ab pec~\citep{deVaucoulers+1991} and a total stellar mass of $\log M_{*}=10.3\,{\rm M}_{\odot}$~\citep{Kennicutt+2011}. In optical images, it shows a bright central bulge and extended irregular arms, in addition to an apparent dust lane that runs along the major axis. NGC~2146 is a Luminous Infrared Galaxy (LIRG) with L$_{8-1000\,\mu m}\sim10^{11}\,{\rm L}_{\odot}$~\citep{Gao&Solomon2004}. As outflows are reported to be common in LIRGs, NGC~2146 shows a superwind outflow along the minor axis which is observed both in optical and X-ray~\citep{Greve+2000, Kreckel+2014}. Moreover, ~\cite{Taramopoulos2001} claimed that NGC~2146 has gone through a merger with a gas-rich low surface brightness spiral companion that occurred about 1\,Gyr ago. The companion has been destroyed during the interaction, as there is no evidence of a double-nucleus in the center of the galaxy, indicating that it is a far-evolved merger~\citep{Tarchi2004}. This merger is believed to be the cause of the starburst activity in NGC~2146, given the high star formation rate of this galaxy, reported to be as high as $\rm{SFR}=39\pm11\,{\rm M}_{\odot}\,{\rm yr}^{-1}$~\citep{Nersesian2019}. The metallicity reported for NGC~2146 is $12+\rm{log}(\rm{O}/\rm{H})=8.78\pm0.16$~\citep[][from calibration of~\cite{PettiniPagel2004}]{DeVis+2019}, equal to $1.2\,Z_{\odot}$ where $Z_{\odot}$ is the solar metallicity~\citep{Asplund+2009}. Properties of NGC~2146 are presented in Table~\ref{table-galaxyproperties}. 

NGC~2976 is a peculiar dwarf galaxy \citep[type SAc pec,][]{deVaucouleurs1991}, a weakly disturbed satellite member of the M81 group. It is a pure disc object with no visible spiral arms, although it is debated whether it has a gas-rich bar and large-scale arms~\citep{Valenzuela+2014}. NGC~2976 has a mass 22 times smaller than that of NGC~2146 and a SFR of 0.13$\pm$0.02\,M$_{\odot}$yr$^{-1}$~\citep{Nersesian2019}. The reported metallicity for this galaxy is $12 + \rm{log} (\rm{O}/\rm{H})=8.39\pm0.03$~\citep[][from calibration of~\cite{Pilyugin+Grebel2016}]{DeVis+2019}, equal to $0.5\,Z_{\odot}$. Properties of NGC~2976 are summarized in Table~\ref{table-galaxyproperties}.

\revise{This paper is organized as follows. After explaining the NIKA2 observations and the complementary data (Sect.~\ref{section-data}), we present the resulting millimeter maps in Sect.~\ref{section-millimeteremission}.} We then describe the framework for modeling the SED of continuum emission from the radio to FIR domains and present the results on both global 
and resolved scales (Sect.~\ref{section-SEDmodeling}). Our results are then discussed in Sect.~\ref{section-discussion} and summarized in Sect.~\ref{section-summary}.

\begin{figure*}[h]
	\centering
	\includegraphics[width=0.4\textwidth]{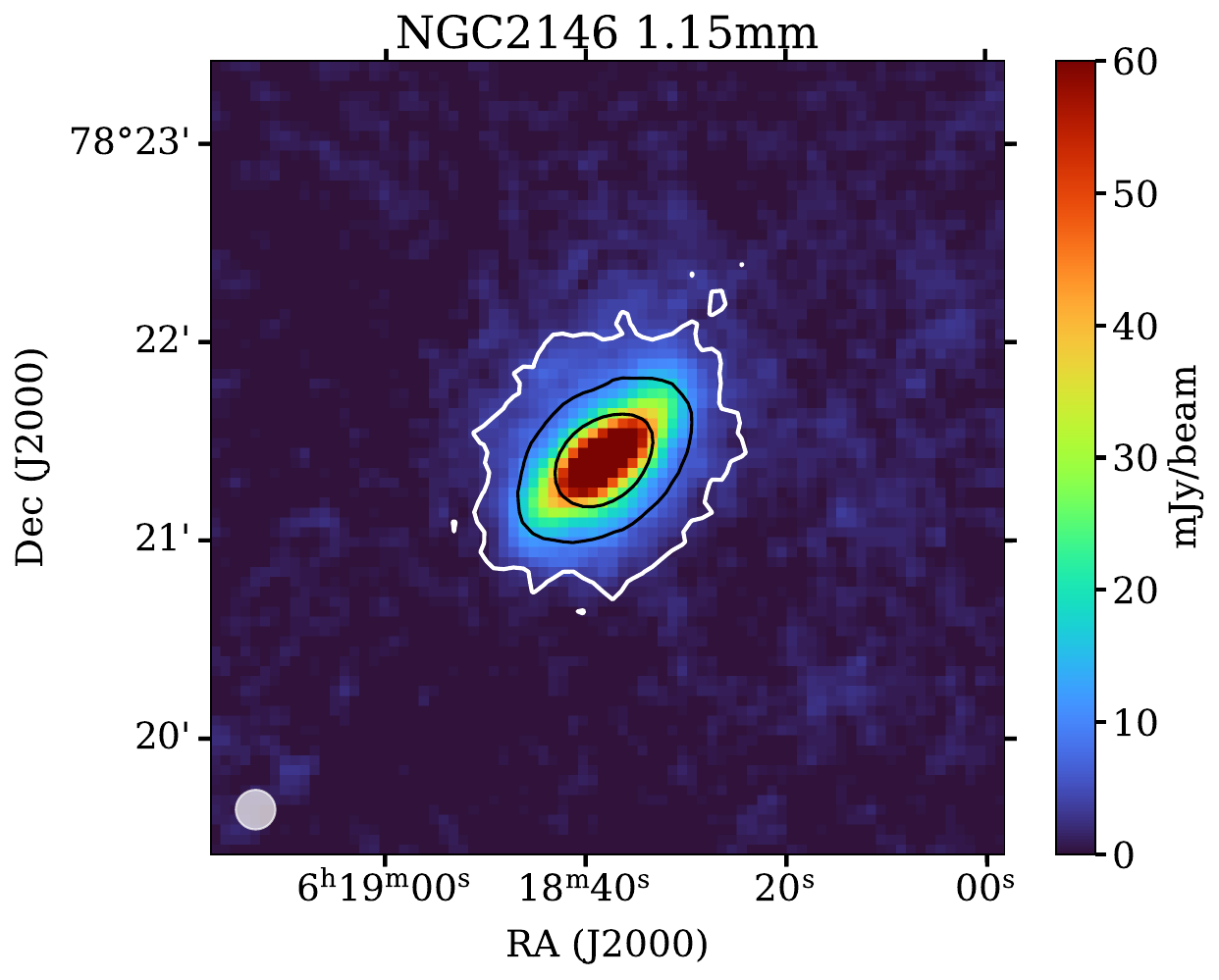}
 \hspace{0.4cm}
 \vspace{0.3cm}
	\includegraphics[width=0.4\textwidth]{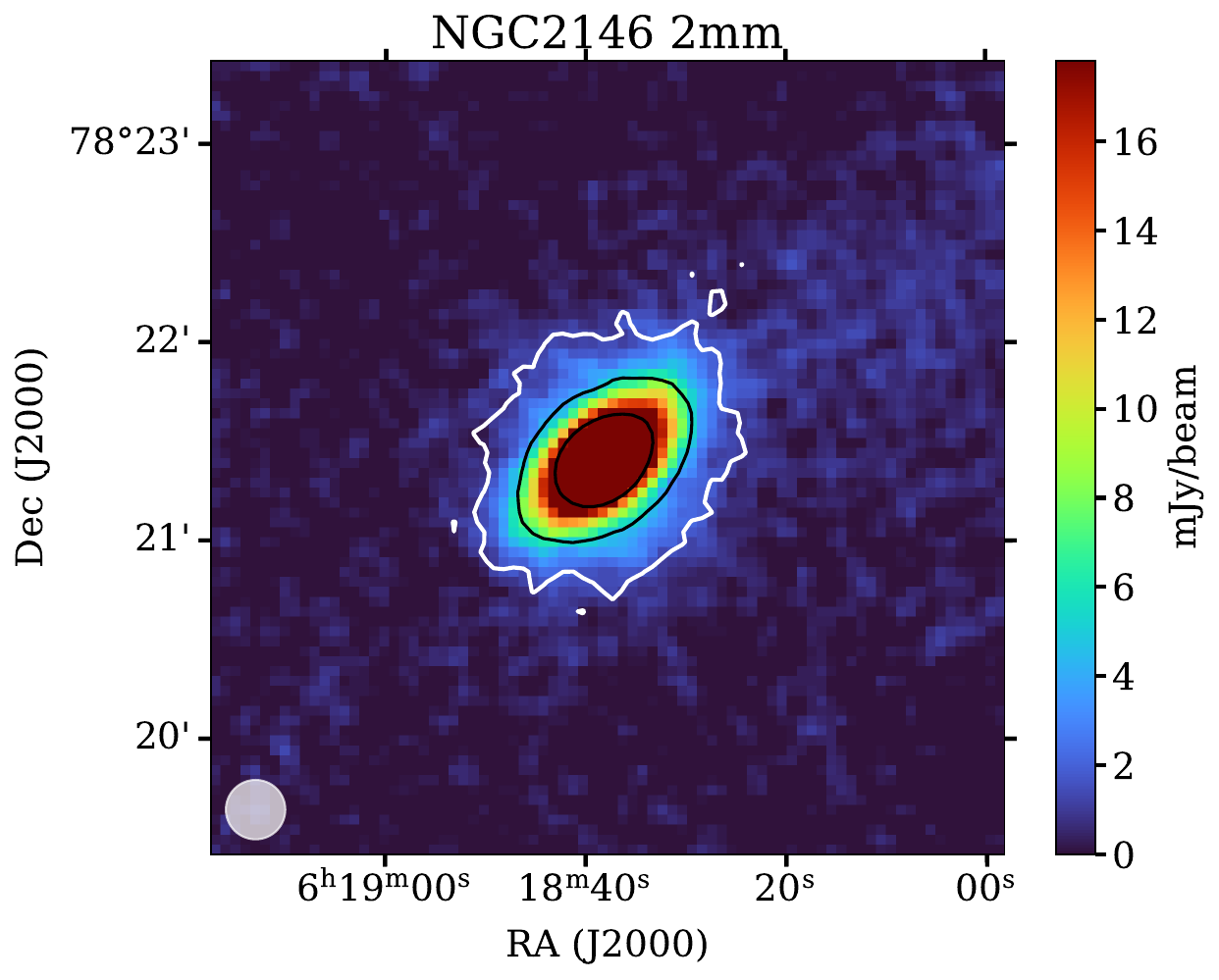}
	\includegraphics[width=0.4\textwidth]{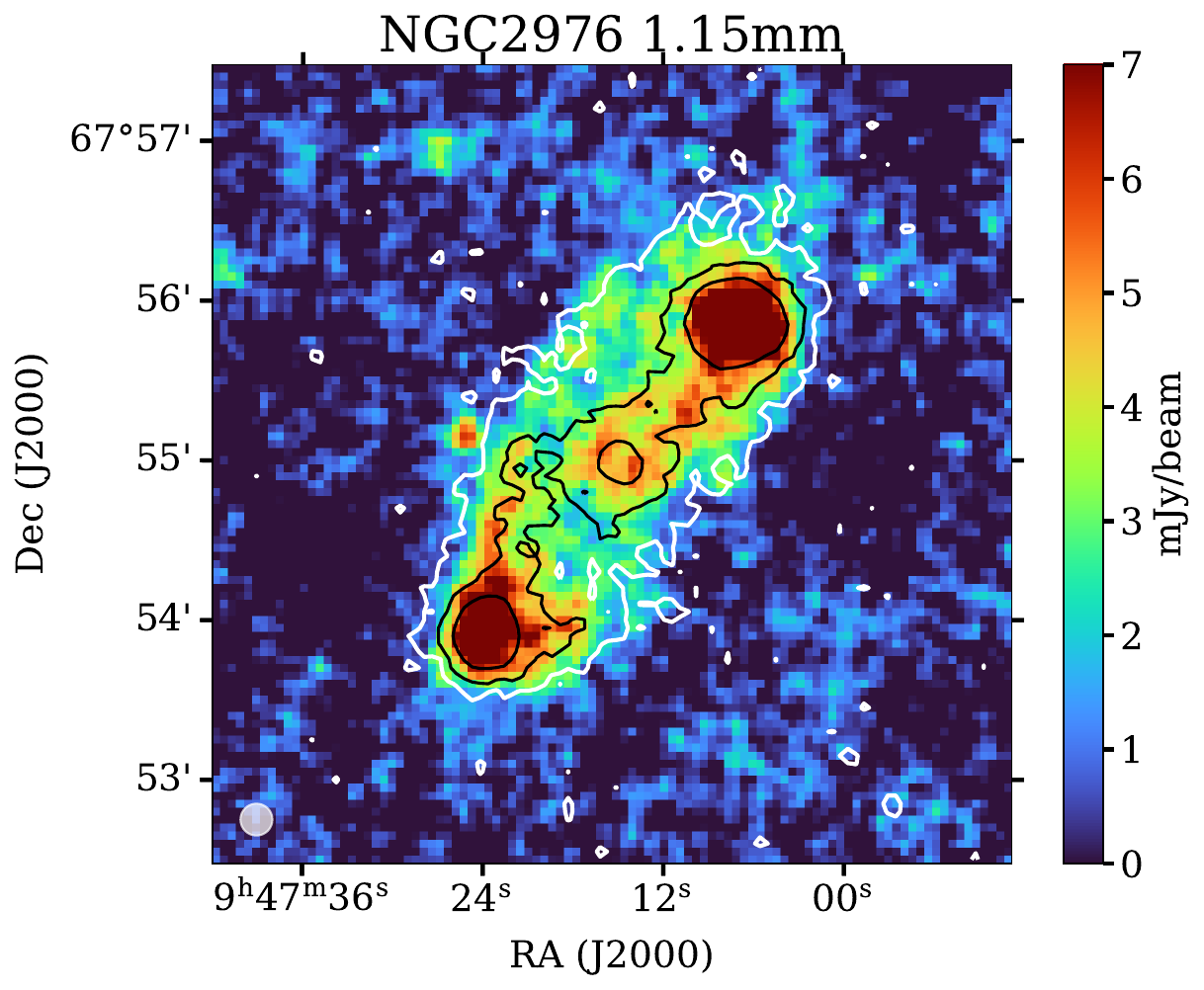}
 \hspace{0.4cm}
	\includegraphics[width=0.4\textwidth]{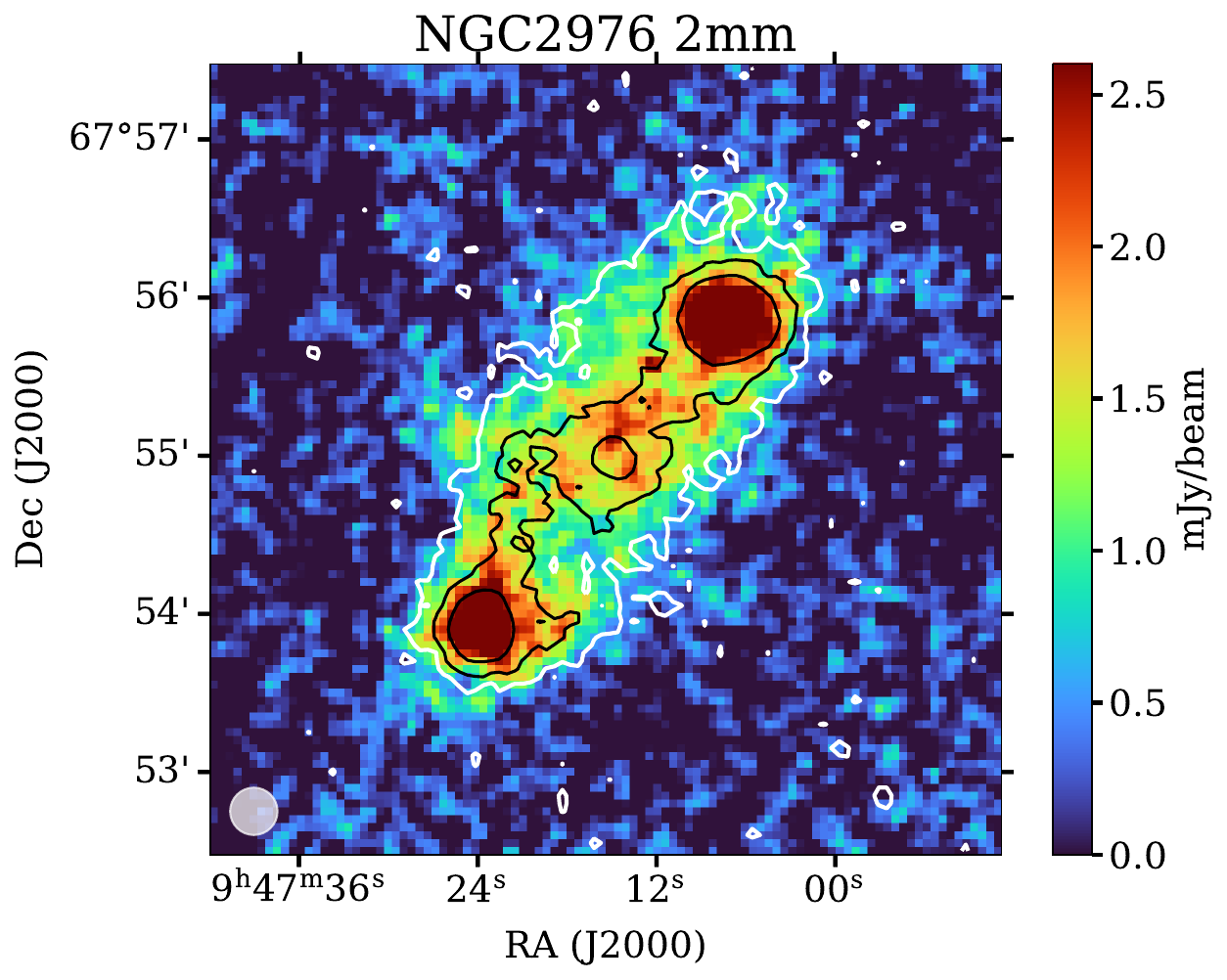}
	\caption{Observed NIKA2 maps of NGC~2146 (\textit{top}) and NGC~2976 (\textit{bottom}) at 1.15\,mm (\textit{left}) and 2\,mm (\textit{right}). The color bars show the flux density in units of mJy/beam. Circled areas in the bottom left corners indicate the angular resolutions (beam widths) which are 12$^{\prime\prime}$ and 18$^{\prime\prime}$ at 1.15\,mm and 2\,mm, respectively. Contours show the MIPS 24\,$\mu$m emission at levels 9, 54, and 324\,Jy/pixel in top and 0.8, 2.4, and 7.2\,Jy/pixel in bottom. FoV is $4^{\prime}\times4^{\prime}$ and $5^{\prime}\times5^{\prime}$ for NGC~2146 and NGC~2976, respectively. }\label{plot-observedmaps}
\end{figure*}

\begin{table}
	\begin{center}
\renewcommand{\arraystretch}{1.1} 
		\begin{tabular}{>{\raggedright}p{2.5cm} p{2.5cm} p{2.5cm}}
			\hline
			Galaxy & NGC~2146 & NGC~2976\\
			\hline
			Position of nucleus (J2000) & RA=$6^{h}18^{m}37.7^{s}$ Dec=78\textdegree 21\arcmin25\arcsec & RA=$9^{h}47^{m}15.3^{s}$ Dec=67\textdegree 55\arcmin00\arcsec \\
			Apparent size $^a$ & $6.0'\times 3.4'$& $5.9'\times2.7'$ \\
			Inclination (0\textdegree=face on) & 37\textdegree & 60\textdegree \\
			Distance\,(Mpc) $^b$ & 17.2 & 3.55 \\
			Physical scale equivalent to $6^{\prime\prime}$ pixel\,(pc) & 524 & 102\\
 $\log M_{*}\,({\rm M}_{\odot})$ $^{b}$ & 10.3 & 8.9\\
 SFR\,$({\rm M}_{\odot}\,{\rm yr}^{-1})$ $^c$ & $39\pm11$ & $0.13\pm0.02$ \\
 $12 + \rm{log} (\rm{O}/\rm{H})$ $^d$ & $8.39\pm0.03$ & $8.78\pm0.16$\\
 $Z\,({\rm Z}_{\odot})$ & 1.2 & 0.5 \\
			\hline
		\end{tabular}
	\end{center}
\caption{Properties of NGC\,2146 and NGC\,2976.\newline
	$^a$ \cite[$D_{25}$ in Blue filter,][]{deVaucoulers+1991}\newline
	$^b$ \cite{Kennicutt+2011}\newline
	$^c$ \cite{Nersesian2019}\newline
	$^d$ \cite{DeVis+2019}}	
 \label{table-galaxyproperties}
\end{table}

\section{Data}\label{section-data}
\subsection{NIKA2 observations}
We observed NGC~2146 and NGC~2976 as a part of IMEGIN with NIKA2, which operates simultaneously at two wavelengths 1.15\,mm and 2\,mm. Observations of NGC~2146 were conducted from 13 to 17 January 2021. A total of 46 scans were taken during 6.2 hours
of on-the-source telescope time. The sky conditions were stable. In addition, the atmospheric opacity $\tau _{225\,\rm{GHz}}$ ranged between 0.11 and 0.28, with an average of $0.23\pm0.05$.
NGC~2976 was observed between 25 October 2020 and 14 February 2021. In total, 66 scans
were taken in 8.3 hours of on-the-source telescope time. During this time, sky opacity $\tau_{225\,\rm{GHz}}$ varied in the range 0.19 and 0.32, on average equal to $0.22\pm 0.06$.\par

In general, scans were conducted along directions at $\pm$45 degrees relative to the major axis of each galaxy. Scan sizes were $17.3^{\prime}\times8.12^{\prime}$ and $17.6^{\prime}\times7.35^{\prime}$ for NGC~2146 and NGC~2976, respectively. The on-the-fly scans were conducted at the maximum possible scanning speed for both galaxies (60$^{\prime\prime}$/sec), given the elevation of the source at the time of the observations and the limitations of the telescope drive system to overcome sky fluctuations as much as possible. Pointing was corrected every hour and the focus was adjusted every three hours.\par

The NIKA2 data were reduced with the \textit{Scanam\_nika} pipeline, as explained in Appendix~\ref{appendix-dataprocessing}. The reduced observed maps at 1.15\,mm and 2\,mm are presented in Fig.~\ref{plot-observedmaps}. \revise{The color bars show the flux density in units mJy\,beam$^{-1}$, and beam sizes are indicated as filled circles in the lower left corner of each map.} The 1.15\,mm maps have an original resolution (FWHM) equal to 12$^{\prime\prime}$ and a pixel size of 3$^{\prime\prime}$. In Fig.~\ref{plot-observedmaps}, the beam of the 1.15\,mm maps corresponds to a physical scale of $\sim$1\,kpc and $\sim$0.2\,kpc in NGC~2146 and NGC~2976, respectively. The original resolution of 2\,mm maps, on the other hand, is 18$^{\prime\prime}$, which corresponds to $\sim$1.6\,kpc and $\sim$0.3\,kpc physical scale in NGC~2146 and NGC~2976, respectively (Table~\ref{table-galaxyproperties}). \revise{The pixel size of 2\,mm maps are also 3$^{\prime\prime}$.} 
 \revise{In addition, we were able to reach a one $\sigma_{\rm rms}$ noise level of 0.98\,mJy/beam at 1.15\,mm and 0.47\,mJy/beam at 2\,mm for NGC~2146. Similarly for NGC~2976, a one $\sigma_{\rm rms}$ noise level of the 1.15\,mm map is 0.97\,mJy/beam while it is 0.37\,mJy/beam at 2\,mm.} \par

\begin{table*}[h]
\centering
\begin{tabular}{ccccccc}
\toprule
 & & Angular & NGC~2146 & NGC~2976 &\\
Telescope & Wavelength & resolution & $\sigma_{\rm rms}$ & $\sigma_{\rm rms}$& Reference\\
\midrule
WSRT & 21\,cm & 12.5$^{\prime\prime}$ & $6.5\times10^{-5}$ & $2.7\times10^{-5}$ & \cite{Braun+2007}\\
WSRT & 18\,cm & 10$^{\prime\prime}$ & $4.7\times10^{-5}$ & $3.8\times10^{-5}$ & \cite{Braun+2007}\\ 
Effelsberg & 6.2\,cm & 144$^{\prime\prime}$ & $5.0\times10^{-4}$ & $2.0\times10^{-3}$& \cite{Tabatabaei+2017}\\
Effelsberg & 2.8\,cm & 69$^{\prime\prime}$ & - & -& \cite{Niklas1995}\\
Planck & 1.38\,mm & 301$^{\prime\prime}$ & $1.8\times10^{-1}$ & $6.8\times10^{-2}$ & \cite{Planck2014, Clark_2018}\\
\textit{Herschel}-SPIRE & 500\,$\mu$m & 36$^{\prime\prime}$ & $5.3\times10^{-2}$ & $1.4\times10^{-2}$& \cite{Kennicutt+2011} \\
\textit{Herschel}-SPIRE & 350\,$\mu$m & 25$^{\prime\prime}$ & $7.4\times10^{-2}$ & $1.8\times10^{-2}$& \cite{Kennicutt+2011}\\
\textit{Herschel}-SPIRE & 250\,$\mu$m & 18$^{\prime\prime}$ & $1.4\times10^{-1}$ & $7.9\times10^{-2}$& \cite{Kennicutt+2011}\\
\textit{Herschel}-PACS & 160\,$\mu$m & 11.4$^{\prime\prime}$ & $9.2\times10^{-1}$ & $3.6\times10^{-1}$& \cite{Kennicutt+2011}\\
\textit{Herschel}-PACS & 100\,$\mu$m & 6.8$^{\prime\prime}$ & $3.7\times10^{0}$ & $9.3\times10^{-1}$ &\cite{Kennicutt+2011}\\\
\textit{Herschel}-PACS & 70\,$\mu$m & 5.6$^{\prime\prime}$ & $3.6\times10^{0}$ & $6.8\times10^{-1}$& \cite{Kennicutt+2011}\\ 
\textit{Spitzer}-MIPS & 24\,$\mu$m & 6$^{\prime\prime}$ & $5.2\times10^{-3}$ & $8.6\times10^{-4}$& \cite{Engelbracht+2008,Kennicutt+2003}\\
GALEX-FUV & 153\,nm & 4.3$^{\prime\prime}$& $6.3\times10^{-7}$ & $4.2\times10^{-7}$ & \cite{GildePaz2007, Clark_2018} \\
IRAM 30\,m HERA & CO(2-1) & 13.4$^{\prime\prime}$& $4.2\times10^{-1}$ & $2.7\times10^{-1}$ & \cite{Leroy+2009}\\
VLA& HI & 5$^{\prime\prime}$& - & $2.7\times10^{0}$ & \cite{Walter+2008}\\
\bottomrule
\end{tabular}
\caption{Complementary data used in this study. The first and second columns show the instrument and wavelength of each band, next to the angular resolution in the third column. The fourth column is the $\sigma_{\rm rms}$ of each band, expressed in unit Jy/beam, except for CO, which is K\,km/s and for HI which is Jy/beam\,m/s. The maps listed above all cover the whole disc of our two galaxies. Resolutions are FWHM on each map. HI data was only available for NGC~2976. The integrated flux at 2.8\,cm and its corresponding uncertainty is quoted from \cite{Niklas1995} and not measured by this work.}
\label{table-alldata}
\end{table*}

\subsection{Complementary data}\label{section-complementarydata}

\begin{figure*}
	\centering
	\includegraphics[width=0.8\textwidth]{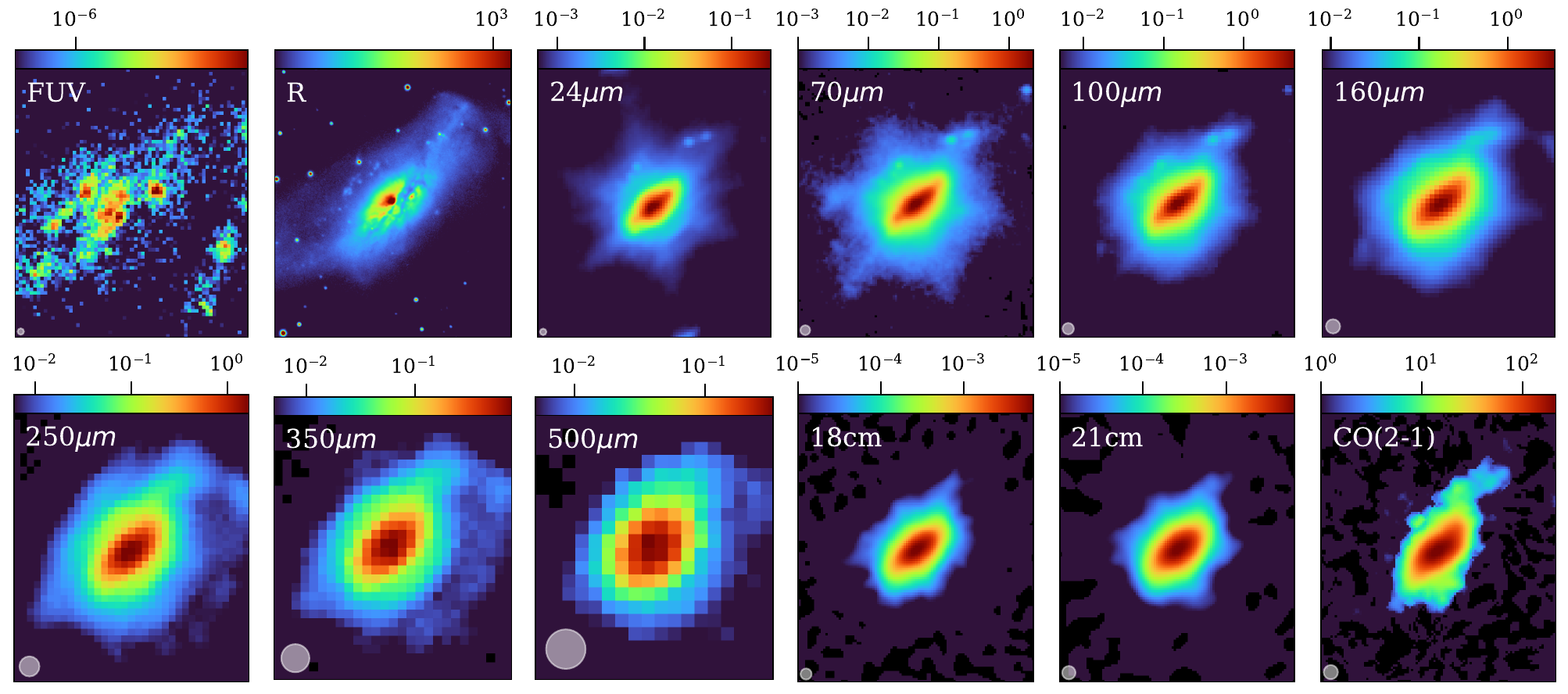}
	\caption{The complementary data used in this work for NGC~2146. GALEX FUV, \textit{Spitzer} MIPS, \textit{Herschel} SPIRE and PACS, radio, CO(2-1) maps are presented. \revise{The optical image in the Red filter displays the optical features of this galaxy and is not used in the analysis.} Also, the Planck 1.38\,mm and radio 6.2\,cm maps are not shown due to poor resolution. \revise{All the demonstrated maps have the same FoV of $3.5^{\prime} \times 4^{\prime}$, centered on the galaxy. The maps are displayed in logarithmic scale and their original resolution, indicated in Table~\ref{table-alldata}, with the beam size represented by a white circle in the bottom left corner of each map.}}\label{plot-NGC2146album}
\end{figure*}

\begin{figure*}
	\centering
	\includegraphics[width=0.8\textwidth]{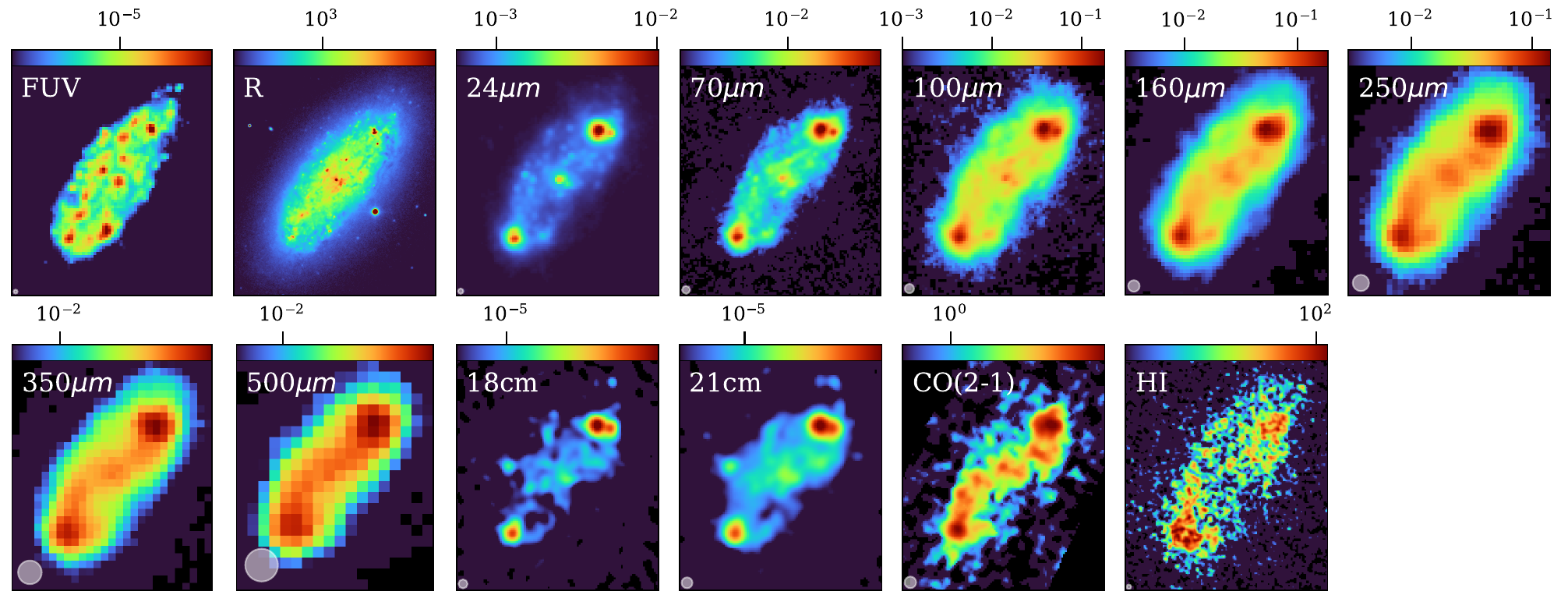}
	\caption{Same as Fig.~\ref{plot-NGC2146album} for NGC~2976. \revise{The HI map, which is only available for this galaxy is also included.} The optical map in the Red filter displays the optical features of this galaxy and is not used in the analysis. Also, the Planck 1.38\,mm and radio 6.2\,cm maps are not shown due to poor resolution. \revise{All the demonstrated maps have the same FoV of $3.7 ^{\prime} \times 4.2^{\prime}$ centered on the galaxy. The maps are all shown in logarithmic scale and at their original resolution, noted in Table~\ref{table-alldata}, and demonstrated by a white circle on the bottom-left corner of each map.}}\label{plot-NGC2976album}
\end{figure*}

Table~\ref{table-alldata} summarizes the complementary data used in this work. A collection of the maps used in this work is presented in Fig.~\ref{plot-NGC2146album} and Fig.~\ref{plot-NGC2976album} for NGC~2146 and NGC~2976, respectively. We further explain the complementary data below.

\revise{Both these galaxies were observed with \textit{Spitzer} MIPS. NGC~2976 was observed as part of the Spitzer Nearby Galaxies Survey~\citep[SINGS,][]{Kennicutt+2003} project and NGC~2146 by \cite{Engelbracht+2008}.} We acquired the MIPS 24\,$\mu$m data from the DustPedia archive~\citep{Clark_2018}. 

These galaxies were also observed with \textit{Herschel} PACS and SPIRE cameras as part of the Key Insight in Nearby Galaxies: A Far-Infrared Survey~\citep[KINGFISH,][]{Kennicutt+2011} project. Moreover, we acquired \textit{Planck} observations of these two galaxies at 1.38\,mm from the DustPedia archive~\citep{Clark_2018}, originally presented in~\cite{Planck2014}.

NGC~2146 and NGC~2976 were observed at 6\,cm with the Effelsberg 100\,m telescope as part of the Key Insight in Nearby Galaxies Emitting in Radio~\citep[KINGFISHER,][]{Tabatabaei+2017} project with 2.5\arcmin~angular resolution. Resolved Radio Continuum (RC) maps of these galaxies were taken with the Westerbork array as part of the WSRT SINGS project~\citep{Braun+2007} with 10\arcsec~and 12.5\arcsec angular resolutions at 18\,cm and 21\,cm wavelengths, respectively. In addition, we use archival photometry from~\cite{Niklas1995} which includes observations at 2.8\,cm with the Effelsberg 100\,m telescope.

To take into account the contamination by the CO line emission in the NIKA2 1.15\,mm observations and to trace molecular gas, we use CO(2-1) data from the HERA CO Line Extragalactic Survey~\citep[HERACLES,][]{Leroy+2009, Schruba+2011, Leroy+2013, Sandstrom+2013}, observed with IRAM 30\,m telescope.

In order to study different neutral gas components, we made use of the VLA HI 21\,cm observations as part of The HI Nearby Galaxy Survey~\citep[THINGS,][]{Walter+2008} project, which is available only for NGC~2976. As explained in~\citep{Schruba+2011}, the HI emission from the disk of NGC~2146 is severely self-absorbed leaving us without useful information on the neutral atomic gas from millimeter-emitting regions of this galaxy.

We also use GALEX Far Ultra-Violet (FUV) data of these galaxies \citep{GildePaz2007} acquired from the DustPedia archive \citep{Clark_2018}. \revise{Just for the sake of comparing with optical structures, Fig.~\ref{plot-NGC2146album} and Fig.~\ref{plot-NGC2976album} also show R-band optical maps observed with the Jacobus Kapteyn Telescope \citep[JKT,][]{James+2004}.} Note, however, that they are not used in this analysis.


\begin{table}
\centering
\begin{tabular}{ccc}
\toprule
Galaxy & NGC~2146 & NGC~2976\\
 \hline
{wavelength} & $S_{\nu}$\,(Jy) & $S_{\nu}\,(Jy)$ \\
\midrule
21\,cm & 
\DTLfetch{global2146}{w}{21}{intgflux} $\pm$ 
\DTLfetch{global2146}{w}{21}{er} & 
\DTLfetch{global2976}{w}{21}{intgflux} $\pm$ 
\DTLfetch{global2976}{w}{21}{er} \\
18\,cm& 
\DTLfetch{global2146}{w}{18}{intgflux} $\pm$ 
\DTLfetch{global2146}{w}{18}{er} &
\DTLfetch{global2976}{w}{18}{intgflux} $\pm$ 		
\DTLfetch{global2976}{w}{18}{er}\\
6.2\,cm& 
\DTLfetch{global2146}{w}{6}{intgflux} $\pm$ 		
\DTLfetch{global2146}{w}{6}{er} &
\DTLfetch{global2976}{w}{6}{intgflux} $\pm$ 
\DTLfetch{global2976}{w}{6}{er} \\
2.8\,cm& 
\DTLfetch{global2146}{w}{28}{intgflux} $\pm$ 
\DTLfetch{global2146}{w}{28}{er}& 
\DTLfetch{global2976}{w}{28}{intgflux} $\pm$ 
\DTLfetch{global2976}{w}{28}{er}\\
2\,mm&
\DTLfetch{global2146}{w}{2}{intgflux} $\pm$ 		
\DTLfetch{global2146}{w}{2}{er} & 	
\DTLfetch{global2976}{w}{2}{intgflux} $\pm$ 	
\DTLfetch{global2976}{w}{2}{er}\\
1.38\,mm& 
\DTLfetch{global2146}{w}{138}{intgflux} $\pm$ 		
\DTLfetch{global2146}{w}{138}{er}	&
\DTLfetch{global2976}{w}{138}{intgflux} $\pm$		
\DTLfetch{global2976}{w}{138}{er}\\
1.15\,mm& 
\DTLfetch{global2146}{w}{1}{intgflux} $\pm$ 	
\DTLfetch{global2146}{w}{1}{er} &
\DTLfetch{global2976}{w}{1}{intgflux} $\pm$	
\DTLfetch{global2976}{w}{1}{er}\\
500\,$\mu$m & 
\DTLfetch{global2146}{w}{500}{intgflux} $\pm$	
\DTLfetch{global2146}{w}{500}{er} & 
\DTLfetch{global2976}{w}{500}{intgflux} $\pm$		
\DTLfetch{global2976}{w}{500}{er}\\
350\,$\mu$m & 
\DTLfetch{global2146}{w}{350}{intgflux} $\pm$		
\DTLfetch{global2146}{w}{350}{er} & 	
\DTLfetch{global2976}{w}{350}{intgflux} $\pm$		
\DTLfetch{global2976}{w}{350}{er}\\
250\,$\mu$m &
\DTLfetch{global2146}{w}{250}{intgflux} $\pm$ 		
\DTLfetch{global2146}{w}{250}{er}& 
\DTLfetch{global2976}{w}{250}{intgflux} $\pm$		
\DTLfetch{global2976}{w}{250}{er}\\
160\,$\mu$m & 
\DTLfetch{global2146}{w}{160}{intgflux} $\pm$	
\DTLfetch{global2146}{w}{160}{er} & 
\DTLfetch{global2976}{w}{160}{intgflux} $\pm$ 	
\DTLfetch{global2976}{w}{160}{er}\\
100\,$\mu$m &
\DTLfetch{global2146}{w}{100}{intgflux} $\pm$	
\DTLfetch{global2146}{w}{100}{er} & 
\DTLfetch{global2976}{w}{100}{intgflux} $\pm$	
\DTLfetch{global2976}{w}{100}{er}\\
70\,$\mu$m &
\DTLfetch{global2146}{w}{70}{intgflux}	$\pm$
\DTLfetch{global2146}{w}{70}{er} & 
\DTLfetch{global2976}{w}{70}{intgflux}	$\pm$
\DTLfetch{global2976}{w}{70}{er}\\
24\,$\mu$m & 
\DTLfetch{global2146}{w}{24}{intgflux} $\pm$
\DTLfetch{global2146}{w}{24}{er}& 
\DTLfetch{global2976}{w}{24}{intgflux} $\pm$
\DTLfetch{global2976}{w}{24}{er}\\
\bottomrule
\end{tabular}
\caption{\revise{Integrated flux densities of the maps used for the global SED analysis.}}
\label{table-photometry}
\end{table}

\section{Millimeter emission from NGC~2146 and NGC~2976} \label{section-millimeteremission}
Fig.~\ref{plot-observedmaps} shows that the millimeter emission emerges from an elliptical disk which is brightest toward the galaxy center in NGC~2146. \revise{No characteristic galactic substructures such as spiral arms are evident at millimeter wavelengths, which is similar to those at FIR/submillimeter and radio wavelengths (Fig.~\ref{plot-NGC2146album}). The maximum flux density is $\simeq$ 95\,mJy/beam at 1.15\,mm and 55\,mJy/beam at 2\,mm at the position of the nucleus at the central source.} The integrated flux densities are $\DTLfetch{global2146}{w}{1-mJy}{intgflux}\pm \DTLfetch{global2146}{w}{1-mJy-er}{intgflux}$\,mJy and $\DTLfetch{global2146}{w}{2-mJy}{intgflux}\pm \DTLfetch{global2146}{w}{2-mJy-er}{intgflux}$\,mJy at 1.15\,mm and 2\,mm, respectively. \par

NGC2976, on the other hand, shows more structural details due to its proximity. Notably, the two bright star-forming regions in the north and south of this galaxy are prominent in the NIKA2 maps, similar to those in the FIR maps (Fig.~\ref{plot-NGC2976album}). 
\revise{Its maximum occurs in the north-western star-forming region with values of 14.0\,mJy/beam and 6.2\,mJy/beam at 1.15\,mm and 2\,mm, respectively. The south-eastern source peaks at 10.8\,mJy/beam at 1.15\,mm and 4.4\,mJy/beam at 2\,mm. The integrated flux of the 1.15\,mm and 2\,mm maps are $\DTLfetch{global2976}{w}{1-mJy}{intgflux}\pm \DTLfetch{global2976}{w}{1-mJy-er}{intgflux}$\,mJy and $\DTLfetch{global2976}{w}{2-mJy}{intgflux}\pm \DTLfetch{global2976}{w}{2-mJy-er}{intgflux}$\,mJy. } \par

\revise{The observed millimeter emission represents a mixture of the thermal dust and free-free emission as well as the non-thermal synchrotron emission. In star-forming galaxies, the free-free and synchrotron radiations are extinction-free tracers of the SFR \citep[e.g.,][]{condon1992,Tabatabaei+2017,Heesen+2022}. The radiation from dust grains is due to heating by different sources and the peak of dust emission depends on the average energy of the radiation field. Warmer dust is heated mainly by the energetic radiation field near young massive stars dominated by UV photons, therefore emitting at shorter wavelengths. \reviseII{Colder dust, which emits at longer wavelengths, is heated mainly by the less energetic ISRF, dominated by optical photons from the old stellar population \citep[see e.g.,][]{Bendo+2015}}. However, the ISRF itself depends on local physical conditions within a galaxy, such as the star formation rate and the ISM density of a galaxy. For instance, the ISRF in a starburst galaxy is expected to be dominated by ionizing UV photons from young/massive stars rather than optical photons from old stars. A similar expectation holds in a dwarf star-forming galaxy with a low-density ISM due to weak shielding and a high probability of the escape of ionizing photons. Despite different dust heating sources, the total IR emission of galaxies (integrated in the wavelength range of, e.g., 8 to 1000\,$\mu$m) is also widely used as a tracer of the SFR \citep{Ken}. On the other hand, cold dust emission can also trace the total gas content of galaxies, including \emph{dark gas} that is not traced by CO or HI \citep[e.g.,][]{Abdo}. Considering that the SFR and the gas surface densities are correlated through the Kennicutt-Schmidt relation, a more indirect correlation between the cold dust emission and the SFR is also expected \citep{Boquien+2011, Bendo+2012}. Hence, the millimeter emission can be strongly correlated with SFR and/or the molecular gas content of galaxies. Fig.~\ref{plot-observedmaps}, which includes contours of 24\,$\mu$m emission as a tracer of SFR overlaid on NIKA2 maps, demonstrates a strong agreement between the two.} This is further investigated in Sect.~\ref{section-tracingSFRandgaswithNIKA2}. We determine the contribution of each millimeter component through modeling the FIR-radio SEDs. This is also needed to study the physical properties of dust. Prior to the SED analyses, we assess and subtract the CO line contamination.

\begin{figure}
	\centering
	\includegraphics[width=0.24\textwidth]{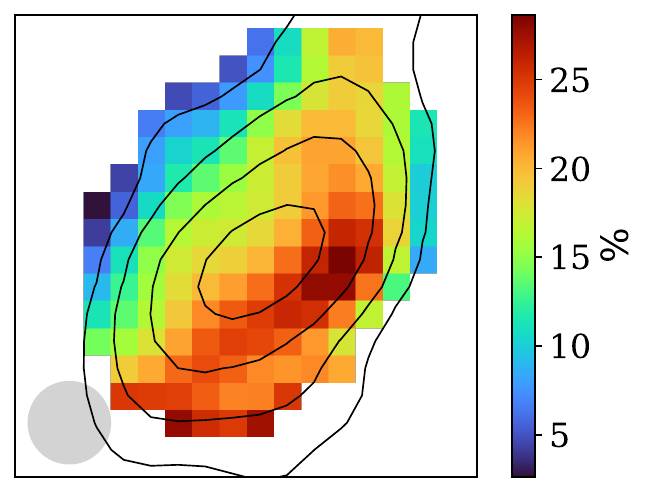}
	\includegraphics[width=0.24\textwidth]{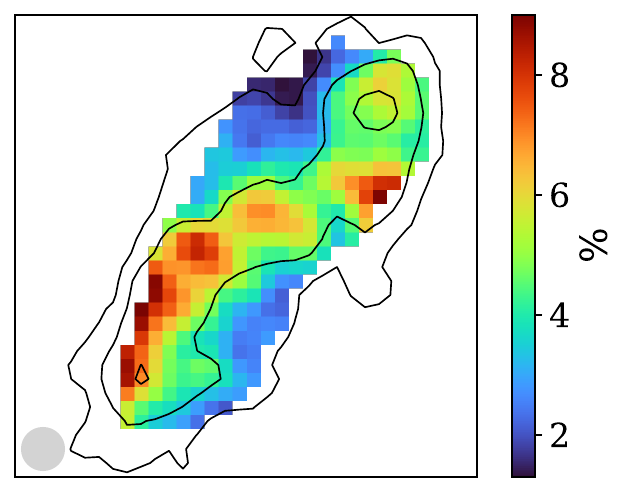}
	\caption{Maps of the CO fraction defined as the ratio of the CO(2-1) line emission to the observed 1.15\,mm emission in NGC~2146 (\textit{left}) and NGC~2976 (\textit{right}) at 18$^{\prime\prime}$ angular resolution. Only pixels above the $3\sigma_{\rm rms}$ level of the observed NIKA2 map are shown. Contours show the intensity of CO emission in each galaxy, starting from $1\sigma_{\rm rms}$ level and increasing in factors of 3. Circles in the bottom-left corners indicate the beam area.}
	\label{plot-COcontribution}
\end{figure}

\section{Contribution of the CO(2-1) emission}\label{section-COcontamination}
The NIKA2 passband at 1.15\,mm overlaps partly with the CO(2-1) molecular line emission at 230.5\,GHz~\citep{Perotto_2020}. Therefore while measuring the continuum flux of our targets at 1.15\,mm with NIKA2, we are inevitably including line emission from CO(2-1). Hence, to study dust continuum emission separately, we have to determine the contamination by the CO(2-1) line and subtract it from the total emission at 1.15\,mm. In Appendix~\ref{appendix-CO} we explain how we have taken the emission of the CO(2-1) line into account.
 
We quantify the contribution of the CO line emission at 18$^{\prime\prime}$ resolution, as this will be our working resolution for the rest of the analysis. Therefore, we convolve the NIKA2 1.15\,mm and CO(2-1) maps to 18$^{\prime\prime}$ resolution. Fig.~\ref{plot-COcontribution} shows the map of the fraction of the NIKA2 1.15\,mm emission that is attributed to CO (2-1), calculated for pixels with values above the $3\sigma_{\rm rms}$ level. \revise{Up to 30\% of the observed 1.15\,mm emission is due to CO line emission in the starburst galaxy NGC~2146. Moreover, in this galaxy, most of the CO line contamination occurs along the forward dusty arm of this galaxy (cf. R-band optical map of Fig.~\ref{plot-NGC2146album}).} In NGC 2976, the CO fraction is smaller with a maximum of only 10\% (Fig.~\ref{plot-COcontribution}), consistent with the fact that dwarf galaxies have smaller reservoirs of molecular gas.

\revise{We find a mean CO(2-1) fraction (within pixels above 3$\sigma_{\rm rms}$ limit) of $\DTLfetch{NGC2146}{variable}{COcontr_mean}{value}\pm \DTLfetch{NGC2146}{variable}{COcontr_std}{value} \%$ (error is the standard deviation) in NGC1246} and $\DTLfetch{NGC2976}{variable}{COcontr_mean}{value} \pm \DTLfetch{NGC2976}{variable}{COcontr_std}{value} \%$ for NGC~2976. We also derive a global value for the CO fraction, by dividing the integrated fluxes of the CO(2-1) and NIKA2 1.15\,mm maps (limited to above $3\sigma_{\rm rms}$ levels). \revise{The global CO fraction amounts to $\DTLfetch{NGC2146}{variable}{COcontr_global}{value}\%$ in NGC~2146,} and lower ($\DTLfetch{NGC2976}{variable}{COcontr_global}{value}\%$) in the dwarf galaxy NGC~2976.

\section{SED modeling}\label{section-SEDmodeling}
\revise{As mentioned in Sect.~\ref{section-millimeteremission}, the emission at millimeter wavelengths is a combination of dust emission and the RC emission components: the thermal free-free and the nonthermal synchrotron emission. The observed continuum flux density at millimeter frequency $\nu$, $S^{\rm mm}_{\nu}$, can be expressed as
\begin{equation}
S^{\rm mm}_{\nu}=S^{\rm dust}_{\nu} + S^{\rm ff}_{\nu}+ S^{\rm syn}_{\nu},
\end{equation}
where $S^{\rm dust}_{\nu}$ is the thermal emission from dust, and $S^{\rm RC}_{\nu}= S^{\rm ff}_{\nu}+S^{\rm syn}_{\nu}$ is the total flux of the RC emission with $S^{\rm ff}_{\nu}$ being the thermal free-free emission and $S^{\rm syn}_{\nu}$ the nonthermal synchrotron emission.} \par

\revise{The synchrotron component has a power-law spectrum, $S^{\rm syn}_{\nu}= A_{\rm syn}\nu^{-\alpha_{\rm n}}$, with $\alpha_{n}$ the synchrotron spectral index and $A_{\rm syn}$ a scaling factor.}\par

\revise{The free-free emission is linearly proportional to the free-free Gaunt factor, $g_{\rm ff}(\nu)$, assuming optically thin emission~\citep{Draine2011}. We model it as $S_{\nu}^{\rm ff}= A_{\rm ff}\,g_{\rm ff}(\nu)$, where $A_{\rm ff}$ is a scaling coefficient which depends weakly on the electron temperature\footnote{We set $T_{e}$ equal to $10^4$\,K and note that changing $T_{e}$ from $10^4$\,K to $5000$\,K results in $\lesssim 3\%$ difference in $A_{1}$.} $T_{e}$.
}\par

\revise{Although the dust emission in the IR to millimeter range can represent a continuum of temperatures, we model the dust emission assuming a double-component Modified Black-Body (MBB) radiation, one from a warm and the other from a cold dust component:
\begin{equation}
S^{\rm dust}_{\nu}= \kappa_{0}\, \left( \frac{\nu}{\nu_{0}}\right)^{\beta_{c}}\frac{M_{c}}{D^{2}} B_{\nu}(T_{c})+\kappa_{0}\, \left( \frac{\nu}{\nu_{0}}\right)^{\beta_{w}}\frac{M_{w}}{D^{2}}B_{\nu}(T_{w}),
\label{eq-2MBB}
\end{equation}
in which $M$, $T$, and $\beta$ are dust mass, temperature, and emissivity index, respectively, and $B_{\nu}(T)$ is the Planck function. The suffixes $c$ and $w$ denote the cold and warm dust components. For the dust absorption coefficient, we use the value of $\kappa_{0}(200\,\mu m)=0.66\,{\rm m^2}\,{\rm kg}^{-1}$ following \cite{Ysard+2024}.}\par

\revise{In summary, we use the following equation to model the continuum emission in millimeter wavelengths,
\begin{multline}
 S^{\rm mm}_{\nu}= \kappa_{0}\, \left( \frac{\nu}{\nu_{0}}\right)^{\beta_{c}}\frac{M_{c}}{D^{2}}\,B_{\nu} (T_{c})\\+\kappa_{0}\, \left( \frac{\nu}{\nu_{0}}\right)^{2}\frac{M_{w}}{D^{2}}\,B_{\nu}(T_{w}) + A_{\rm ff}\,g_{\rm ff}(\nu) + A_{\rm syn}\nu^{-\alpha_{\rm n}}
\label{eq-main}
\end{multline}
Eq.~\ref{eq-main} includes 9 free parameters, namely $M_{c}$, $T_{c}$, $\beta_{c}$, $M_{w}$, $T_{w}$, $\beta_{w}$, $A_{\rm ff}$, $A_{\rm syn}$, and $\alpha_{\rm n}$. 
To reduce the number of free parameters and considering that warm dust constitutes only a small fraction of the total dust content of galaxies, we set $\beta_{w}=2$ throughout this work~\citep{Tabatabaei+2014}.}\par

\subsection{Global SED modeling}\label{section-globalSEDmodeling}
\revise{The SED model presented in Sect.~\ref{section-SEDmodeling} is used to fit the observed integrated flux densities from the IR to the radio wavelengths. This is required for a precise decomposition of the millimeter emission components and to obtain the global properties of dust in NGC~2146 and NGC~2976. }

\subsubsection{Photometry}\label{section-photometry}

We fit our SED model (Eq.~\ref{eq-main}) to the integrated flux \revise{densities 
at $24\,\mu$m}, $70\,\mu m$, $100\,\mu m$, $160\,\mu m$, $250\,\mu m$, $350\,\mu m$ and $500\,\mu m$, as well as the millimeter and centimeter domain including NIKA2 1.15\,mm and 2\,mm, Planck 1.38\,mm, RC 2.8\,cm, 6.2\,cm, 18\,cm, and 21\,cm. We obtain the integrated fluxes using the CASA software~\citep{CASApaper}. We integrate the maps in a circular aperture, using the task \textit{imstat} in CASA. \revise{The radius of the reference circular aperture used for the photometry is 160$^{\prime\prime}$ and 175$^{\prime\prime}$ for NGC~2146 and NGC~2976, respectively, which is almost equal to $R_{25}$ of each galaxy.
}\par

\revise{The resulting photometry values are listed in Table~\ref{table-photometry}.} The uncertainty of integrated fluxes in this table is computed following the method in Appendix \ref{appendix-errorestimation}. Additionally, the integrated fluxes reported for NIKA2 1.15\,mm and Planck 1.38\,mm are corrected for the CO contamination, as explained in Sect.~\ref{section-COcontamination} and Appendix~\ref{appendix-CO}. 

\subsubsection{Fitting method and results}\label{section-resultsglobal}
\revise{We employ a Bayesian approach to model the IR to radio SEDs on a global scale, utilizing all available maps in this range. We use the Markov Chain Monte Carlo (MCMC) method to implement the Bayesian approach. For each free parameter, we report the median of the posterior probability distribution functions as the resulting value and 20\%-80\% percentiles as error margins. During the fitting process, we accounted for the transmission functions of the various telescopes from which we obtained the data. The details of the fitting process are explained in Appendix \ref{appendix-MCMC}.}\par

\begin{figure}
 \includegraphics[width=0.48\textwidth]{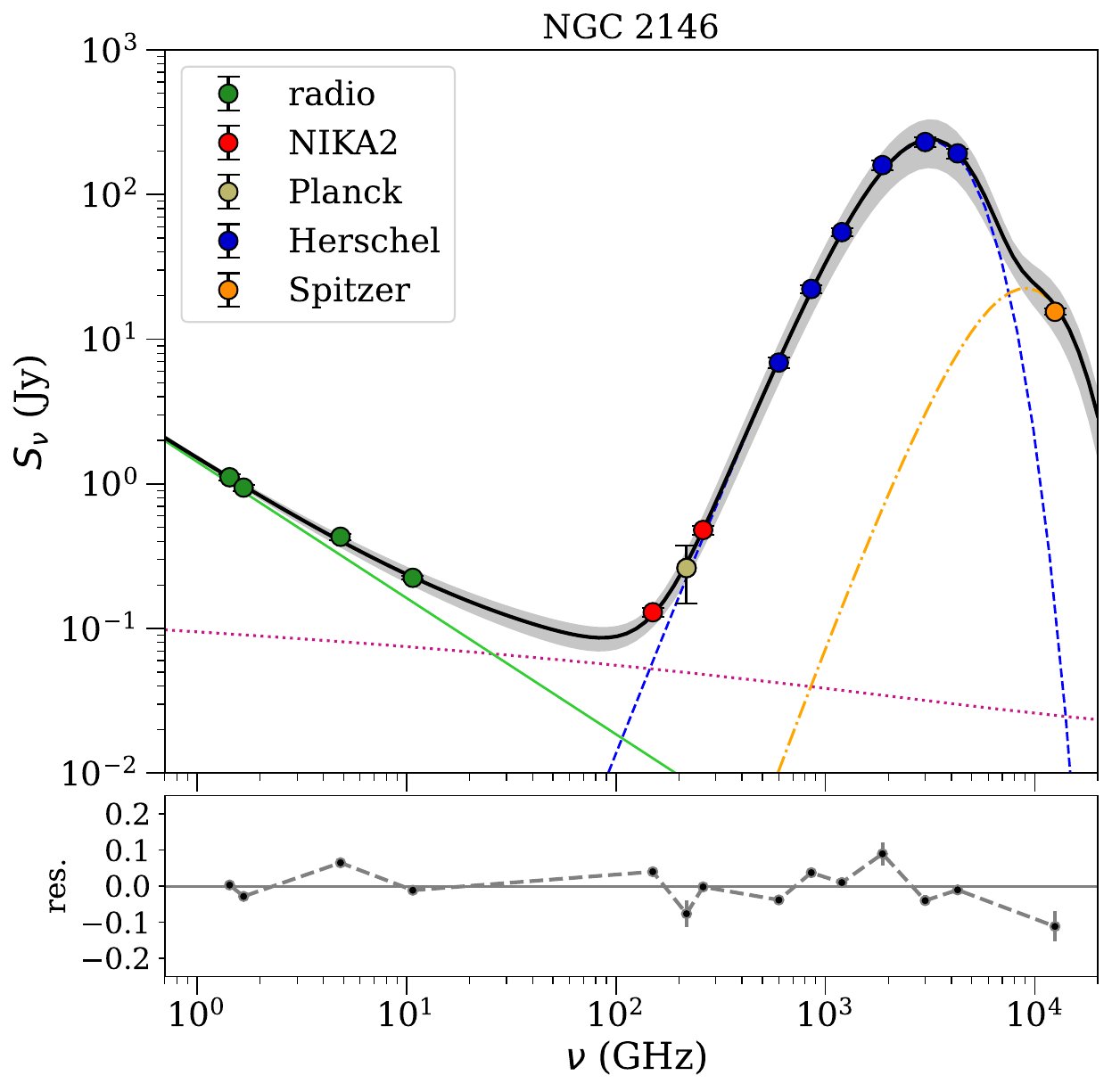}
 \includegraphics[width=0.48\textwidth]{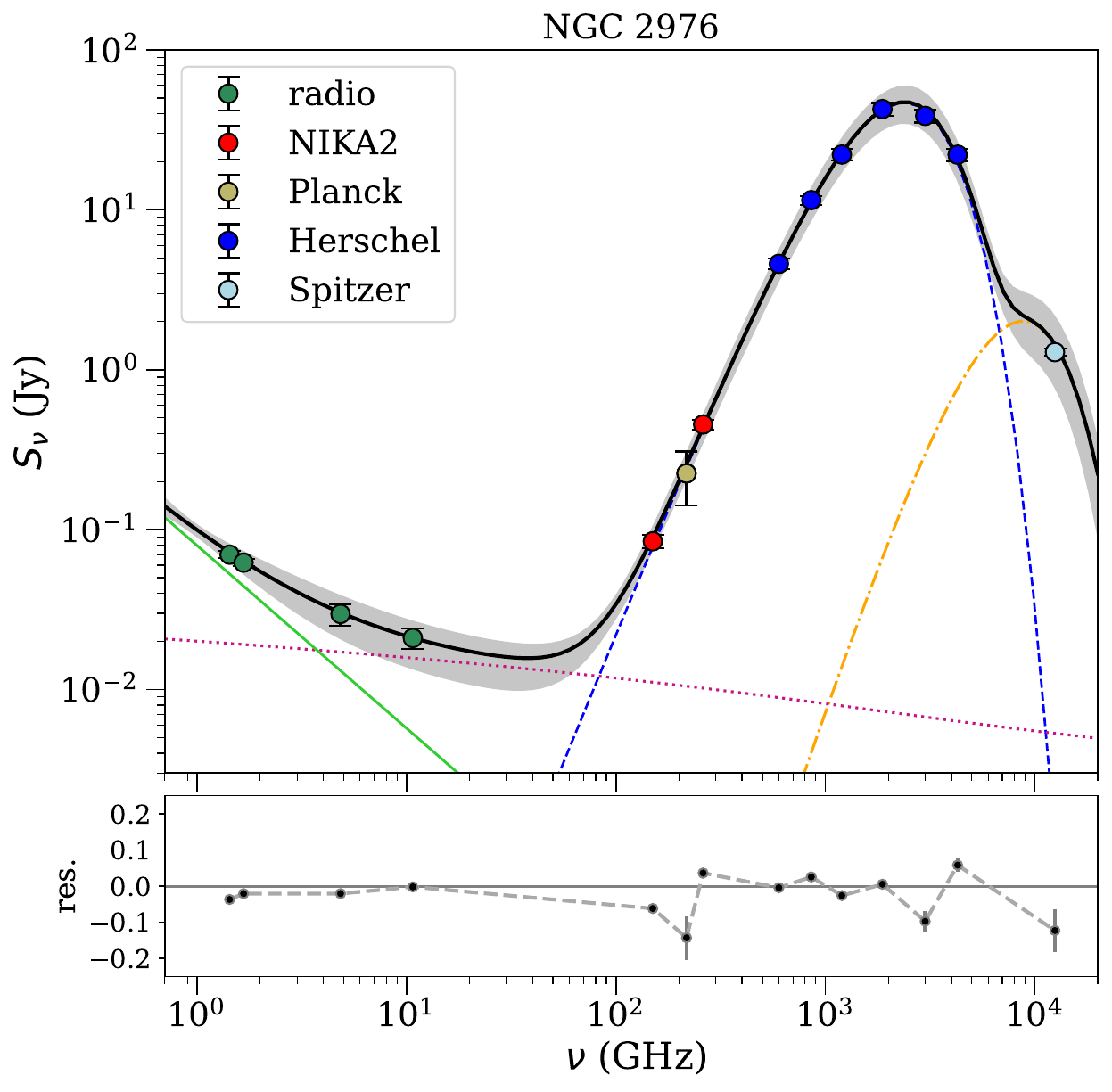}
 \caption{\revise{Global IR-to-radio SED ({\it solid black curve}) and its decomposed individual components of the cold dust ({\it blue dashed curve}), warm dust ({\it orange dot-dashed curve}), free-free ({\it purple dotted curve}), and synchrotron emission ({\it green solid curve}) in NGC~2146 (\textit{top}) and NGC~2976 (\textit{bottom}). The shaded grey area shows the uncertainty of the model. The relative residual of the model and observations with respect to the observation is shown at the bottom of each plot. The resulting SED parameters are listed in Table \ref{table-parameters}. }}
 \label{plot-globalSED}
\end{figure}

\revise{The SEDs modeled are illustrated in Fig.~\ref{plot-globalSED}, including the uncertainty of the modeled flux as a shaded area. The relative residual (observation-model/observation) of each data point is shown in the bottom panels of each SED in Fig.~\ref{plot-globalSED}. The relative residuals are $\DTLfetch{NGC2146}{variable}{relres_1}{value}\%$ and $\DTLfetch{NGC2146}{variable}{relres_2}{value}\%$ at 1.15\,mm and 2\,mm, respectively, in NGC~2146. 
For NGC~2976, they are $\DTLfetch{NGC2976}{variable}{relres_1}{value}\%$ and $\DTLfetch{NGC2976}{variable}{relres_2}{value}\%$ 
at 1.15\,mm and 2\,mm, respectively. At other wavelengths, the relative residuals are less than 10\% (20\%) in NGC~2146 (NGC2976).
} \par

\revise{The resulting parameters listed in Table~\ref{table-parameters} show that a large part of the dust content in NGC~2146 has a temperature of $\simeq 32$\,K, while only about 0.1\% of it can be attributed to the warm dust component with a temperature of $\simeq 88$\,K. A similar condition holds in NGC~2976, although, the major part of the dust content is slightly colder ($\simeq$27\,K) than that in NGC~2146. 
%
The total dust mass listed in Table~\ref{table-parameters} is consistent with previous findings reported by \cite{Hunt+2019} and \cite{Nersesian2019} for both of these galaxies. \cite{Galliano+2021} accounted for the mixing of the physical conditions in their dust modeling and obtained a larger dust mass (by a factor of 2). This disparity, as explained in \cite{Galliano+2021}, arises from the fact that the MBB model is an isothermal approximation and does not consider the coldest, less emissive, yet substantial regions in the galaxy, leading to a potential systematic underestimation of mass.}\par
%
%
%
\revise{Considering the RC components, we find a free-free fraction of $f_{\rm ff}=\DTLfetch{NGC2146}{variable}{thfr6}{value} ^{+\DTLfetch{NGC2146}{variable}{thfr6_re}{value}}_{-\DTLfetch{NGC2146}{variable}{thfr6_le}{value}} $ at 6\,cm and $f_{\rm ff}=\DTLfetch{NGC2146}{variable}{thfr21}{value} ^{+\DTLfetch{NGC2146}{variable}{thfr21_re}{value}}_{-\DTLfetch{NGC2146}{variable}{thfr21_le}{value}} $ at 21\,cm for NGC~2146. Higher fractions are obtained for NGC~2976 with $f_{\rm ff}$ equal to $\DTLfetch{NGC2976}{variable}{thfr6}{value} ^{+\DTLfetch{NGC2976}{variable}{thfr6_re}{value}}_{-\DTLfetch{NGC2976}{variable}{thfr6_le}{value}}$ and $\DTLfetch{NGC2976}{variable}{thfr21}{value} ^{+\DTLfetch{NGC2976}{variable}{thfr21_re}{value}}_{-\DTLfetch{NGC2976}{variable}{thfr21_le}{value}}$ at 6\,cm and 21\,cm, respectively, as expected for dwarf galaxies \citep{Tabatabaei+2017}.
Fig.~\ref{plot-globalSED} shows that, in NGC~2146, the free-free emission is dominating the SED between 25 and 240 GHz. The contribution of the free-free emission to total emission peaks at 67\% at 76\,GHz. Similar values have been reported in other starburst galaxies such as M~82 \citep{condon1992, Peel+2011} and NGC~3256 \citep{Michiyama+2020}, or in the central starburst regions of normal star-forming galaxies such as NGC~4945 \citep{Bendo+2016}, NGC~253 \citep{Peel+2011, Bendo+2015}, and NGC~1808 \citep{Salak+2017, Chen+2023}.}

The global continuum emission of these two galaxies is decomposed into its three components at NIKA2 wavelengths, i.e., the emission from the free-free, synchrotron, and dust (Eq.~\ref{eq-main}). The results of this decomposition are reported in Table~\ref{table-globaldecomposition}. As expected, moving from 1.15\,mm to 2\,mm, the contribution of dust emission decreases, while the contributions of free-free and synchrotron emission increase.

\revise{We note that at 2\,mm, \DTLfetch{NGC2976}{variable}{dustcontr_2}{value}\,\% of the emission is due to dust in NGC~2976, \revise{which agrees with that reported by \cite{Katsioli+2023} for the normal star-forming galaxy NGC~891}. However, this contribution is smaller in NGC~2146 (\DTLfetch{NGC2146}{variable}{dustcontr_2}{value}\,\%), indicating that this starburst galaxy has stronger radio emission at 2\,mm. Therefore, we emphasize that assuming a dominant dust component at 2\,mm is not always correct and it can depend on the ISM and star formation properties.}\par

\begin{table}
	\begin{center}
 \renewcommand{\arraystretch}{1.5} 
		\begin{tabular}{| l| l| l|} 
			\hline
			Galaxy & NGC~2146 & NGC~2976 \\
			\hline
			$T_{\rm{c}}$\,(K)& 
 $\DTLfetch{NGC2146}{variable}{T_c}{value}
 ^{+\DTLfetch{NGC2146}{variable}{T_c_re}{value}}
 _{-\DTLfetch{NGC2146}{variable}{T_c_le}{value}}$ &
 $\DTLfetch{NGC2976}{variable}{T_c}{value}
 ^{+\DTLfetch{NGC2976}{variable}{T_c_re}{value}}
 _{-\DTLfetch{NGC2976}{variable}{T_c_le}{value}}$\\
 		$T_{\rm{w}}$\,(K)& 
 $\DTLfetch{NGC2146}{variable}{T_w}{value}
 ^{+\DTLfetch{NGC2146}{variable}{T_w_re}{value}}
 _{-\DTLfetch{NGC2146}{variable}{T_w_le}{value}}$ &
 $\DTLfetch{NGC2976}{variable}{T_w}{value}
 ^{+\DTLfetch{NGC2976}{variable}{T_w_re}{value}}
 _{-\DTLfetch{NGC2976}{variable}{T_w_le}{value}}$ \\
 $M_{\rm{c}}$\,(${\rm M}_{\odot}$)	& 
 $\DTLfetch{NGC2146}{variable}{M_c}{value}
 ^{+\DTLfetch{NGC2146}{variable}{M_c_re}{value}}
 _{-\DTLfetch{NGC2146}{variable}{M_c_le}{value}} \times 10^{7}$ &
 $ \DTLfetch{NGC2976}{variable}{M_c}{value}
 ^{+\DTLfetch{NGC2976}{variable}{M_c_re}{value}}
 _{-\DTLfetch{NGC2976}{variable}{M_c_le}{value}} \times 10^{5}$ \\
 $M_{\rm{w}}$\,(${\rm M}_{\odot}$)	& 
 $\DTLfetch{NGC2146}{variable}{M_w}{value}
 ^{+\DTLfetch{NGC2146}{variable}{M_w_re}{value}}
 _{-\DTLfetch{NGC2146}{variable}{M_w_le}{value}} \times 10^{4}$ &
 $ \DTLfetch{NGC2976}{variable}{M_w}{value}
 ^{+\DTLfetch{NGC2976}{variable}{M_w_re}{value}}
 _{-\DTLfetch{NGC2976}{variable}{M_w_le}{value}} \times 10^{1}$ \\
			$\beta_{c}$	& 
 $\DTLfetch{NGC2146}{variable}{beta}{value}
 ^{+\DTLfetch{NGC2146}{variable}{beta_re}{value}}
 _{-\DTLfetch{NGC2146}{variable}{beta_le}{value}} $ & 
 $\DTLfetch{NGC2976}{variable}{beta}{value}
 ^{+\DTLfetch{NGC2976}{variable}{beta_re}{value}}
 _{-\DTLfetch{NGC2976}{variable}{beta_le}{value}} $ \\
			$A_{1}$ & 
 $\DTLfetch{NGC2146}{variable}{A1}{value}
 ^{+\DTLfetch{NGC2146}{variable}{A1_re}{value}}
 _{-\DTLfetch{NGC2146}{variable}{A1_le}{value}} $ &
 $ \DTLfetch{NGC2976}{variable}{A1}{value}
 ^{+\DTLfetch{NGC2976}{variable}{A1_re}{value}}
 _{-\DTLfetch{NGC2976}{variable}{A1_le}{value}} $\\
			$A_{2}$			& $ \DTLfetch{NGC2146}{variable}{A2}{value}
 ^{+\DTLfetch{NGC2146}{variable}{A2_re}{value}}
 _{-\DTLfetch{NGC2146}{variable}{A2_le}{value}} $ &
 $ \DTLfetch{NGC2976}{variable}{A2}{value}
 ^{+\DTLfetch{NGC2976}{variable}{A2_re}{value}}
 _{-\DTLfetch{NGC2976}{variable}{A2_le}{value}} $ \\
			$\alpha_{\rm n}$	& 
 $\DTLfetch{NGC2146}{variable}{alpha}{value}
 ^{+\DTLfetch{NGC2146}{variable}{alpha_re}{value}}
 _{-\DTLfetch{NGC2146}{variable}{alpha_le}{value}} $ &
 $ \DTLfetch{NGC2976}{variable}{alpha}{value}
 ^{+\DTLfetch{NGC2976}{variable}{alpha_re}{value}}
 _{-\DTLfetch{NGC2976}{variable}{alpha_le}{value}} $ \\
 \hline
 \noalign {\medskip}
 \end{tabular}
 \caption{{\revise{Resulting} SED parameters in the global modeling. The reported values are the median and the errors are 20 and 80 percent quantiles of the posterior probability distribution function of each parameter.}}
 \label{table-parameters}
	\end{center}
\end{table}

\begin{table}
\begin{center}
\renewcommand{\arraystretch}{1.1} 
\begin{tabular}{|c|cc|cc|} 
\hline
 Galaxy & \textbf{\,\,NGC~2146} & & \textbf{\,\,NGC~2976} \\
\cline{1-5} 
Wavelength& 1.15\,mm & 2\,mm & 1.15\,mm & 2\,mm\\ \hline
free-free & \DTLfetch{NGC2146}{variable}{ffcontr_1}{value}\,\% & \DTLfetch{NGC2146}{variable}{ffcontr_2}{value}\,\% & \DTLfetch{NGC2976}{variable}{ffcontr_1}{value}\,\% & \DTLfetch{NGC2976}{variable}{ffcontr_2}{value}\,\% \\
synchrotron & \DTLfetch{NGC2146}{variable}{synccontr_1}{value}\,\% & \DTLfetch{NGC2146}{variable}{synccontr_2}{value}\,\% & \DTLfetch{NGC2976}{variable}{synccontr_1}{value}\,\% & \DTLfetch{NGC2976}{variable}{synccontr_2}{value}\,\% \\
dust & \DTLfetch{NGC2146}{variable}{dustcontr_1}{value}\,\% & \DTLfetch{NGC2146}{variable}{dustcontr_2}{value}\,\% & \DTLfetch{NGC2976}{variable}{dustcontr_1}{value}\,\% & \DTLfetch{NGC2976}{variable}{dustcontr_2}{value}\,\% \\
\hline
\end{tabular}
\caption{Contribution of the three constituents of continuum emission, namely free-free, synchrotron, and dust emissions, with respect to the modeled flux at 1.15\,mm and 2\,mm.}
\label{table-globaldecomposition}
\end{center}
\end{table}

\subsection{Resolved SED modeling}\label{section-resolvedSEDmodeling}
\revise{As follows, we use a pixel-wise analysis of the IR-to-radio SED to map the millimeter emission constituents (dust, free-free, and synchrotron), dust physical parameters, and the free-free fraction in NGC~2146 and NGC~2976.}
\subsubsection{\revise{Photometry}} 
We chose 18$^{\prime\prime}$ \reviseII{(the FWHM at 2\,mm and 250\,$\mu$m)} as our working resolution to study the SED on spatial scales of $\sim$1.6\,kpc and $\sim$0.3\,kpc in NGC~2146 and NGC~2976, respectively. Therefore, the Planck 1.38\,mm, SPIRE 350\,$\mu$m, SPIRE 500\,$\mu$m, radio 2.8\,cm, and radio 6\,cm maps {are excluded from} this analysis due to their poorer angular resolutions (Table~\ref{table-alldata}). We convolve the \textit{Spitzer} MIPS, \textit{Herschel} PACS and 1.15\,mm map to the 18$^{\prime\prime}$ resolution using Gaussian kernels.
\reviseII{The various bands sampling the dust SED are observed with different instruments, from space or on the ground, with very different final performances regarding flux conservation and noise subtraction. To homogenize the spatially-resolved dust SED (both in terms of recovered flux fraction and noise properties), and in order to obtain reliable parameters from the SED fitting, as unbiased as possible, we build simulations as explained in Appendix~\ref{appendix-dataprocessing}.}

The radio maps are convolved similarly with a Gaussian kernel with CASA software, using task \textit{imsmooth}. Ultimately, we reproject all the maps to the same geometry and coordinate system and regrid them to 6$^{\prime\prime}$ pixel size. We note that the pixel size is chosen to be $1/3$ of the beam size to ensure Nyquist sampling. \revise{Our analysis is limited to pixels with a signal-to-noise ratio, SNR$>3$, for all maps used over the entire analysis. }

\subsubsection{Fitting method and results} \label{section-resultsresolved}
\revise{The lack of resolved RC maps at 2.8 and 6\,cm prevents constraining the synchrotron spectral index if chosen as a variable. Therefore, we opt to use the globally obtained $\alpha_{\rm n}$ (Table~\ref{table-parameters}) for different pixels.} We should note that this assumption is not entirely realistic, because studies show that $\alpha_{\rm n}$ can vary in different parts of a galaxy~\citep[e.g.][]{Tabatabaei+2013a, Baes+2010}. However, to constrain the SEDs locally, more resolved maps are needed covering the radio domain, which is not available currently. \revise{Moreover, to reduce parameter distribution biases caused by degeneracy and noise-injected correlations between the dust SED parameters, particularly that of the dust temperature and emissivity index \citep[e.g.,][]{Shetty2009, Juvela+2012, Kelly+2012}, we also fix $\beta_{c}$ to its global value (see Table~\ref{table-parameters}). This choice is then assessed by repeating the analysis using different fixed $\beta_{c}$ values around its global value in each galaxy (Table ~\ref{table-pvalues}). The goodness of the SED fits is then evaluated for each $\beta_{c}$ selected. In the context of the Bayesian approach, posterior predictive \textit{p}-values are used for assessing the quality of a fit \citep{Galliano+2021, Galliano2022}. In this method, a \textit{p}-value close to 0.5 suggests that the model is a good fit for the data, while a \textit{p}-value close to 0 or 1 suggests that the model may not be a good fit. The \textit{p}-values per pixel for each $\beta_{c}$ are calculated and reported in Table~\ref{table-pvalues}. 
The data is best reproduced taking $\beta_{c}=1.8$ for NGC~2146 and $\beta_{c}=1.4$ for NGC~2976 which agree with the global results (Table~\ref{table-parameters}) taking into account their uncertainties. We note that fixing $\beta_{c}$ is a first approximation, as it can spatially vary inside a galaxy\citep{Tabatabaei+2014}.}\par

\begin{table}
 \centering
 \textbf{NGC~2146}\\
 \begin{tabular}{cccccc}
 \hline
 $\beta_{c}$ & 1.6 & 1.7 & 1.8 & 1.9 & 2.0\\
 \textit{p}-value& 0.38& 0.44 & 0.49 & 0.53 & 0.56 \\
 \hline
 \end{tabular}
 \textbf{\\ NGC~2976}\\
 \begin{tabular}{ccccccc}
 \hline
 $\beta_{c}$ & 1.1 & 1.2 & 1.3 & 1.4 & 1.5 & 1.6\\
 \textit{p}-value& 0.22 & 0.32 & 0.41 & 0.48 & 0.53 & 0.55\\
 \hline
 \end{tabular}
 \caption{\revise{ Mean of \textit{p}-values in the pixel-wise SED modeling with a fixed $\beta_{c}$.}}
 \label{table-pvalues}
\end{table}

\begin{figure}
 \includegraphics[width=0.45\textwidth]{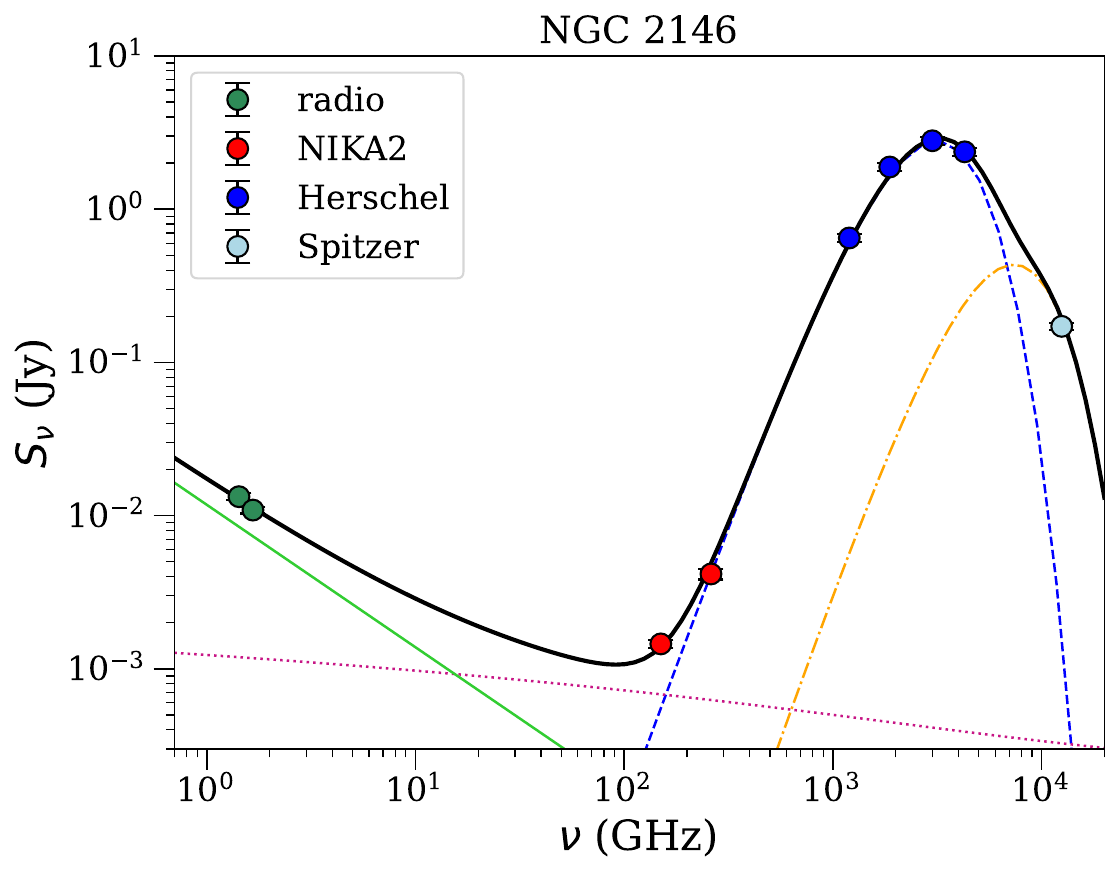}
 \includegraphics[width=0.45\textwidth]{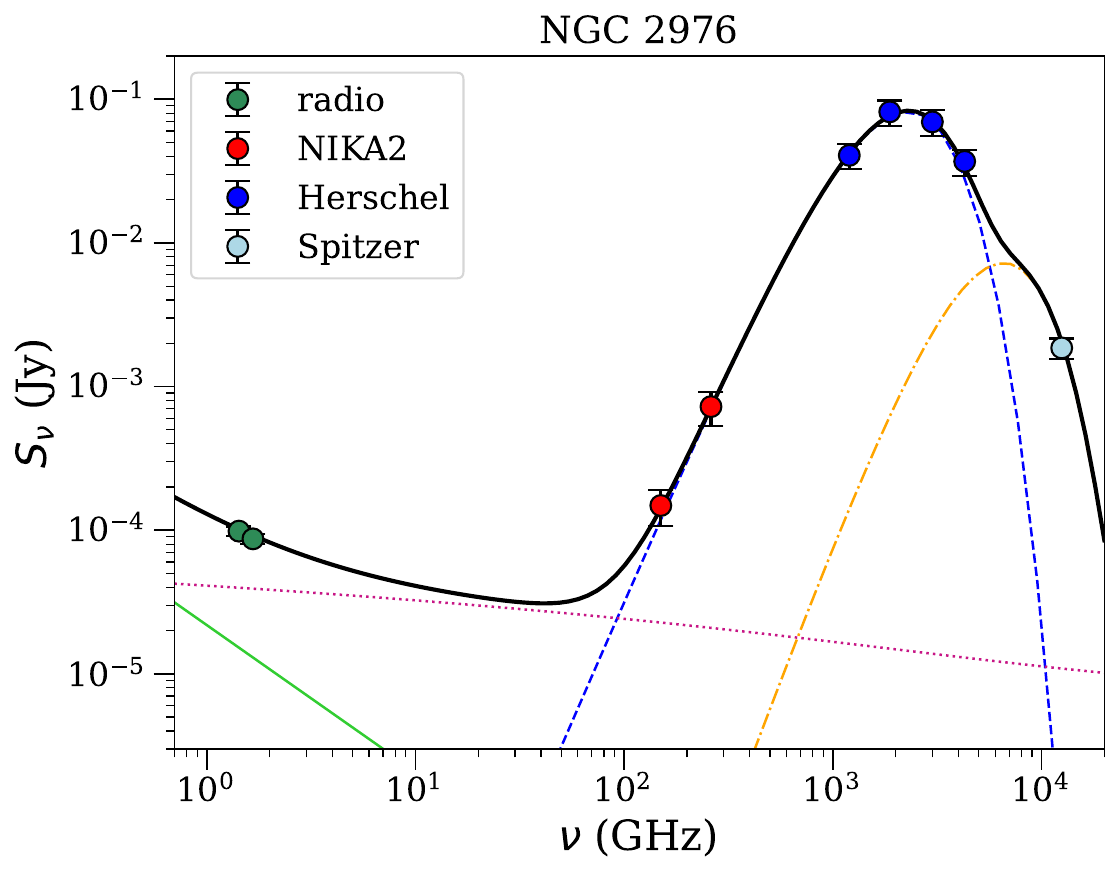}
 \caption{\revise{The IR-to-radio SEDs of a selected pixel in NGC~2146 (\textit{top}) and NGC~2976 (\textit{bottom}). Curves show the cold dust ({\it blue dashed curve}), warm dust ({\it orange dot-dashed curve}), free-free ({\it purple dotted curve}), and synchrotron ({\it green solid curve}) emissions.}}
 \label{plot-sampleSED}
\end{figure}

\begin{figure*}
	\centering
	\includegraphics[width=0.7\textwidth]{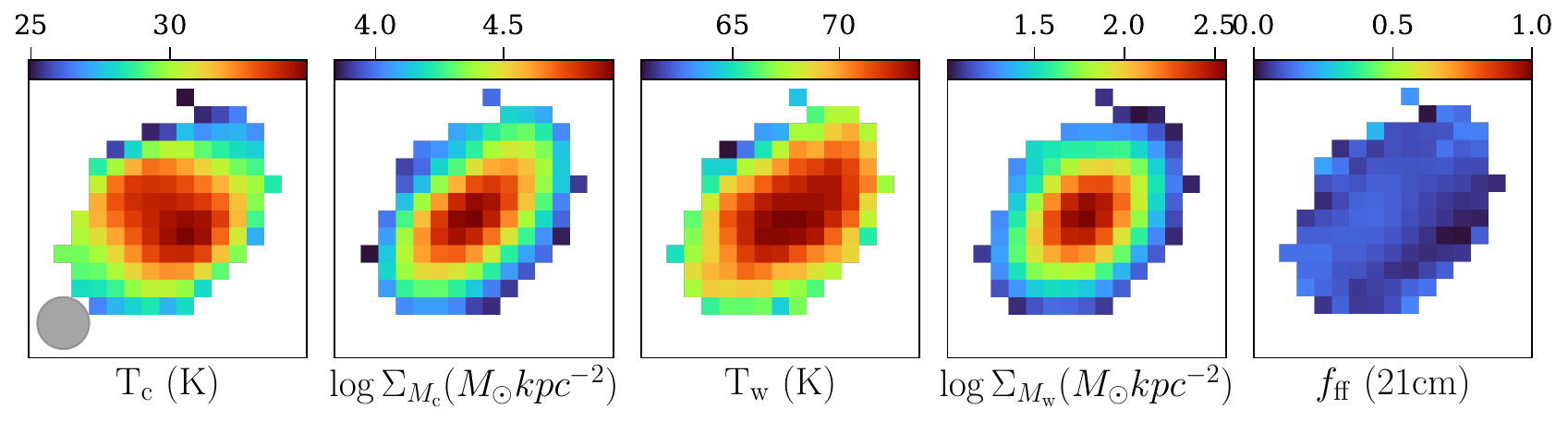}
 \includegraphics[width=0.7\textwidth]{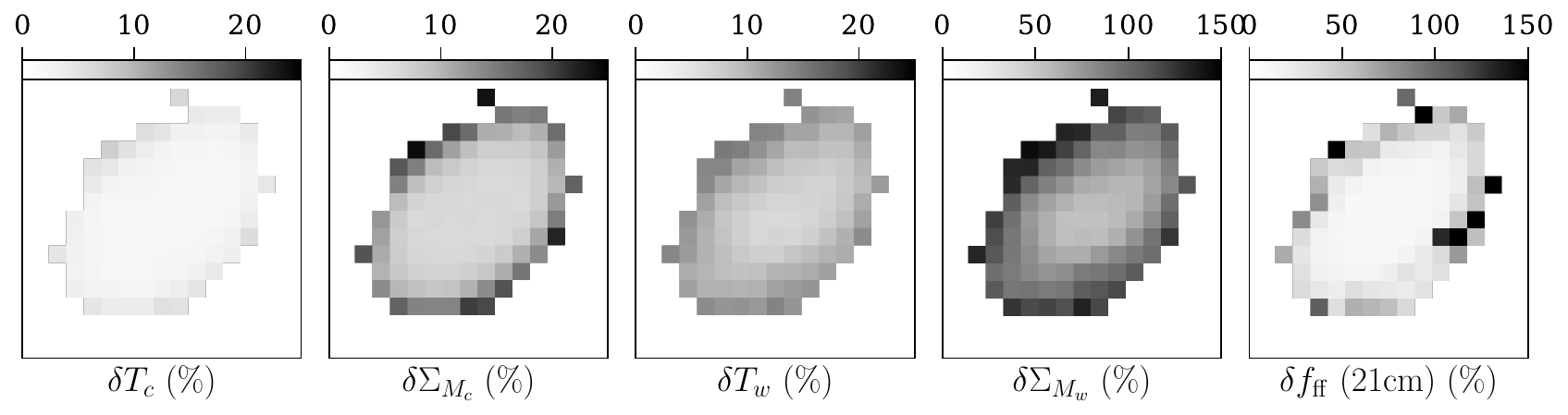}
	\caption{\revise{ Maps of the resulting SED parameters for NGC~2146 
 for $6^{\prime\prime}$ pixels ($\simeq$~525\,pc). From left-to right: cold dust temperature $T_{\rm c}$\,(K), cold dust mass surface density $\Sigma_{M_{\rm c}}\,({\rm M}_{\odot}{\rm kpc}^{-2})$, warm dust temperature $T_{\rm c}$\,(K), warm dust mass surface density $\Sigma_{M_{\rm c}}\,({\rm M}_{\odot}{\rm kpc}^{-2})$, and thermal free-free fraction at 21\,cm $f_{th}(21\,cm)$. The SED analysis uses only pixels with intensities $>3\sigma_{\rm rms}$ at all wavelengths. The beam width of $18^{\prime\prime}$ is demonstrated in the lower-left corners. Also shown is the percentage of relative error of each parameter (\it{bottom row}).} }
	\label{plot-2146parametermaps}
\end{figure*}

\begin{figure*}
	\centering
	\includegraphics[width=0.7\textwidth]{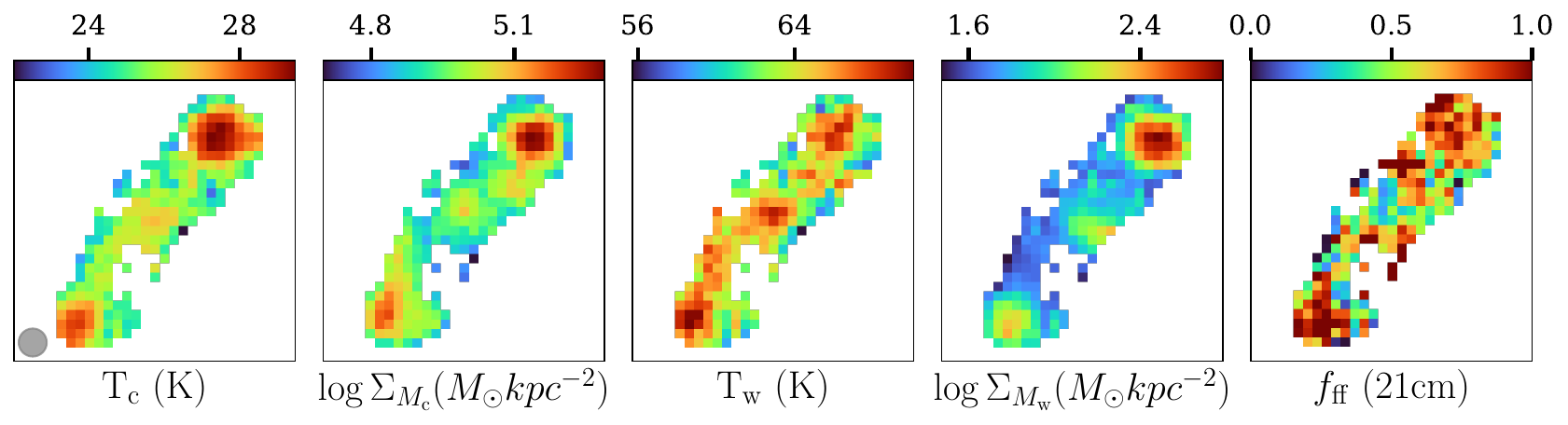}
 \includegraphics[width=0.7\textwidth]{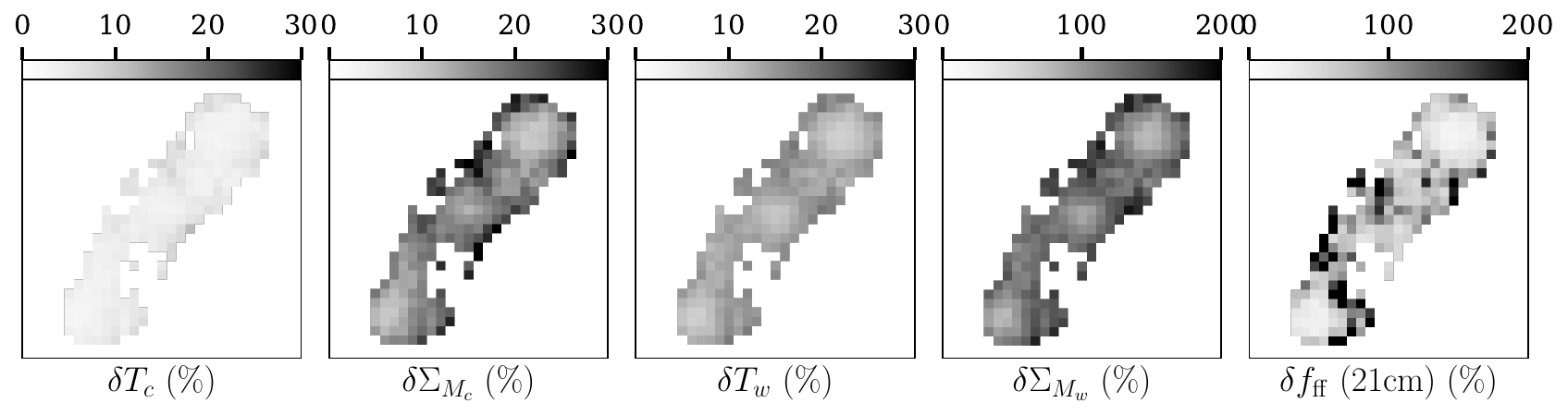}
	\caption{\revise{ Same as Fig.~\ref{plot-2146parametermaps} for NGC~2976. The SED analysis uses only pixels with intensities $>1\sigma_{\rm rms}$ at all wavelengths for this dwarf galaxy.} }
	\label{plot-2976parametermaps}
\end{figure*}

\revise{For the resolved study, Eq(4) is applied using the flux densities per pixel instead of the integrated flux densities. 
Similar to the global study, we implement the MCMC Bayesian inference. The fitting method and uncertainties are explained in detail in Appendix~\ref{appendix-errorestimation} and~\ref{appendix-MCMC}. Figure~\ref{plot-sampleSED} shows the resulting SEDs of a selected pixel for each galaxy. The resulting parameter maps, i.e., medians of the posterior distribution functions for $T_{\rm c}$, $\Sigma_{M_{\rm c}}$, $T_{\rm w}$, $\Sigma_{M_{\rm w}}$, $f_{\rm{ff}}(21\,cm)$ along with their relative uncertainty maps, are shown in Figs.~\ref{plot-2146parametermaps}~and~\ref{plot-2976parametermaps}.} We note that the surface density of the dust mass is corrected for the inclination of each galaxy. The mean value and dispersion of each parameter are reported in Table~\ref{table-paramstat}. These maps are discussed as follows.\par

\revise{In NGC~2146, no spiral arms or star-forming regions are evident in the resulting SED parameter maps at the resolution of this study ($\sim 1.6$~kpc) as expected from the observed maps (Fig.~\ref{plot-observedmaps}). With $T_{\rm{c}}$ varying mainly in a narrow range of $\DTLfetch{NGC2146}{variable}{temp_min_c}{value}^{+\DTLfetch{NGC2146}{variable}{temp_min_c_er}{value}}_{-\DTLfetch{NGC2146}{variable}{temp_min_c_el}{value}}$ to $\DTLfetch{NGC2146}{variable}{temp_max_c}{value}^{+\DTLfetch{NGC2146}{variable}{temp_max_c_er}{value}}_{-\DTLfetch{NGC2146}{variable}{temp_max_c_el}{value}}$\,K, the cold dust component has a higher surface density in the inner disc of NGC~2146 reaching a maximum of $\Sigma_{M_{\rm c}}=\DTLfetch{NGC2146}{variable}{mdens_max_c}{value}^{\DTLfetch{NGC2146}{variable}{mdens_max_c_er}{value}}_{-\DTLfetch{NGC2146}{variable}{mdens_max_c_el}{value}}\times10^{6}\,{\rm M}_{\odot}{\rm kpc}^{-2}$ in the center of this galaxy. 
The warm dust component ($\DTLfetch{NGC2146}{variable}{temp_min_w}{value}^{+\DTLfetch{NGC2146}{variable}{temp_min_w_er}{value}}_{-\DTLfetch{NGC2146}{variable}{temp_min_c_el}{value}} <T_{\rm{w}}< \DTLfetch{NGC2146}{variable}{temp_max_w}{value}^{+\DTLfetch{NGC2146}{variable}{temp_max_w_er}{value}}_{-\DTLfetch{NGC2146}{variable}{temp_max_w_el}{value}}$\,K) becomes denser towards the center reaching a peak of $\Sigma_{M_{\rm w}}=\DTLfetch{NGC2146}{variable}{mdens_max_w}{value}^{\DTLfetch{NGC2146}{variable}{mdens_max_w_er}{value}}_{-\DTLfetch{NGC2146}{variable}{mdens_max_w_el}{value}}\times10^{4}\,{\rm M}_{\odot}{\rm kpc}^{-2}$. The mass surface density of the warm dust is lower than that of the cold dust by more than 2 orders of magnitude while its variation over the galaxy is 3 times larger. Both $T_{\rm{c}}$ and $T_{\rm{w}}$ are higher in the inner part of this galaxy, with slightly different distributions, but overall consistent within error margins.}\par 
%
%
\revise{ At about 0.3\,kpc resolution, the resulting SED parameter maps exhibit resolved structures in NGC~2976. Cold dust with temperatures ranging between $\DTLfetch{NGC2976}{variable}{temp_min_c}{value}_{-\DTLfetch{NGC2976}{variable}{temp_max_c_el}{value}}^{+\DTLfetch{NGC2976}{variable}{temp_max_c_er}{value}}$\,K and $\DTLfetch{NGC2976}{variable}{temp_max_c}{value}_{-\DTLfetch{NGC2976}{variable}{temp_min_c_el}{value}}^{+\DTLfetch{NGC2976}{variable}{temp_min_c_er}{value}}$\,K 
is denser in the star-forming regions particularly the one in the north-west hosting the maximum surface density of $\Sigma_{M_{\rm c}}=\DTLfetch{NGC2976}{variable}{mdens_max_c}{value}^{\DTLfetch{NGC2976}{variable}{mdens_max_c_er}{value}}_{-\DTLfetch{NGC2976}{variable}{mdens_max_c_el}{value}}\times10^{5}\,{\rm M}_{\odot}{\rm kpc}^{-2}$. 
The warm component with temperatures ranging between $\DTLfetch{NGC2976}{variable}{temp_min_w}{value}_{-\DTLfetch{NGC2976}{variable}{temp_max_w_el}{value}}^{+\DTLfetch{NGC2976}{variable}{temp_max_w_er}{value}}$\,K and $\DTLfetch{NGC2976}{variable}{temp_max_w}{value}_{-\DTLfetch{NGC2976}{variable}{temp_min_w_el}{value}}^{+\DTLfetch{NGC2976}{variable}{temp_min_w_er}{value}}$\,K follows a similar distribution, although it is more concentrated in the star-forming regions compared to the cold dust (the star-forming vs non-star-forming contrast is 5 times larger than that of the cold dust). The cold and warm components have similar temperature distributions across NGC~2976, although $T_{\rm w}$ has a higher fluctuation than $T_{\rm c}$. This indicates that both components are effectively heated by young massive stars in this galaxy.}\par 
%
%
Figures~\ref{plot-2146parametermaps} and~\ref{plot-2976parametermaps} also present the maps of the free-free fraction, $f_{\rm ff}$, (Eq.~\ref{eq-thermalfractioneq}) at 21\,cm. \revise{We note that large relative errors $\delta f_{\rm ff}$ is due to poor sampling of the SED in the radio domain particularly in regions of lower SNR. }
 \revise{In NGC~2146, $f_{\rm ff}$ varies between 10\% in the outskirts to 20\% in the inner disk of this galaxy but with $\delta f_{\rm ff}$ as large as 50\%. The mean $f_{\rm ff}$ obtained is the same as the global value reported in Sect.~\ref{section-globalSEDmodeling} and agrees with that of spiral galaxies \citep{Tabatabaei+2017}.
In NGC~2976, about 100\% of the RC emission is due to the free-free emission in its star-forming regions in the north and south, with a small relative error. In other regions, $f_{\rm ff}$ is about 50\%, but largely uncertain. This is also consistent with the global analysis showing that the free-free emission has an important contribution to the observed RC emission from this dwarf galaxy.}

\begin{figure}
	\includegraphics[width=0.3\textheight]{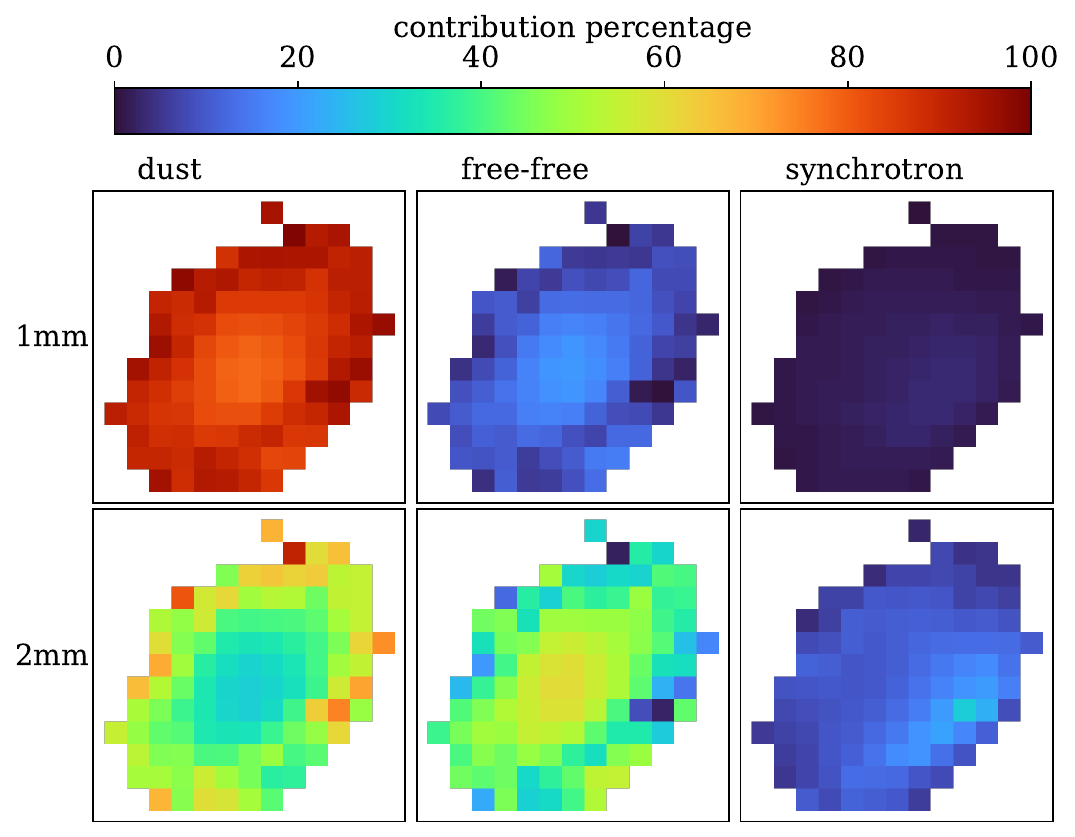}
	\includegraphics[width=0.3\textheight]{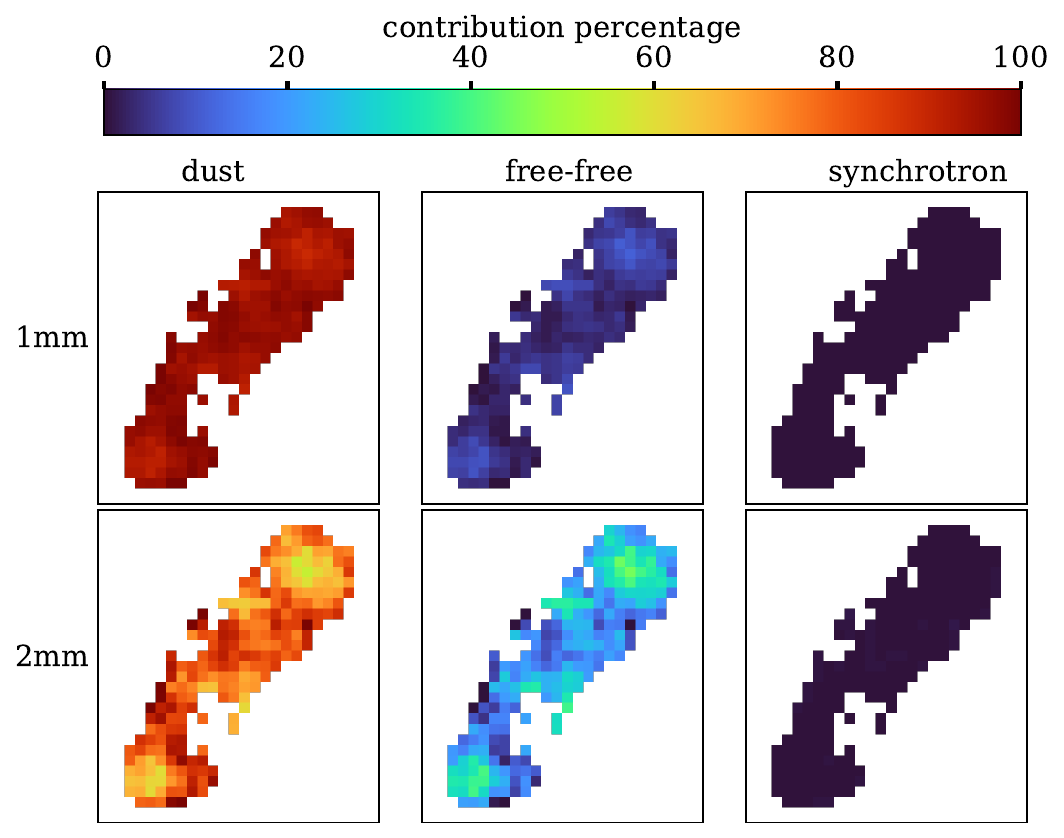}
	\caption{Contributions from dust (\textit{left column}), free-free (\textit{middle column}), and synchrotron (\textit{right column}) emission in the total modeled flux at 1.15\,mm (\textit{top row}) and 2\,mm (\textit{bottom row}) are shown in percentage for NGC~2146 (\textit{top panel}) and NGC~2976 (\textit{bottom panel}).} \label{plot-decomposition}
\end{figure}

\revise{The contributions of different components of the millimeter emission at 1.15\,mm and 2\,mm are shown in Fig.~\ref{plot-decomposition}. The mean of each map is consistent with its globally determined contribution listed in Table~\ref{table-globaldecomposition}.
Several known facts are demonstrated in this figure; for example, moving from 1.15\,mm to 2\,mm, the contribution of dust emission decreases while that of the RC emission increases, as expected from the model. Besides, Fig.~\ref{plot-decomposition} shows that in both galaxies and at both millimeter wavelengths, the RC emission is dominated by the free-free emission. }
\revise{In NGC~2976, the 1.15\,mm emission is almost entirely produced by dust, but at 2\,mm, the free-free emission becomes as strong as dust emission in the star-forming regions, resulting in equal contributions. This trend is more pronounced in NGC~2146, in which the contribution of the free-free emission increases more than that of dust in the inner disc, moving from 1.15\,mm to 2\,mm. In addition, synchrotron emission is negligible at both wavelengths in NGC~2976, whereas it has a contribution as large as 30\% at 2\,mm in NGC~2146. }
\revise{Note that a larger contribution of dust emission in NGC~2976 does not mean that a larger amount of dust is present in the ISM of this dwarf galaxy; it simply shows that NGC~2146 emits the free-free and synchrotron radiation much more intensely than NGC~2976 at millimeter wavelengths.This is clearly seen in Figs.~\ref{plot-NGC2146album} and~\ref{plot-NGC2976album} and Table~\ref{table-parameters} and~\ref{table-paramstat}. Consequently, we emphasize that dust may account for only 30\% of the observed 2 mm emission in certain regions of a starburst galaxy.}

\begin{figure}
 \centering
	\includegraphics[width=0.24\textwidth]{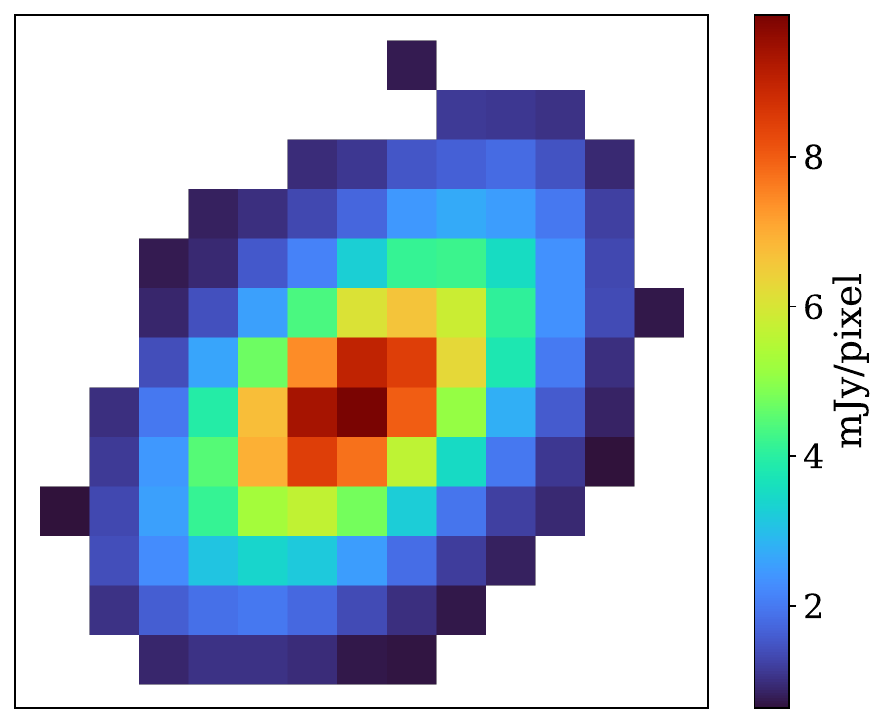}
	\includegraphics[width=0.23\textwidth]{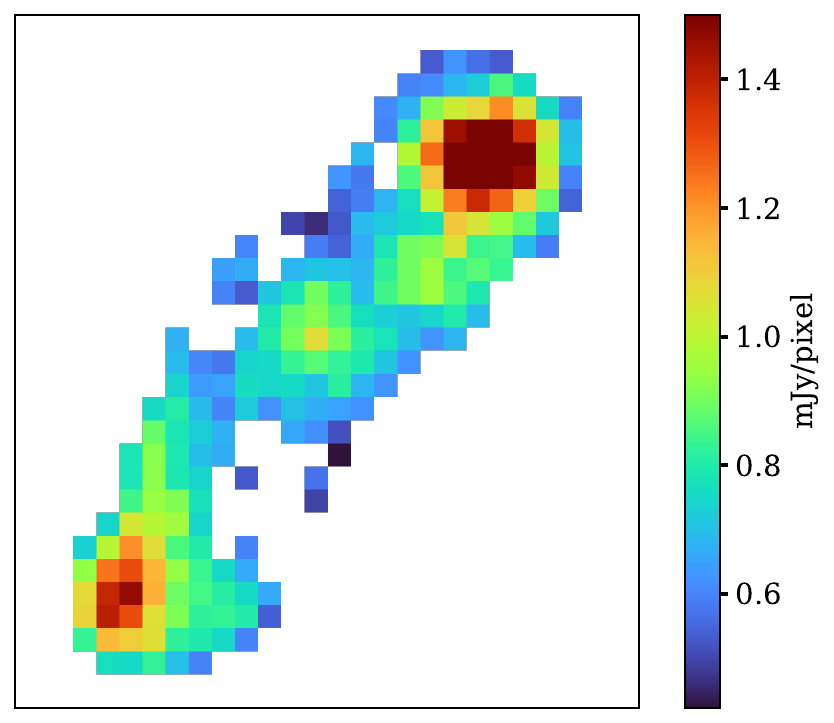}
	\caption{Map of pure dust emission at 1.15\,mm (after subtraction of the CO line contamination, free-free, and synchrotron emission) in mJy/pixel in NGC2146 (\textit{ top}) and NGC~2976 (\textit{bottom}).} \label{plot-dust}
\end{figure}

\revise{Finally, subtracting the contributions of the free-free and synchrotron emission from the observed millimeter emission, we present maps of the pure dust emission (sum of both warm and cold dust emission) at 1.15\,mm in Fig.~\ref{plot-dust} (see Eq.~\ref{eq-2MBB}). We remind the reader that the 1.15\,mm emission was already corrected for the contamination by the CO(2-1) line emission in Sect.~\ref{section-COcontamination}.}

\begin{table}
\begin{center}
\begin{tabular}{|l| l| l|}
\hline
Galaxy & NGC~2146 & NGC~2976 \\
\hline
T$_{\rm c}$ (K)&	$\DTLfetch{NGC2146}{variable}{temp_mean_c}{value} 
\pm \DTLfetch{NGC2146}{variable}{temp_std_c}{value}$ 
& $\DTLfetch{NGC2976}{variable}{temp_mean_c}{value} 
\pm \DTLfetch{NGC2976}{variable}{temp_std_c}{value}$ \\
$\Sigma_{M_{\rm c}}$ (${\rm M}_{\odot}{\rm kpc}^{-2}$)	&
$(\DTLfetch{NGC2146}{variable}{mdens_mean_c}{value} 
\pm \DTLfetch{NGC2146}{variable}{mdens_std_c}{value}) 
\times 10^{5}$ 
& $(\DTLfetch{NGC2976}{variable}{mdens_mean_c}{value} 
\pm \DTLfetch{NGC2976}{variable}{mdens_std_c}{value}) \times 10^{4}$ \\

T$_{\rm w}$ (K)&	$\DTLfetch{NGC2146}{variable}{temp_mean_w}{value} 
\pm \DTLfetch{NGC2146}{variable}{temp_std_w}{value}$ 
& $\DTLfetch{NGC2976}{variable}{temp_mean_w}{value} 
\pm \DTLfetch{NGC2976}{variable}{temp_std_w}{value}$ \\
$\Sigma_{M_{\rm w}}$ (${\rm M}_{\odot}{\rm kpc}^{-2}$)	&
$(\DTLfetch{NGC2146}{variable}{mdens_mean_w}{value} 
\pm \DTLfetch{NGC2146}{variable}{mdens_std_w}{value}) 
\times 10^{3}$ 
& $(\DTLfetch{NGC2976}{variable}{mdens_mean_w}{value} 
\pm \DTLfetch{NGC2976}{variable}{mdens_std_w}{value}) \times 10^{2}$ \\

$f_{\rm ff}$	&	$\DTLfetch{NGC2146}{variable}{thfr_mean}{value} 
\pm \DTLfetch{NGC2146}{variable}{thfr_std}{value}$ 
& $\DTLfetch{NGC2976}{variable}{thfr_mean}{value} 
\pm \DTLfetch{NGC2976}{variable}{thfr_std}{value}$ \\
\hline
\noalign {\medskip}
\end{tabular}
\caption{\revise{ Mean of the resulting SED parameter maps with their standard deviation as errors. The free-free fraction $f_{\rm th}$ refers to the wavelength of 21\,cm.}}
\label{table-paramstat}
\end{center}
\end{table}

\section{Discussion} \label{section-discussion}
\revise{In previous sections, we showed that the NIKA2 observations enable constraining the IR-to-radio SED of galaxies in the millimeter domain. We determined the contribution of dust emission to millimeter emission at 18\arcsec angular resolution and presented maps of the mass and temperature of the cold and warm dust components in NGC~2976 and NGC~2146. } 
In this section, \revise{we explore any correlation between the NIKA2 maps and the SFR and molecular gas tracers in NGC~2146 and NGC~2976 as prototypes for starburst and dwarf galaxies, respectively. Then, we discuss the interplay between dust and total neutral gas in NGC~2976.}

\begin{figure*}[h]
 \centering
 \includegraphics[width=0.47\textwidth]{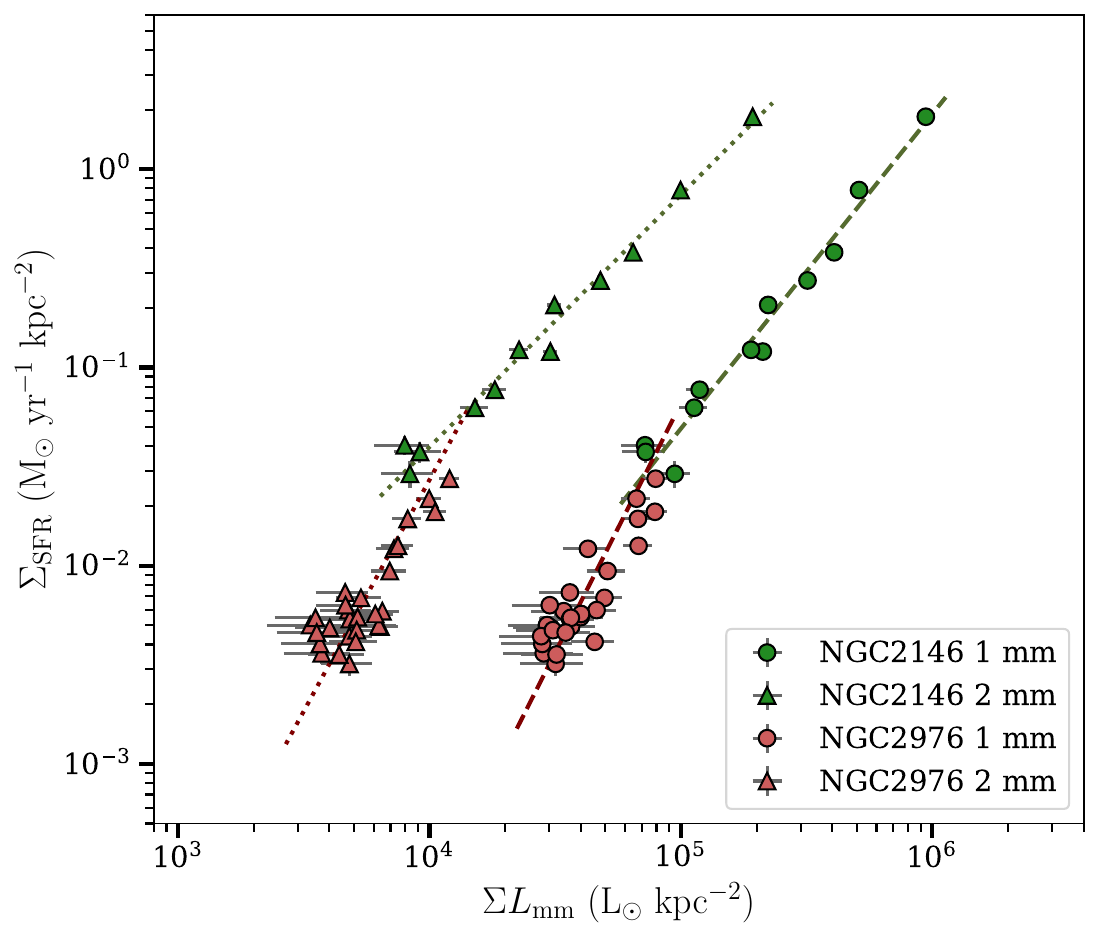}
 \includegraphics[width=0.47\textwidth]{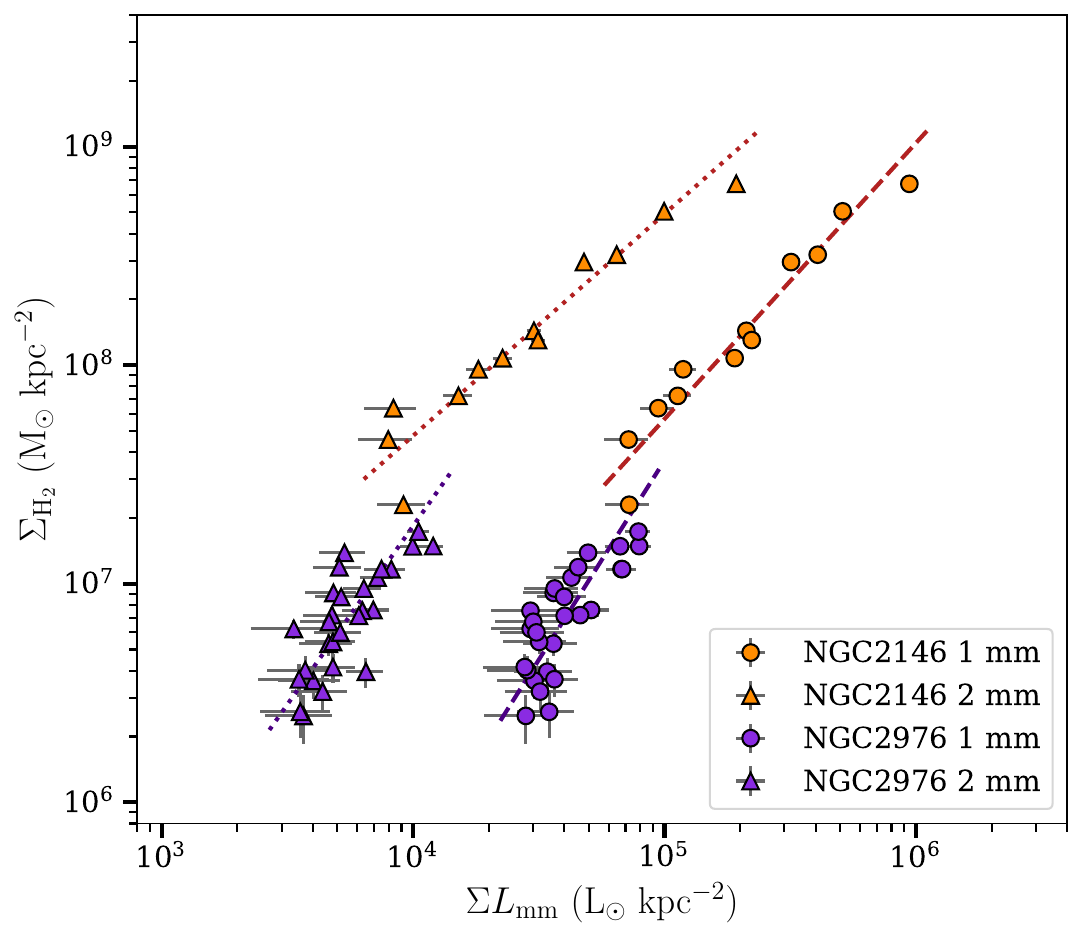}
	\caption{ \revise{\textit{Left}- Correlations between the surface densities of the SFR, $\Sigma_{\rm SFR}$, and the observed millimeter luminosity, $\Sigma{\rm L}_{\rm mm}$, at 1.15\,mm (\textit{circles}) and 2\,mm (\textit{triangles}) in NGC~2146 (\textit{green}) and NGC~2976 (\textit{red}). Lines show the OLS bisector fits (see Table~\ref{table-correlations}). \textit{Right}- Same as in the left but for the surface density of molecular gas, $\Sigma_{\rm H_2}$, instead of $\Sigma_{\rm SFR}$ as Y-axis. The maps are smoothed to one pixel per 18$^{\prime\prime}$ beam width to avoid internal correlation between pixels. Error bars indicate one $\sigma_{\rm rms}$ noise of each map, and when not shown, it means that they are smaller than the symbol size. }} \label{plot-correlations}
\end{figure*}

\subsection{\revise{Relating millimeter emission to SFR and molecular gas}}\label{section-tracingSFRandgaswithNIKA2}
\revise{As discussed in Sect.~\ref{section-millimeteremission}, the millimeter continuum emission can be strongly correlated with both the SFR and the molecular gas. The correlation with SFR is indicated by an agreement between the millimeter emission and the 24\,$\mu$m emission contours in both NGC~2146 and NGC~2976 in Fig.~\ref{plot-observedmaps}. We further investigate the correlations between the 1.15\,mm and 2\,mm emission observed with NIKA2 and the surface density of SFR $\Sigma_{\rm SFR}$ as well as the molecular gas surface density $\Sigma_{\rm H_2}$ in more detail.}

\revise{To compute $\Sigma_{\rm SFR}$, we use the dust emission at 24\,$\mu$m (assuming heating by young massive stars as its main source) in combination with dust-obscured FUV emission as a hybrid SFR tracer.} The calibration proposed by~\cite{Hao+2011} combines \textit{Spitzer} MIPS 24\,$\mu$m observations with GALEX FUV observations to compute the SFR with the following relation: ${\rm SFR}=3.36\times 10^{-44}(L_{\rm FUV} + 3.89 L_{\rm 24\,\mu m})$. \revise{In this relation, the SFR is expressed in M$_{\odot}$\,yr$^{-1}$ and luminosity in erg\,s$^{-1}$ and a Kroupa IMF~\citep{Kroupa2003} is assumed.} We additionally corrected for the inclination of each galaxy.

To compute the molecular hydrogen gas mass, we use CO(2-1) line emission as a tracer. We use the relation given in~\cite{Leroy+2009}, which is $\Sigma_{{\rm H}_2} [{\rm M}_{\odot}{\rm pc}^{-2}] =5.5\,I_{\rm CO}\,[{\rm K}\,{\rm km}\,{\rm s}^{-1}]$. In this relation,~\cite{Leroy+2009} have assumed $X_{\rm{CO}}=2\times10^{20}\,{\rm cm}^{-2} ({\rm K}\,{\rm km}\,{\rm s}^{-1})^{-1}$ equal to the solar neighborhood and the CO $J=2\rightarrow 1$ to $J=1\rightarrow 0$ ratio of $R_{21}=0.8$. This equation also includes a factor of 1.36 to account for Helium. Additionally, we corrected for the inclination of each galaxy. \revise{In the previous sections, we used a Nyquist sampling to avoid aliasing, choosing our working pixel size (6$^{\prime\prime}$) to be $1/3$ of our working resolution (18$^{\prime\prime}$). For cross-correlation analyses, to ensure that we are working with independent data points, we regrid the maps to one pixel per beam.} In addition, we limit our analysis to pixels above the $3\sigma_{\rm rms}$ limit of the NIKA2 maps.\par

We explore the correlations between the surface density of the observed luminosity at 1.15\,mm and 2\,mm (in the form of $\nu L_{\nu}$ and without subtracting the RC and CO contributions), $\Sigma{\rm L}_{\rm mm}$, and $\Sigma_{\rm SFR}$ in these galaxies. The left panel of Figure \ref{plot-correlations} shows the resulting correlations. The error bars indicate the $\sigma_{\rm rms}$ noise in the corresponding maps. We find the best-fit line in each case with the ordinary least squares bisector (OLS-bisector) method~\citep{Akritas+1996}. Slopes of the fitted lines in the $\log-\log$ plane and Pearson correlation coefficients $r_{\rm P}$ are reported in Table~\ref{table-correlations}.\par 

\revise{As shown in Fig.~\ref{plot-correlations} and Table~\ref{table-correlations}, a strong correlation holds between $\Sigma_{\rm SFR}$ and $\Sigma{\rm L}_{\rm mm}$ at both wavelengths and in both galaxies, although the correlations are tighter in NGC~2146. The relation of $\Sigma_{\rm SFR}$ vs the millimeter emission is super-linear, particularly in the case of the dwarf galaxy NGC~2976. In other words, the NIKA2 millimeter maps do not increase linearly with the SFR.} 
\revise{Exploring possible sources of this super-linearity, the correlations with the different components of the millimeter emission (RC and dust) are also investigated separately. In all cases, a strong correlation holds, however, a linear correlation is found only with the emission from the RC component, $\Sigma{\rm L}_{\rm RC}$ at both millimeter wavelengths and in both galaxies (Table~\ref{table-correlations}). Linear $\Sigma_{\rm SFR}-\Sigma{\rm L}_{\rm RC}$ is expected since the millimeter RC emission is dominated by the free-free emission, which is a linear tracer of the SFR. On the other hand, the correlation of $\Sigma_{\rm SFR}$ with the dust component of the millimeter emission, $\Sigma{\rm L}_{\rm dust}$, is not only super-linear, but also has the steepest fitted slope (in the log-log plane) in each galaxy. The super-linearity occurs as $\Sigma{\rm L}_{\rm SFR}$ drops faster than $\Sigma{\rm L}_{\rm dust}$ towards the low surface density tail of dust emission (or an excess of $\Sigma{\rm L}_{\rm dust}$-to-$\Sigma{\rm L}_{\rm SFR}$). Hence, the millimeter dust emission from galaxies must be partly produced by a diffuse or extended component that is heated by a diffuse ISRF instead of young, massive stars. 
The super-linearity of the $\Sigma_{\rm SFR}-\Sigma{\rm L}_{\rm mm}$ correlation is then linked to such an extended dust component. \reviseII{We note that the relation is much steeper in NGC~2976 than in NGC~2146 (Table~\ref{table-correlations}).} The more significant contribution of dust than the RC emission (as SFR tracer) in NGC~2976 than NGC~2146 (Sect.~\ref{section-resultsglobal} and~\ref{section-resultsresolved}) can lead to this difference if a significant portion of dust is in the form of diffuse, extended in NGC~2976 compared to that in NGC~2146. A relative excess of dust-to-RC emission in this dwarf galaxy is also evident by comparing its global SED in the centimeter-to-submillimeter range to that of NGC~2164 (Fig.~\ref{plot-globalSED}). An excess of extended cold dust emission was also reported in the LMC by \cite{Galliano+2011}. }

\revise{In fact, cold dust emission is often used to trace the gas content of galaxies~\citep{Boulanger+1996, Bianchi2022}.} Although the millimeter correlation with the SFR already implies a correlation with gas (due to the Kennicutt-Schmidt relation), we also investigated the relations between NIKA2 emission at 1.15 and 2\,mm with molecular gas. 

\revise{Fig.~\ref{plot-correlations}-right shows the molecular gas mass surface density, $\Sigma_{\rm{H}_2}$, plotted against $\Sigma{\rm L}_{\rm mm}$ (in from $\nu L_{\nu}$). The Pearson correlation coefficients and slopes of the best-fit lines are listed in Table~\ref{table-correlations}. 
In NGC~2146, the correlation coefficients between $\Sigma_{\rm H_2}$ and the millimeter components agree within errors. The correlation is always linear with $\Sigma{\rm L}_{\rm RC}$ and super-linear with $\Sigma{\rm L}_{\rm dust}$ at both millimeter wavelengths. In NGC~2976, $\Sigma_{\rm H_2}$ is well correlated with the millimeter emission components, although the scatter is larger with dust than with RC. }

\revise{Overall, the slopes of the $\Sigma_{\rm H_2}$ correlations with millimeter emission and its components are consistent with those of $\Sigma_{\rm SFR}$ within errors as expected from the Kennicutt-Schmidt relation. We note that, in NGC~2146, the correlations of both the $\Sigma_{\rm SFR}$ and $\Sigma_{\rm{H}_2}$ with $\Sigma{\rm L}_{\rm mm}$ are flatter and closer to linearity at 2\,mm than at 1.15\,mm which can be explained by a larger contribution of the RC emission at that longer wavelength in this starburst galaxy.} 

\revise{As discussed, the different correlations in these two galaxies simply reflect their very different properties such as the ISM density and dustiness, radiation field, and metallicity. A more general overview of the effect of these properties on the millimeter emission in galaxies, including normal star-forming spirals, will be presented in the forthcoming IMEGIN papers.}

\begin{table*}[h]
\renewcommand{\arraystretch}{1.3} 

\begin{center}
\begin{tabular}{|c|c|ccc|ccc|}
\hline
\textbf{NGC~2146}& X & Y & $a$ & $r_{\rm P}$ & Y & $a$ & $r_{\rm P}$\\
\hline
\multirow{3}{*}{1.15\,mm}& $\Sigma$L$_{\rm mm}$ &$\Sigma_{\rm SFR}$ & 1.5$\pm$0.1 & 0.99$\pm$0.01 & $\Sigma_{\rm H_2}$ & 1.3$\pm$0.1 & 0.97$\pm$0.02 \\
 &$\Sigma$L$_{\rm dust}$ &$\Sigma_{\rm SFR}$& 1.6$\pm$0.1 & 0.99$\pm$0.01 & $\Sigma_{\rm H_2}$ & 1.4$\pm$0.2 & 0.97$\pm$0.02\\
 &$\Sigma$L$_{\rm RC}$ &$\Sigma_{\rm SFR}$& 1.1$\pm$0.1 & 0.97$\pm$0.02 & $\Sigma_{\rm H_2}$ & 0.9$\pm$0.1 & 0.96$\pm$0.03 \\
\hline
\multirow{3}{*}{2\,mm}&$\Sigma$L$_{\rm mm}$ &$\Sigma_{\rm SFR}$ & 1.3$\pm$0.1 & 0.99$\pm$0.01 & $\Sigma_{\rm H_2}$ & 1.1$\pm$0.1 & 0.97$\pm$0.02 \\
&$\Sigma$L$_{\rm dust}$ &$\Sigma_{\rm SFR}$& 1.6$\pm$0.1 & 0.98$\pm$0.01 & $\Sigma_{\rm H_2}$ & 1.4$\pm$0.2 & 0.97$\pm$0.02 \\
&$\Sigma$L$_{\rm RC}$ &$\Sigma_{\rm SFR}$& 1.1$\pm$0.1 & 0.98$\pm$0.02 & $\Sigma_{\rm H_2}$ & 0.9$\pm$0.1 & 0.96$\pm$0.02\\

\hline
\end{tabular}
\\

\begin{tabular}{|c|c|ccc|ccc|}
\hline
\textbf{NGC~2976}& X & Y & $a$ & $r_{\rm P}$ & Y & $a$ & $r_{\rm P}$\\
\hline
\multirow{3}{*}{1.15\,mm}& $\Sigma$L$_{\rm mm}$ &$\Sigma_{\rm SFR}$ & 2.5$\pm$0.8 & 0.89$\pm$0.04 & $\Sigma_{\rm H_2}$ & 2.5$\pm$1.0 & 0.78$\pm$0.08 \\
 &$\Sigma$L$_{\rm dust}$ &$\Sigma_{\rm SFR}$& 2.9$\pm$0.7 & 0.92$\pm$0.03 & $\Sigma_{\rm H_2}$ & 2.7$\pm$0.9 & 0.87$\pm$0.05 \\
 &$\Sigma$L$_{\rm RC}$ &$\Sigma_{\rm SFR}$& 0.9$\pm$0.1 & 0.78$\pm$0.08 & $\Sigma_{\rm H_2}$ & 0.9$\pm$0.2 & 0.57$\pm$0.13\\
\hline
\multirow{3}{*}{2\,mm}&$\Sigma$L$_{\rm mm}$ &$\Sigma_{\rm SFR}$ & 2.3$\pm$0.5 & 0.86$\pm$0.05 & $\Sigma_{\rm H_2}$ & 2.3$\pm$0.7 & 0.78$\pm$0.08 \\
&$\Sigma$L$_{\rm dust}$ &$\Sigma_{\rm SFR}$& 2.9$\pm$0.7 & 0.91$\pm$0.03 & $\Sigma_{\rm H_2}$ & 2.8$\pm$0.8 & 0.87$\pm$0.05 \\
&$\Sigma$L$_{\rm RC}$ &$\Sigma_{\rm SFR}$& 1.0$\pm$0.1 & 0.79$\pm$0.08 & $\Sigma_{\rm H_2}$ & 0.9$\pm$0.2 & 0.57$\pm$0.13\\

\hline
\end{tabular}

\end{center}
\caption{\revise{ Parameters of the linear fits in logarithmic scales ($\log{\rm Y} = a\log {\rm X}+b$) in Fig.~\ref{plot-correlations} obtained
using the OLS-Bisector method \citep{Akritas+1996} with $r_{\rm P}$ the Pearson correlation coefficient.}}
\label{table-correlations}

\end{table*}

\subsection{Interplay of dust and gas in NGC~2976}\label{section-gasNGC2976}
In this section, we obtain the total neutral gas content of NGC~2976, explore the correlation of dust and neutral gas components, and investigate the role of dust in the formation of molecular gas in the ISM of this galaxy. As explained in Sect.~\ref{section-complementarydata}, this analysis is not done for NGC~2146 due to a lack of suitable HI data.

\subsubsection{Neutral gas and distribution of DGR}
The total gas mass is a sum of the molecular gas mass $M_{{\rm H}_2}$ and the atomic gas mass $M_{\rm HI}$, that is $M_{\rm gas}=M_{\rm HI}+M_{{\rm H}_2}$ (we neglect the relatively small amount of ionized gas). To compute $M_{{\rm H}_2}$ in NGC~2976, we use the emission of the CO(2-1) line as a well-known tracer and use the relation explained in Sect.~\ref{section-tracingSFRandgaswithNIKA2}. We report a total molecular gas mass of $7.3\times10^{7}\,{\rm M}_{\odot}$ throughout the map, consistent with the value reported in~\cite{Leroy+2009}. \revise{Inside our working aperture, i.e., for pixels above $3\sigma_{\rm rms}$ noise level of the NIKA2 1.15\,mm map, we find a total molecular hydrogen mass of $M_{{\rm H}_2}=\DTLfetch{NGC2976}{variable}{H2mass}{value}\times 10^{7}\,{\rm M}_{\odot}$ in NGC~2976.}

\revise{To estimate the atomic gas mass, we first convolved and regrided the HI map to our working resolution and geometry.} Then we convert the flux density of HI to the mass of HI in each pixel using $M_{\rm HI}\,[{\rm M}_{\odot}]=2.36\times 10^{5}\,D^{2}\,I_{\rm HI}\,{\rm [Jy/beam\,m/s]}$, with $D$ the distance to NGC~2976 in Mpc~\citep{Walter+2008}. 
We compute total HI mass over the whole map to be $1.4\times 10^8\,{\rm M}_{\odot}$ (consistent with the value reported in \cite{Walter+2008}). \revise{We find a total atomic hydrogen mass of $M_{\rm HI}=\DTLfetch{NGC2976}{variable}{HImass}{value}\times10^{7}\,{\rm M}_{\odot}$ in the same regions of the H$_2$ mass determination. Having computed masses of both molecular and atomic gas, the total neutral gas mass is derived in each pixel for NGC~2976.}

\begin{figure}[ht]
	\centering
	\includegraphics[width=0.3\textwidth]{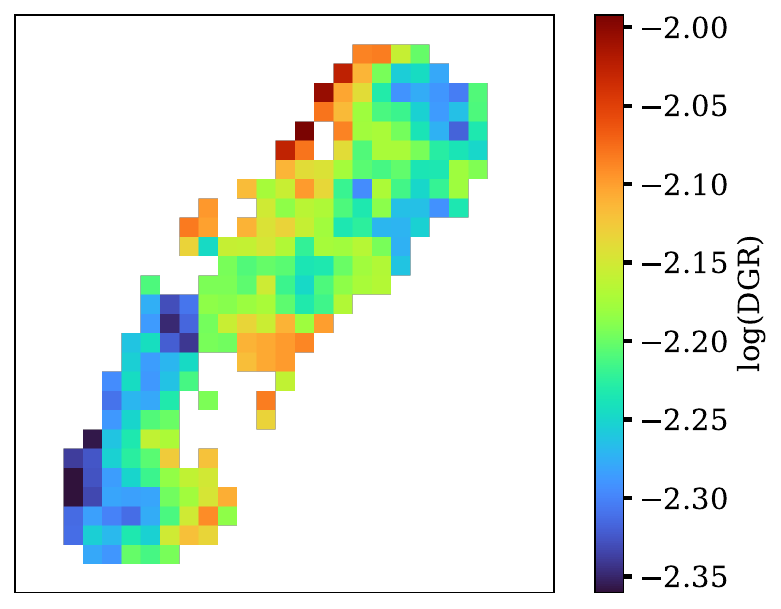}
	\caption{\revise{ Map of the Dust-to-Gas mass Ratio (DGR) in logarithmic scale, $log(M_{\text{dust}}/M_{\text{gas}})$ at $>3\sigma_{\rm rms}$ level at 18\arcsec angular resolution in NGC~2976.}}
	\label{plot-2976DGR}
\end{figure}

Knowing the gas and dust mass per pixel, we create a map of the Dust-to-Gas Ratio (DGR) as shown in Fig.~\ref{plot-2976DGR}. \revise{DGR varies from 0.004 to 0.009 in NGC~2976. On average, DGR is $ \DTLfetch{NGC2976}{variable}{DGRmean}{value} \pm \DTLfetch{NGC2976}{variable}{DGRstd}{value}$ (error is the standard deviation). Integrating the dust mass in the same way as for the gas mass explained above and dividing them by each other, we find a global value of DGR equal to \DTLfetch{NGC2976}{variable}{globalDGR}{value}, which equals the mean DGR. These results 
 agree with the DGR values reported by \cite{Sandstrom+2013} for NGC~2976 within error margins.} 

\subsubsection{Molecular and atomic gas fractions}
\begin{figure*}
 \centering
	\includegraphics[width=0.75\textwidth]{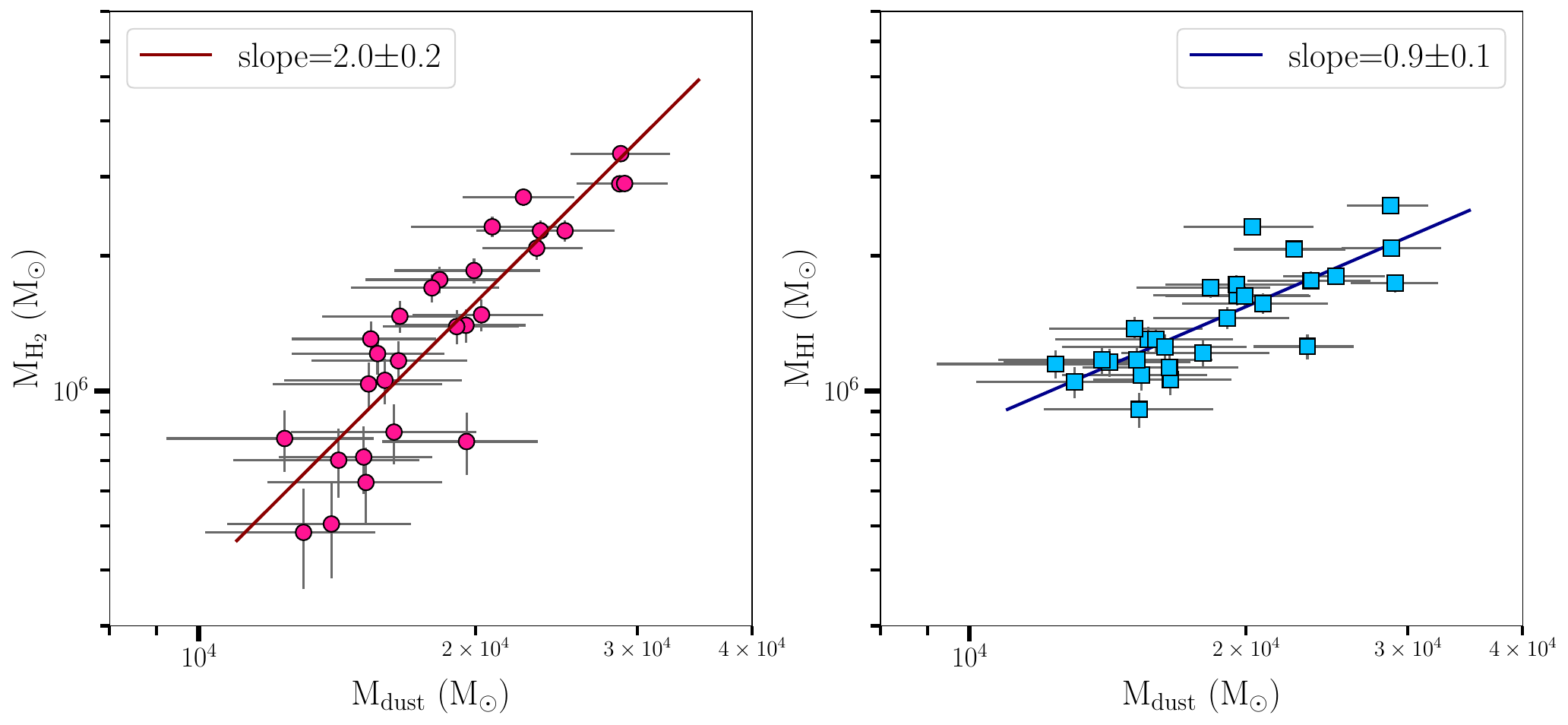}
	\caption{Relation of atomic gas mass $M_{\rm HI}$ (\textit{blue}) and molecular gas mass $M_{{\rm H}_2}$ (\textit{pink}) with dust mass $M_{\rm{dust}}$ in NGC~2976. A $3\sigma_{\rm rms}$ cutoff is applied on the included pixels. \revise{Lines show Bisector-OLS fits with slopes indicated as legends (see also Table \ref{table-gascorrelations})}.} \label{plot-gasdustmass1}
\end{figure*}

\begin{figure}
 \centering
	\includegraphics[width=0.45\textwidth]{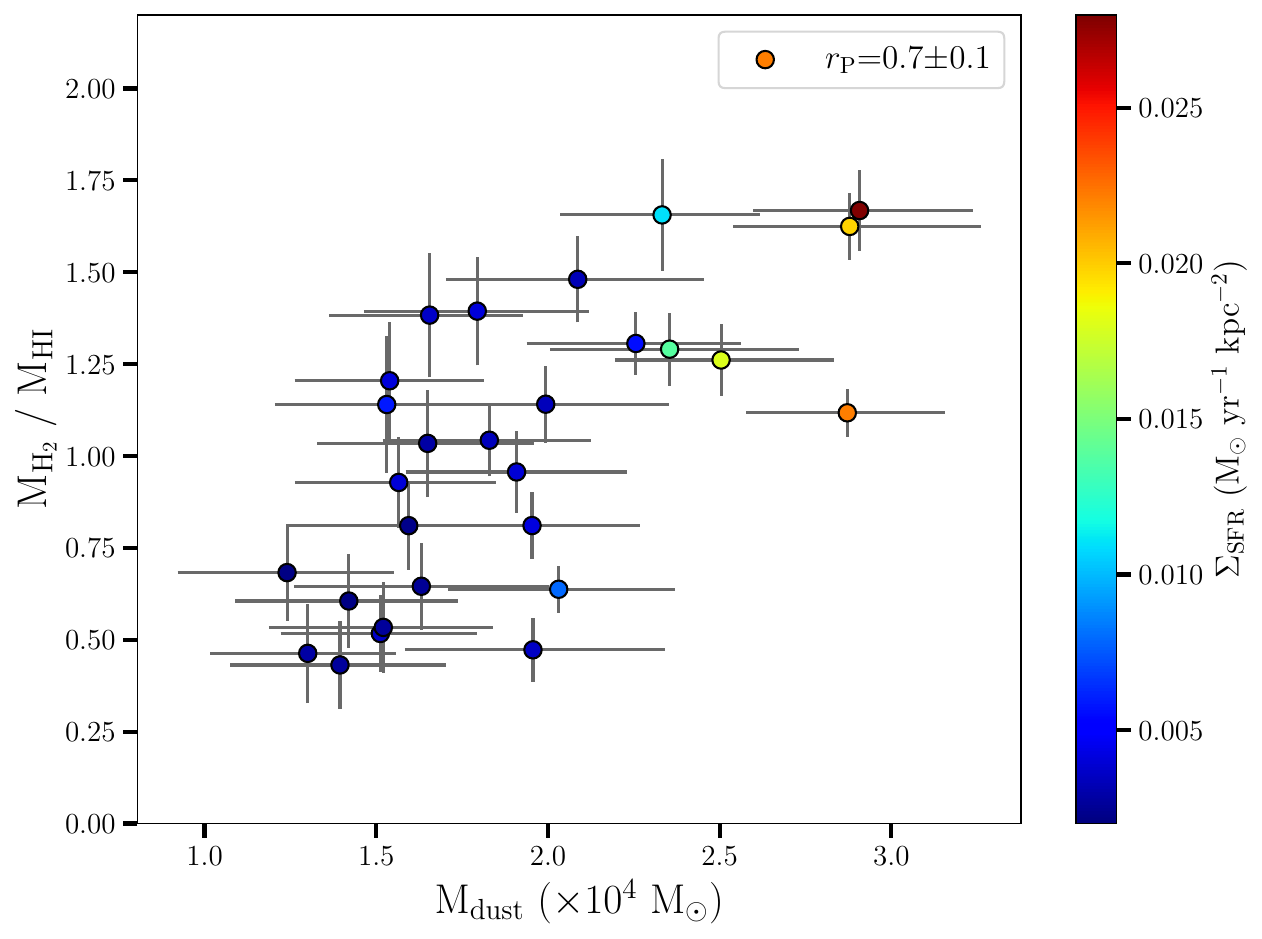}
	\caption{\revise{ Ratio of molecular to atomic gas mass, $M_{\rm H_2}/M_{\rm HI}$, vs dust mass, ${\rm M_{\rm{dust}}}$, color-coded with the SFR surface density in NGC~2976.}} \label{plot-gasdustmass2}
\end{figure}

\revise{In this section, we investigate the relations between the different gas components and dust. The analysis uses independent pixels of beam size ($>3\sigma_{\rm rms}$) to avoid internal correlation between the data points.
Fig.~\ref{plot-gasdustmass1} shows that the mass of both atomic $M_{\rm HI}$ and molecular $M_{\rm H_2}$ gas increase as the total (cold + warm) dust mass $M_{\rm dust}$ increases. However, $M_{\rm HI}$ increases with a smaller slope in the log-log plane (slope${\sim}$1) than $M_{\rm H_2}$ (slope${\sim}$2).} Fig.~\ref{plot-gasdustmass1} shows that as we move from regions with low dust masses to regions with high dust masses, $M_{\rm HI}$ increases by a factor of about 3, while $M_{{\rm H}_2}$ increases by a factor of approximately 10, which is three times larger. It is inferred from Fig.~\ref{plot-gasdustmass1} that although gas is mixed with dust in the ISM of galaxies and hence shows a positive correlation, different gas components have different relations with dust. \revise{Therefore, moving from low dust mass regions to high dust mass regions, the amount of molecular gas increases faster with the amount of dust.} 
Fig.~\ref{plot-gasdustmass2} shows this relation more evidently, in which the ratio of molecular gas to atomic gas, $M_{\rm H_2}/M_{\rm HI}$, is plotted against the dust mass, ${\rm M_{\rm dust}}$. $\Sigma_{\rm SFR}$, is shown as the color scale. 
\revise{At the bottom-left of the plot, where dust masses are minimal, atomic gas mass is almost twice the molecular gas mass. However, moving from low dust mass regions to high dust mass areas in the top-right corner of the plot, the ratio $M_{\rm H_2}/M_{\rm HI}$ increases from 0.5 to more than 1.5. This trend is accompanied by an increase in star formation activity. This indicates that in regions characterized by high dust mass and intense star-forming activity, the contribution from molecular gas is almost twice that of atomic gas. }

\subsubsection{Formation and dissociation of H$_{2}$}
\cite{Reach&Boulanger1998} proposed that in relatively cold and dense phases of atomic gas, where the formation and destruction of ${\rm H}_2$ molecules are balanced, the column density of ${\rm H}_2$ should change with the column density of HI to a power of 2. \cite{Nieten2006} and \cite{Tabatabaei2010} found such a relationship in M31 suggesting that a balance in the formation and dissociation of hydrogen molecules holds on the average in the ISM of this galaxy. 
Investigating this in NGC~2976,
we find that H$_2$ is proportional to HI to a power smaller than 2 (\DTLfetch{NGC2976}{variable}{HIH2_a}{value}$\pm$\DTLfetch{NGC2976}{variable}{HIH2_ae}{value}, see Table~\ref{table-gascorrelations}). \revise{This indicates that the formation and destruction of hydrogen molecules are imbalanced in NGC~2976. This refers to a different ISM condition than that found in M~31 because of which a relatively cold and dense phase of atomic gas in balance with H$_2$ is not easily maintained. This agrees with other studies showing that, in dwarf galaxies, the fraction of the H$_2$ mass fraction, which is sensitive to the dust-to-gas ratio and the strength of the interstellar radiation field is out of chemical equilibrium \citep{Hu} and the formation of hydrogen molecules is less efficient than in dustier spirals~\citep{Hirashita+2017, Hunter+2024}.}\par

\begin{table}
	\begin{center}
		\begin{tabular}{l l l l l}
			\hline
			X & Y & a & b & $r_{P}$\\
			\hline
			$M_{dust}$ & $M_{{\rm H}_2}$ 
 & \DTLfetch{NGC2976}{variable}{dustH2_a}{value}
 $\pm$ 
 \DTLfetch{NGC2976}{variable}{dustH2_ae}{value} 
 &	\DTLfetch{NGC2976}{variable}{dustH2_b}{value} 
 $\pm$ 
 \DTLfetch{NGC2976}{variable}{dustH2_be}{value} 
 &	\DTLfetch{NGC2976}{variable}{dustH2_r}{value} 
 $\pm$ 
 \DTLfetch{NGC2976}{variable}{dustH2_re}{value} \\
			$M_{dust}$ & $M_{\rm HI}$ 
 & \DTLfetch{NGC2976}{variable}{dustHI_a}{value} 
 $\pm$ 
 \DTLfetch{NGC2976}{variable}{dustHI_ae}{value} 
 & \DTLfetch{NGC2976}{variable}{dustHI_b}{value}
 $\pm$ 
 \DTLfetch{NGC2976}{variable}{dustHI_be}{value} 
 & \DTLfetch{NGC2976}{variable}{dustHI_r}{value} 
 $\pm$ 
 \DTLfetch{NGC2976}{variable}{dustHI_re}{value} \\
			$M_{\rm HI}$ & $M_{{\rm H}_2}$ 
 & \DTLfetch{NGC2976}{variable}{HIH2_a}{value} 
 $\pm$ 
 \DTLfetch{NGC2976}{variable}{HIH2_ae}{value} 
 & \DTLfetch{NGC2976}{variable}{HIH2_b}{value} 
 $\pm$ 
 \DTLfetch{NGC2976}{variable}{HIH2_be}{value} 
 & \DTLfetch{NGC2976}{variable}{HIH2_r}{value} 
 $\pm$ 
 \DTLfetch{NGC2976}{variable}{HIH2_re}{value} \\
			\hline
		\end{tabular}
		\caption{Relation between different gas components and dust, considering $logY=alogX+b$ with $r_{P}$ the Pearson correlation coefficient.}
		\label{table-gascorrelations}
	\end{center}
\end{table}

\section{Summary} \label{section-summary}
In this work, we present the first NIKA2 millimeter observations of two nearby galaxies, NGC~2146 (starburst spiral) and NGC~2976 (peculiar dwarf), at 1.15\,mm and 2\,mm as part of the Guaranteed Time Large Project IMEGIN. These observations provide robust resolved information about the dust physical properties by constraining their FIR-radio SED in the millimeter domain. After subtracting the contribution from the CO line emission, the SEDs are modeled both globally and locally within galaxies using a Bayesian approach. \revise{We utilized a double-component MBB model with a variable $\beta_{c}$ in the global analysis while keeping it constant for the resolved analysis, as a first-order approximation. This resulted in generating cold and warm dust mass, temperature, and free-free fraction maps of these galaxies, allowing us to decompose the components of the millimeter emission and study the relations between the properties of dust and gas.} We summarize our main findings below.

\begin{itemize}
\item \revise{The millimeter emission at 1.15\,mm is contaminated by the CO(2-1) line emission by $\DTLfetch{NGC2146}{variable}{COcontr_global}{value}\%$ ($\DTLfetch{NGC2976}{variable}{COcontr_global}{value}\%$) in NGC~2146 (NGC~2976) on average with a maximum variation of $\sim$30\% ($\sim$10\%) across these galaxies.}

\item \revise{Studying the global IR to radio SEDs, we find that a dust component with a temperature of $\sim$32\,K constitutes more than 99\% of total dust mass in NGC~2146. A similar portion of the dust is found at lower temperatures of $\sim$27\,K prevailing the total dust in NGC~2976. 
With a total dust mass lower than that of NGC~2146 by 2 orders of magnitude, NGC~2976 has a dust emissivity index $\beta_c\simeq 1.3$ which is flatter than that of NGC~2146 $\beta_c\simeq 1.7$ as expected for a low-metallicity dwarf galaxy. }

\item \revise{ The global millimeter emission of these galaxies is dominated by dust ($\sim88\%$) at 1.15\,mm. At 2\,mm, this contribution decreases to less than 50\% in NGC~2146, while it remains almost unchanged in NGC~2976.}

\item \revise{The pixel-by-pixel SED analysis shows that dust temperature is generally higher in the inner disc in NGC~2146 and star-forming regions in NGC~2976. Cold and warm dust components exhibit similar distributions in both galaxies, although the warm (minor) component shows a larger fluctuation. }

\item \revise{By computing the total gas content of NGC~2976, we present a map of DGR in this galaxy. DGR ranges between 0.004 and 0.1, while its mean value is $ \DTLfetch{NGC2976}{variable}{DGRmean}{value} \pm \DTLfetch{NGC2976}{variable}{DGRstd}{value}$. We show that the contribution from molecular gas is almost twice that of atomic gas in regions characterized by high dust mass and intense star-forming activity. In addition, no global balance holds between the formation and dissociation of H$_2$ in this dwarf galaxy.}
\item \revise{Tight but super-linear correlations hold between $\Sigma_{\rm SFR}$ and the millimeter emission in general. Similarly, $\Sigma_{\rm H_2}$ is well correlated with millimeter emission in both galaxies. As $\Sigma_{\rm SFR}-$RC correlation is almost linear, the super-linear correlation of $\Sigma_{\rm SFR}$ with the millimeter emission is linked to its super-linear correlation with dust. A diffuse distribution of the mm dust emission can explain the super-linearity of its correlation with $\Sigma_{\rm SFR}$.} 
\end{itemize}

\revise{Finally, we stress that the different results obtained for NGC2146 (starburst) and NGC2976 (peculiar dwarf) can be linked to their very different natures, stellar mass, SFR, and ISM properties. Similar studies in normal and intermediate-mass spirals will shed light on this in the forthcoming IMEGIN papers.}

\begin{acknowledgements} 
 We thank the IRAM staff for their support during the observation campaigns. The NIKA2 dilution cryostat has been designed and built at the Institut N\'eel. In particular, we acknowledge the crucial contribution of the Cryogenics Group, in particular Gregory Garde, Henri Rodenas, Jean-Paul Leggeri, and Philippe Camus. The Foundation Nanoscience Grenoble and the LabEx FOCUS ANR-11-LABX-0013 have partially funded this work. This work is supported by the French National Research Agency under the contracts "MKIDS", "NIKA" and ANR-15-CE31-0017 and in the framework of the "Investissements d'avenir" program (ANR-15-IDEX-02). This work is supported by the Programme National Physique et Chimie du Milieu Interstellaire (PCMI) and the Programme National Cosmology et Galaxies (PNCG) of the CNRS/INSU with INC/INP co-funded by CEA and CNES. This work has benefited from the support of the European Research Council Advanced Grant ORISTARS under the European Union's Seventh Framework Programme (Grant Agreement no. 291294). A. R. acknowledges financial support from the Italian Ministry of University and Research - Project Proposal CIR01$\_00010$. S. K. acknowledges support from the Hellenic Foundation for Research and Innovation (HFRI) under the 3rd Call for HFRI PhD Fellowships (Fellowship Number: 5357). M.B., A.N., and S.C.M. acknowledge support from the Flemish Fund for Scientific Research (FWO-Vlaanderen, research project G0C4723N). MME acknowledges the support of the French Agence Nationale de la Recherche (ANR), under grant ANR-22-CE31-0010. This work made use of HERACLES, The HERA CO-Line Extragalactic Survey ~\citep{Leroy+2009}. DustPedia is a collaborative focused research project supported by the European Union under the Seventh Framework Programme (2007-2013) call (proposal no. 606847). The participating institutions are Cardiff University, UK; National Observatory of Athens, Greece; Ghent University, Belgium; Universit\'e Paris Sud, France; National Institute for Astrophysics, Italy and CEA, France.
\end{acknowledgements}

\bibliography{allref.bib}

\begin{appendix}
	
\section{Data processing}
\label{appendix-dataprocessing}
\subsection{Re-calibration of detector gains and beam solid angles}
\label{appendix-gains_beams}
The flux calibration is applied through the NIKA2 pipeline according to the analysis presented in~\cite{Perotto_2020} which is strictly valid only for point sources. In addition, the response of individual detectors to extended emission is corrected, by using the atmospheric emission in the raw data of the science observations themselves as a template of very extended emission. This is possible whenever the scan duration is longer than about 5 min and the sources (to be masked in the process) occupy a small fraction of the map. Both conditions are met in the IMEGIN observations.
The gains are very stable on the timescale of an observing pool.

All Uranus beam maps (large maps in the azimuth direction, with optimal redundancy provided by a fine sampling in elevation) acquired within 19 runs from October 2017 to January 2023, spanning all IMEGIN observations, were reduced and analyzed to monitor variations of the beam solid angle during that period. \revise{With an aperture size of 2.5 FWHM (2\,mm), corresponding to twice the half-power beam width (HPBW) in maps of both bands convolved to the same angular resolution (a compromise for maps containing a mixture of compact sources and diffuse emission), and excluding the outlier run 50 (December 2019), we found effective beam solid angles
$ \Omega_{\rm eff}(1.15\,{\rm mm}) = (188 \pm 11)\,{\rm arcsec}^2$
and $\Omega_{\rm eff}(2\,{\rm mm}) = (381 \pm 11)\, {\rm arcsec}^2$, corresponding to rms flux calibration uncertainties of 5.9\% at 1.15\,mm and 2.9\% at 2\,mm. The rms uncertainty is in line with that~\cite{Perotto_2020} derived from three reference runs 
at both wavelengths.}

\subsection{Processing of NIKA2 data}
The NIKA2 data were processed with the \textit{Scanam\_nika} software, adapted from \textit{Scanamorphos} originally developed for \textit{Herschel} on-the-fly imaging data. The underlying principle is to fully exploit the redundancy built in the observations (each sky position within the nominal-coverage area being sampled by multiple detectors at multiple times, and in variable atmospheric conditions) while making minimal assumptions about the atmospheric and instrumental noise, as described by~\cite{Roussel_2013} and~\cite{Roussel+2020}. The low-frequency noise, or drift, is decomposed into the following components:
\begin{itemize}
\item the average drift time series averaged over all valid detectors, each one weighted by its inverse variance at the highest frequencies;
\item the drift per electronic box time series, averaged over all valid detectors of each of the 4 (2\,mm) or 16 (1.15\,mm) electronic boxes; and the equivalent component obtained by artificially slicing the detector arrays into the same number of boxes, but orthogonal to the physical boxes;
\item the individual noise time series (one function of time for each detector);
\item the noise series not as a function of time, but binned per position on the sky along the array trajectory, projected on the long axis of the electronic boxes; and the equivalent component projected on the short axis of the electronic boxes.
\end{itemize}
The latter components were introduced to remove a significant residual noise remaining after subtraction of the other correlated noise components and found empirically by the resemblance of their spatial pattern in the map with that of standing waves. They correspond to a genuine electronic noise best seen in instrumental coordinates, arising in the cryostat (internal NIKA2 communication from F.X. D\'esert and N. Ponthieu, March 2023).\par
The minimum timescale of the drifts, $t_c$, is determined by the stability length and the scanning speed (see~\cite{Roussel_2013}). The average drift is subtracted on this timescale, while the other drift components are subtracted on successively smaller or larger timescales, multiples of $t_c$.\par
The process is iterative and uses a source mask that is automatically updated at each iteration, as the sources gradually emerge from the decreasing residual noise. The mask is always constrained to be confined within a $5\sigma_{\rm rms}$ isophote in the SPIRE 500\,$\mu$m map, to avoid contamination by spurious structures created by noise.\par
At the end of the processing, the median surface brightness of the sky and $\sigma_{\rm rms}$ are calculated in the area of the field with optimal coverage, excluding the pixels contained within the final source mask. The sky level is then subtracted from the map.

\subsection{Simulations for homogeneous dust SEDs from instruments with dissimilar transfer functions}
\label{appendix-simulations}
\begin{figure}
\centering
	\includegraphics[width=0.5\textwidth]{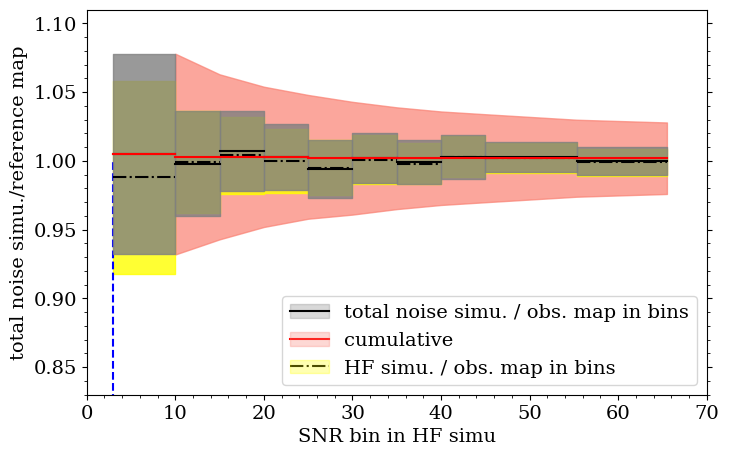}
	\includegraphics[width=0.5\textwidth]{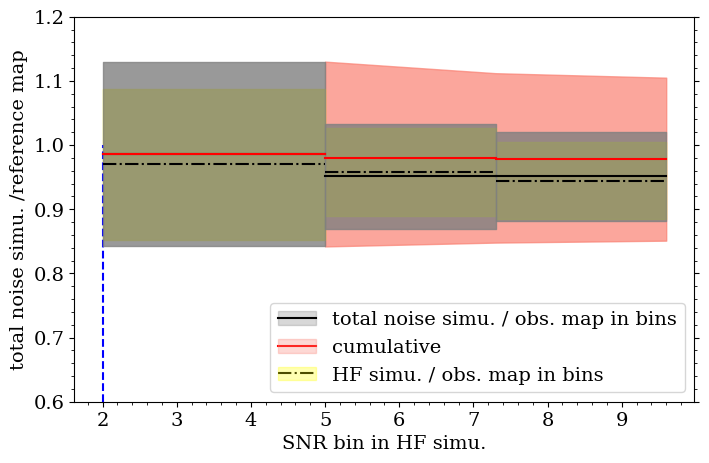}
	\caption{Transfer functions at 250\,$\mu$m (with a FWHM of 18$^{\prime\prime}$) binned as a function of signal-to-noise ratio (SNR) where the noise standard deviation $\sigma_{\rm HF}$ is from a simulation including only white and high-frequency noise. The red curve and red shaded area show the cumulative fraction of the flux that is recovered and the associated uncertainties, from the lower SNR limit ($3 \sigma_{\rm HF}$ for NGC\,2146 and $2 \sigma_{\rm HF}$ for NGC\,2976) up to the maximum of the current SNR bin.} \label{fig:transferfunctions}
\end{figure}

The astrophysical signal at 1.15 and 2\,mm is filtered in a complex way, depending on the atmospheric conditions and stability at the time of observations, source brightness and geometry, observation parameters such as scanning speed, scan geometry, and the processing algorithm.
\reviseII{In general, sources bigger than the instantaneous FoV of NIKA2 (6.5$^{\prime}$) as well as regions with low signal-to-noise ratios are filtered due to those effects. By contrast, all spatial scales smaller than the map size can be accurately recovered with the \textit{Herschel} and \textit{Spitzer} space observatories.
Hence, to build coherent spatially-resolved dust SEDs traced from 24\,$\mu$m to 2\,mm, 
this discrepancy has to be mitigated: the \textit{Herschel} and \textit{Spitzer} maps have to be filtered in the same way as the NIKA2 maps. The ratio of the sky map that is obtained to the map that would be obtained by a perfect instrument outside the atmosphere is hereafter called the transfer function.} 
We chose to apply the transfer function of NIKA2 at 2\,mm (assuming that it is equivalent to that at 1.15\,mm, since the observations in both bands are simultaneous) to all observations of dust from space.
For a given \textit{Herschel}/\textit{Spitzer} band, the process is the following:
\begin{itemize}
\item convolve the map with a Gaussian kernel to bring it approximately to the same FWHM as that of the 2\,mm map (except at 350 and 500\,$\mu$m);
\item \revise{correct for any small astrometric offset between the maps, by fitting the brightest compact source with a Gaussian for each band;}
\item rescale the map so that the ratio of the MIPS or \textit{Herschel} band to the 2\,mm band has a median of 1 within the galaxy extent;
\item simulate time series from the map as if it had been observed with NIKA2 with the same scan geometry and duration;
\item add to these time series the low-frequency and high-frequency noise that was extracted from the 2\,mm data (drifts, glitches, etc.) and a realization of the white noise present in the 2\,mm data;
\item process these time series in the same way as the NIKA2 data and project the result on the same spatial grid;
\item rescale the map back to its original unit.
\end{itemize}
The comparison of the simulated output map with the initial map directly gives the spatially resolved transfer function, and it can be checked that it is very similar for all \textit{Spitzer} and \textit{Herschel} bands.
In addition, for these small galaxies, it is very close to unity in all signal-to-noise ratio bins, as shown in Figure~\ref{fig:transferfunctions}. 

\reviseII{The main benefit of applying the NIKA2 transfer function to the other bands, instead of using the original maps, is then to homogenize the noise properties across the dust SED,
so that the parameters extracted from the SED fits are as unbiased as possible.} 
{The NIKA2 transfer function was not applied to the UV, optical, and gas maps, because these data are not used in SED fitting. Therefore, it is irrelevant that their noise properties are inhomogeneous, since for these two galaxies, the only significant effect of applying the NIKA2 transfer function is to alter the noise properties, and not to alter the flux.}

\section{ CO contamination} \label{appendix-CO}
The emission measured at 1.15\,mm with NIKA2 consists mainly of continuum emission and partly of the CO(2-1) line emission. This is caused by the fact that the NIKA2 passband includes the CO (2-1) emission line at 230.5\,GHz. In this section, we explain how we compute the amount of this contamination and subtract it from the total observed continuum emission to obtain pure dust emission in the 1.15\,mm NIKA2 band for the SED modeling.\par
The CO(2-1) line intensity is measured in units of the main beam temperature, while the continuum flux density is measured in units of mJy/beam. We follow the procedure explained in~\cite{Drabek_2012} to convert from K\,km/s to mJy/beam
\begin{equation}
\frac{C}{\text{mJy beam}^{-1} \text{per K\,km\,s}^{-1}}=\frac{F_{\nu}}{\int T_{\text{MB}} \,dv} ,
\end{equation}
where $F_{\nu}$ is the line flux density and $I_{CO}=\int T_{\text{MB}} \,dv$ is velocity integrated main beam temperature of the CO(2-1) line emission in units $\text{K\,km s}^{-1}$, and $C$ is the conversion factor. This is equivalent to
\begin{equation}
\frac{C}{\text{mJy beam}^{-1} \text{per K\,km s}^{-1}}=\frac{2k\nu ^{3}}{c^{3}}\frac{g_{\nu}(\text{line})}{\int g_{\nu} \, d\nu}\Omega_{\text{B}}I_{CO}.
\label{COfactor}
\end{equation}
In the above equation, $k$ is Boltzmann's constant, $\nu$ is the line frequency, $\Omega_{\text{B}}$ is the solid angle corresponding to NIKA2 beam area, and $g_\nu(\text{line})$ is the transmission function at the molecular line frequency and $\int g_{\nu} \, d\nu$ is the integral of the transmission function of the NIKA2 1.15\,mm detector over the whole range of filter frequencies,~\cite{Perotto_2020}. We find the value of $\frac{g_{\nu}(\text{line})}{\int g_{\nu} \, d\nu}$ by integrating on the NIKA2 1.15\,mm pass-band and find it to be $7.85 \times 10^{-3}\,\text{GHz}^{-1}$. Also, the beam area corresponding to our working resolution of 18$^{\prime\prime}$ is $8.63 \times 10^{-9}$ steradians. We find the conversion factor $C$ to be equal to 0.085\,mJy/beam per K\,km/s for NIKA2 1.15\,mm pass-band.\par
The Planck pass-band at 1.38mm is also affected by line emission contamination ~\citep{Planck2014}. We follow a similar procedure to subtract CO(2-1) contamination from the Planck 1.38\,mm map before including it in the SED fitting process. Using equation \ref{COfactor} and taking into account Planck's pass-band and resolution (301$^{\prime\prime}$), we find $C=50.6$\,mJy/beam per K\,km/s. \par

\section{Error estimation}\label{appendix-errorestimation}
The measured integrated flux in each waveband can be subject to different stochastic or systematic uncertainties. We estimate the stochastic error by measuring the rms noise level, $\sigma_{\rm rms}$, of the maps. The calibration uncertainty of the observations (see Sect.~\ref{table-alldata}) is used to estimate the systematic error. The total error is then
\begin{equation}\label{errorformula}
	\delta_{er} =\sqrt{\delta_{cal}^{2}+\delta_{\rm rms}^{2}}.
\end{equation} 
For the integrated flux densities, $\delta_{\rm rms}$ is given by 
\begin{equation}
	\delta_{\rm rms}=\sigma_{\rm rms}\sqrt{N_{beam}}=\sigma_{\rm rms} \frac{a}{\theta}\sqrt{\frac{N}{1.133}},
\end{equation}
where $\sigma_{\rm rms}$ is the rms noise level in Jy/beam, $N_{beam}$ and $N$ is the number of beams and number of pixels in the aperture, $\theta$ the angular resolution and $a$ the pixel size, both in units of arcsec$^2$. Note that because the maps have different resolutions and pixel sizes, $N$, $\theta$, and $a$ vary for each map (see Table~\ref{table-alldata}).\par

\section{Model fitting using Bayesian approach} \label{appendix-MCMC}
To model the SED, we fit a model with $n_{dim}$ free parameters to $n_{data}$ data points. \revise{For the two galaxies, we have studied, in the global case $n_{dim}=8$ and $n_{data}=14$, and in the resolved case $n_{dim}=6$ and $n_{data}=9$.} In the context of the Bayesian approach, we look for the Probability Distribution Function (PDF) of the posterior $p(x|d)$, which is the PDF of the free parameters, conditional on the data points. The Markov Chain Monte-Carlo (MCMC) method is an algorithm that can sample a PDF -here the posterior. To implement the MCMC method to sample the posterior distribution, we use the \textit{emcee} Python package~\citep{Foreman_Mackey_2013}. This package uses a pre-specified number of random walkers ($n_{\text{walkers}}$) in the parameter space. The parameter space has $n_{dim}$ dimensions and is defined by the prior PDFs. The random walkers explore the parameter space. In each step, the likelihood PDF is computed, which is a function of the absolute difference between the model and the data points at different wavelengths. After the random walkers move a total of $n_{steps}$ steps, we have an array of length $n_{steps}$ of the values that each random walker has acquired in each step. This array should be refined in two ways. Firstly, the initial $n_{burn}$ steps are deleted, because they incorporate the initial steps that the random walkers are not close to the best value. Second, this array is filtered by every $n_{thin}$ number to resolve the autocorrelation between the steps. We chose $n_{thin}$ such that the length of the time series produced by random walkers is at least 40 times the autocorrelation function of the time series. \revise{We found the value ranges $n_{walkers}=40$, $n_{steps}=60000-100000$, and $n_{thin}=200-1000$ are appropriate for our problem.} We assume flat prior PDFs for all free parameters. \revise{Also, to be able to sweep all the parameter space, we set the boundaries of our free parameters to be $5K<T_{c}<35\,K$, $35<T_{w}<95$, $10^{1}\,{\rm M}_{\odot}<M<10^{8}\,{\rm M}_{\odot}$, $0.5<\beta<3.0$, $-0.01<A_{1}<0.1$, $-0.001<A_{2}<1$, and $0.2<\alpha_{\rm n}<2.5$.}

\begin{figure}
\centering
	\includegraphics[width=0.47\textwidth]{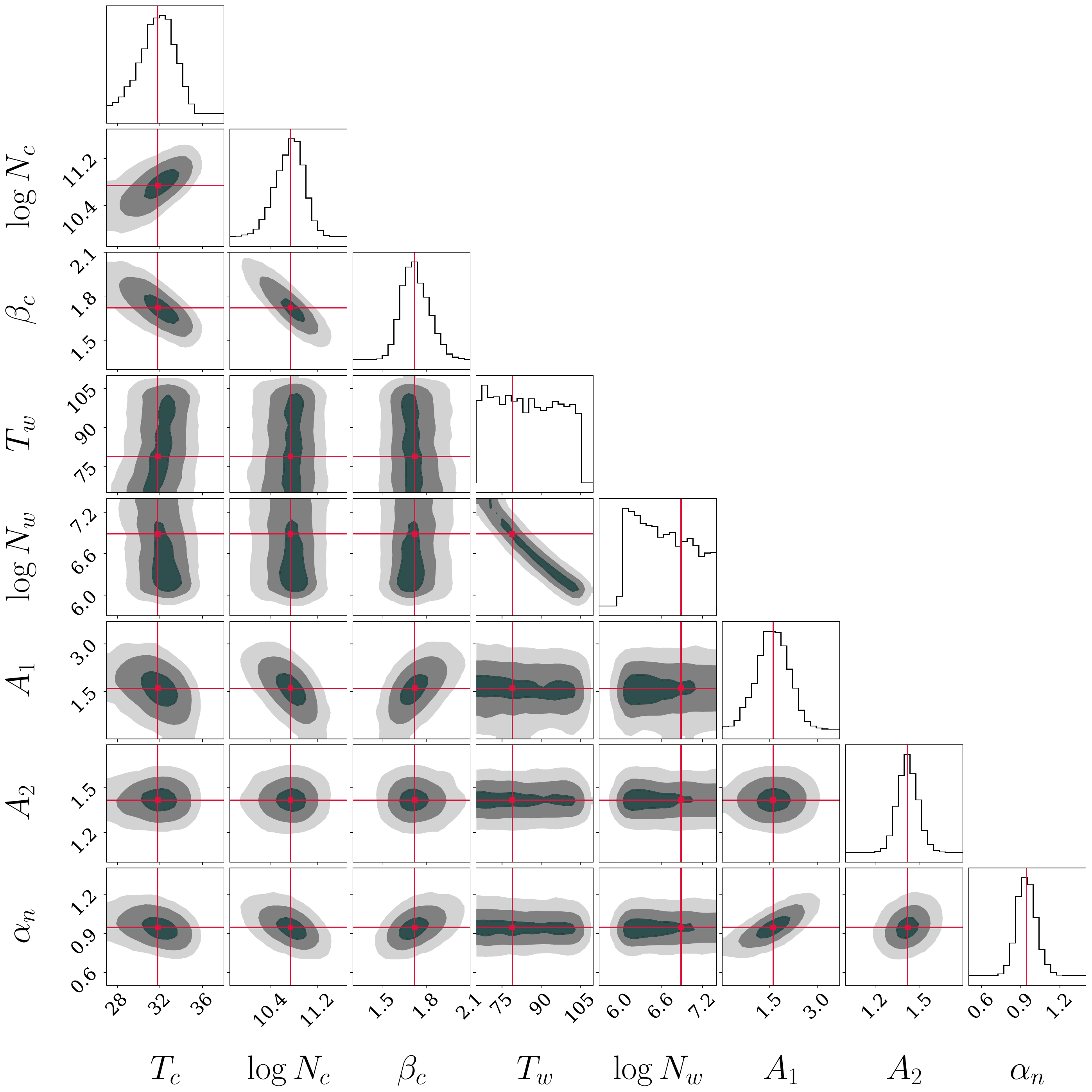}
	\includegraphics[width=0.47\textwidth]{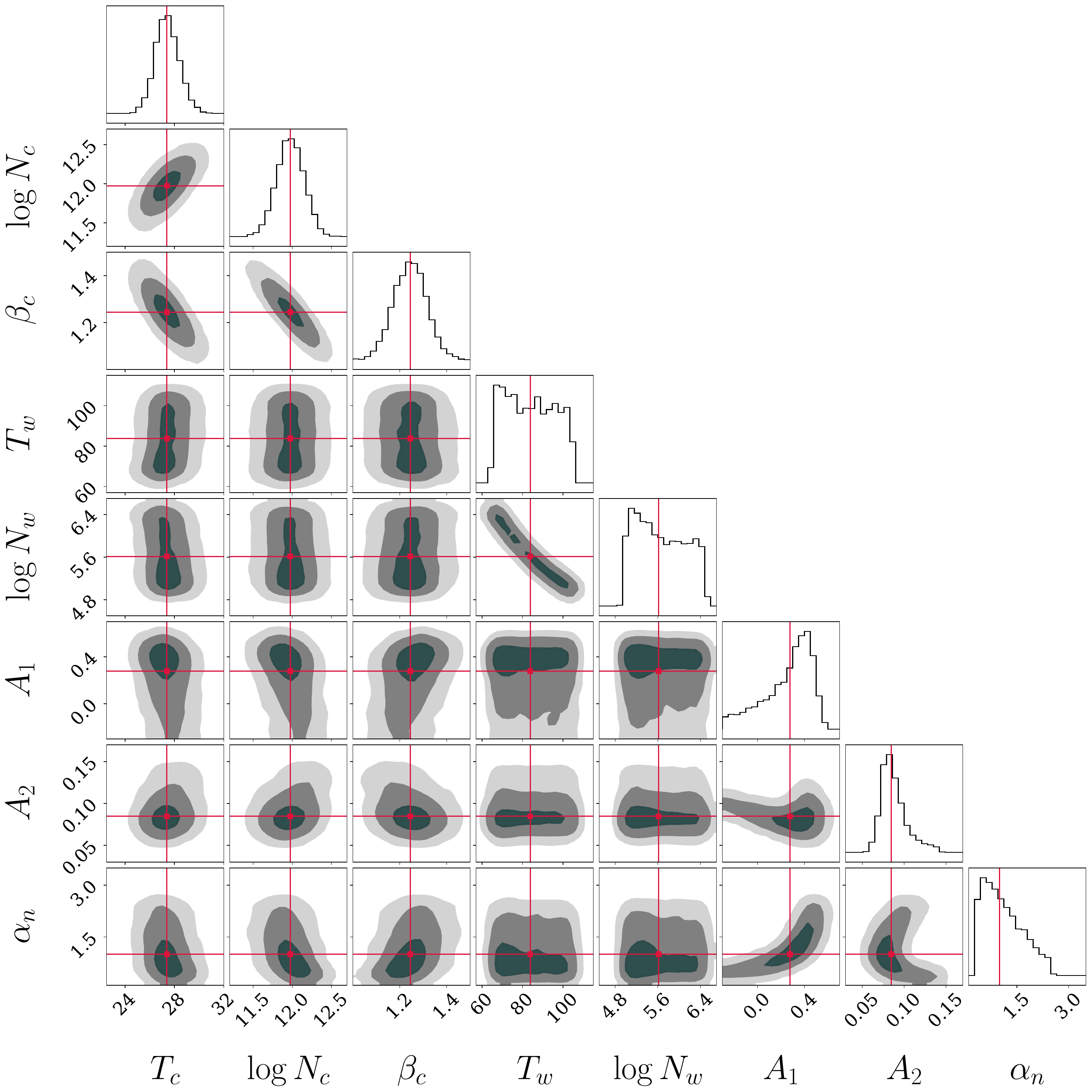}
	\caption{Corner plots of joint posterior PDFs as a result of the MCMC process, carried out for global SED modeling of (top) NGC~2146 and (bottom) NGC~2976. The median of each distribution function is shown with the red lines. Note that the axis for $A_{1}$ is in units of $10^2$ for clarity.} \label{globalcornerplots}
\end{figure}

The joint posterior PDF of random walkers moving in the parameters space is then plotted. This joint PDF is a sample of the posterior PDF of the free parameters, which can be displayed in a corner plot. We report the median value as the resulting value for each free parameter and the $20\%$ and $80\%$ percentiles as the error margins. Since the posterior distributions are not necessarily symmetrical, we report asymmetrical error margins for each parameter in Table~\ref{table-parameters}. Fig.~\ref{globalcornerplots} shows the corner plots of the global SED modeling of the two galaxies. 

Each step of the random walkers is assigned to a set of values for the $n_{dim}$ free parameters. This set of values results in a value for the model function in that step. Then prediction of this particular model value for each $n_{data}$ wavelength is compared to the actual observed data points. In other words, in each step, the prediction of the model for $n_{data}$ wavelengths is computed. To correctly compute the prediction of the model for each wavelength, the likelihood function should be convolved with the transmission function of each instrument. This is essential because we need to know the prediction of the model \textit{as observed} with each instrument at each wavelength. To have a correct prediction of each instrument in their nominal wavelengths, we implement the transmission function of all the involved instruments in the likelihood function. This process makes the core slower, and we compensate for that by implementing multiprocessing methods.\par 

\begin{figure}
 \centering
 \includegraphics[width=0.45\textwidth]{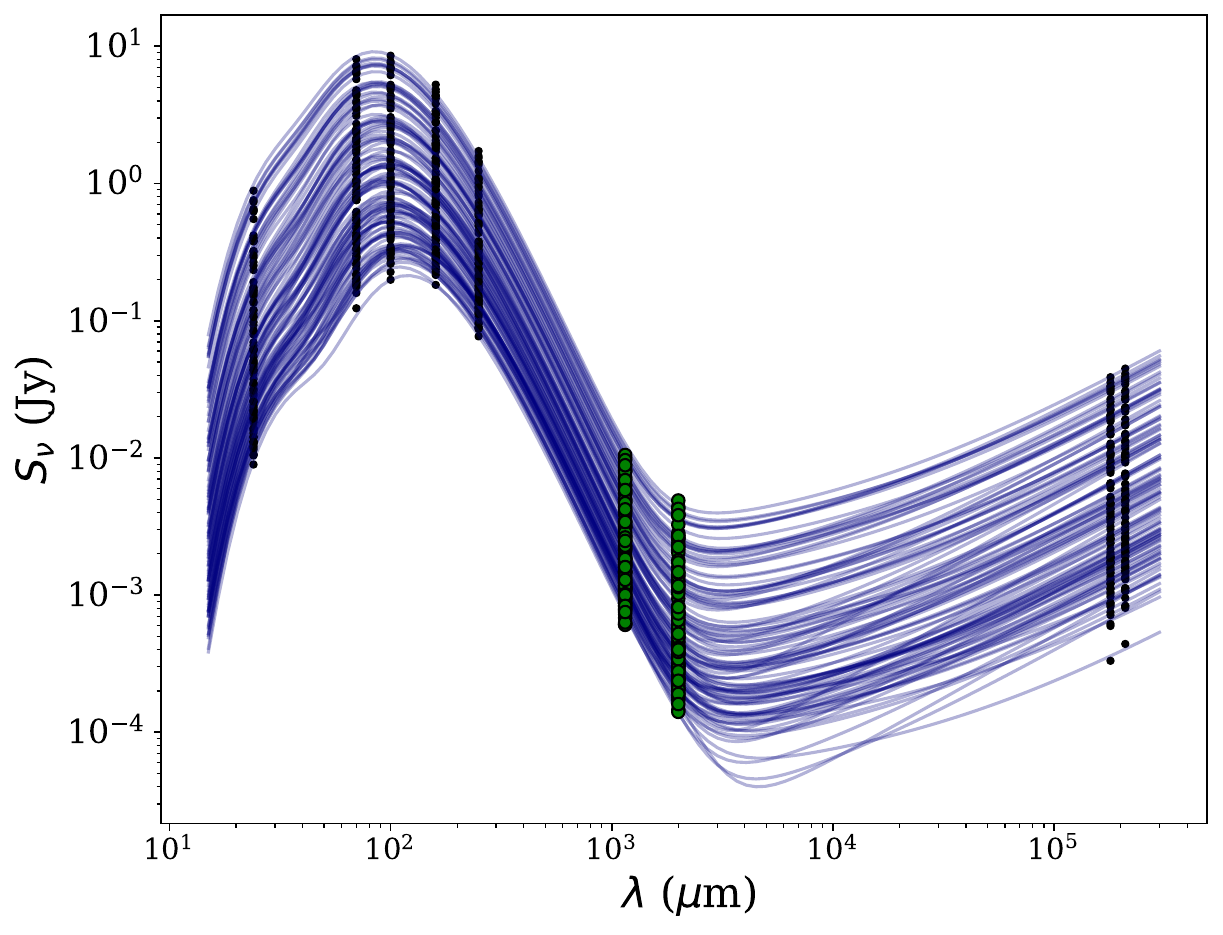}
 \includegraphics[width=0.45\textwidth]{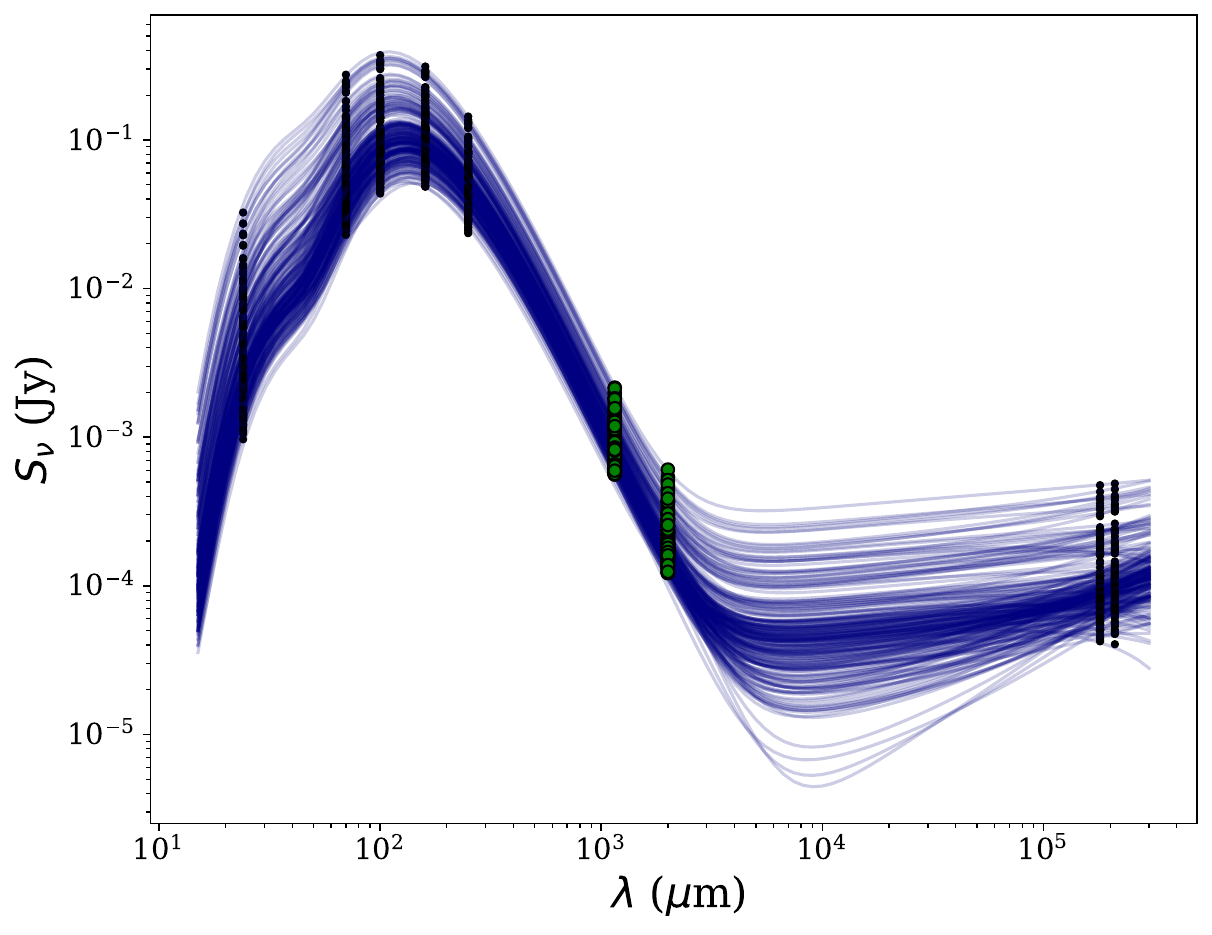}
 \caption{A compilation of all the SEDs for individual pixels (above $3\sigma_{\rm rms}$ limit) in NGC~2146 (\textit{top}) and NGC~2976 (\textit{bottom}). The green dots show NIKA2 observations.}
 \label{plot-allSEDs}
\end{figure}

Figure~\ref{plot-allSEDs} presents a compilation of all SEDs for all the pixels, which shows the significance of NIKA2 observations, represented by green dots, in resolved studies. It is evident that without NIKA2 observations, it would be impossible to constrain the trough of the SED in the millimeter regime.\par

\end{appendix}

\end{document}